\documentclass[prb,amsmath,amssymb,aps,showpacs,twocolumn]{revtex4-1}

\usepackage{graphicx}
\usepackage{dcolumn}
\usepackage{bm}

\usepackage{balance}
\usepackage{amsmath}
\usepackage{amsfonts}
\usepackage[applemac]{inputenc}
\usepackage{bbm}
\usepackage{braket}
\usepackage{float} 

\usepackage{natbib}

\begin{document}
\title{Realizing and detecting a topological insulator in the AIII symmetry class}

\author{Carlos G. Velasco}
\author{Bel\'en Paredes}

\affiliation{Arnold Sommerfeld Center for Theoretical Physics	\\		
Ludwig-Maximilians-Universit\"at M\"unchen, 80333 M\"unchen, Germany}
\affiliation{Instituto de F\'{i}sica Te\'{o}rica CSIC/UAM \\
C/Nicol\'{a}s Cabrera, 13-15 Cantoblanco, 28049 Madrid, Spain}

\begin{abstract} 
Topological insulators in the AIII symmetry class lack experimental realization.
Moreover, fractionalization in one-dimensional topological insulators has not been yet directly observed.
Our work might open possibilities for both challenges. 
We propose a one-dimensional model realizing the AIII symmetry class which can be realized in current experiments with ultracold atomic gases.
We further report on a distinctive property of topological edge modes in the AIII class: in contrast to those in the well studied BDI class, they have {\em non-zero momentum}. 
Exploiting this feature we propose a path for the detection of fractionalization. A fermion added to an AIII system splits into two halves localized at opposite momenta, which can be detected by imaging the momentum distribution.

\end{abstract}





\maketitle

In one dimension the topological or trivial character of an insulator is completely determined by the presence or absence of chiral symmetry \cite{Ryu2010, Altland1997}. Since chiral symmetry is the composition of time reversal (T) and charge conjugation (C) symmetries, two distinct classes of one-dimensional topological insulators arise: those invariant under T and C, and those breaking both symmetries. 
The first class, called the BDI symmetry class, is represented by polyacetylene \cite{Jackiw1976, Goldstone1981, Su1979, Heeger1988}, which has been the focus of major experimental and theoretical attention. 
The second class, dubbed the AIII class, has been in contrast rarely explicitly discussed. In a recent interesting connection  \cite{Sierra2014}, AIII topological insulators have been proposed to open a physical pathway to Riemann's conjecture, for one-dimensional models realizing the Rieman zeros seem to belong to the AIII symmetry class. However, AIII topological insulators lack to our knowledge experimental realization.

The topological character of one-dimensional insulators is manifested in the emergence of topologically protected zero energy modes \cite{Teo2010}, leading to the fascinating phenomena of particle number fractionalization \cite{Jackiw1976, Goldstone1981, Su1979, Kivelson2002}. This spectacular property was predicted for polyacetylene by the seminal model of Su, Schrieffer, and Heeger (the SSH model) \cite{Su1979}. At filling factor $\nu=1/2$, a fermion added to a flat background density splits into two quasiparticles with $1/2$ charge, which are spatially localized at the edges of the one-dimensional system. 
Though many rather striking properties predicted by the SSH model were confirmed in experiments with polyacetylene, 
a direct measurement of fractionalization has not been realized yet in one-dimensional topological insulators. 

The unique detection possibilities opened up recently with ultracold atoms in optical lattices \cite{Bloch2012, Bakr2009, Sherson2010} promise to allow for the direct observation of key signatures of topological states such as the Chern number \cite{Jotzu2014, Aidelsburger2015}, the Berry curvature \cite{Duca2015, Flaeschner2016} or the topological edge modes \cite{Mancini2015, Stuhl2015}. For instance, an atomic version of the SSH model has been recently realized \cite{Atala2013}.

Here, we propose a physical model that realizes a one-dimensional topological insulator in the AIII symmetry class. It can be realized in current experiments with ultracold atoms by combining a superlattice structure \cite{Atala2013} with artificial gauge fields \cite{Dalibard2011}. 
We further report on a distinctive property of topological edge modes in the AIII symmetry class. In contrast to those in the BDI class, which must necessarily have zero (or $\pi$) momentum, edge modes in the AIII class exhibit {\em non-zero momentum}. This feature is the fingerprint of the AIII class. It is a direct manifestation (in a bulk-edge correspondence) of the breaking of time reversal symmetry. Exploiting this property we propose a path for the direct observation of fractionalization in the AIII model. A fermion added to a flat background density splits into two halves with opposite {\em chirality}, which can be directly observed by imaging the momentum distribution of the system. This offers an alternative route with respect to previous proposals for the detection of topological edge modes in atomic systems, which are based on the in situ observation of the spatial density \cite{Goldman2012, Goldman2013, Goldman2013b}. 



\textit{The model}.
We consider a dimerized lattice model with two sites per unit cell:
\begin{equation}\label{eq:hamiltoniano}
H=-\sum_{n}^{N}\left(J^\prime\,\hat{a}_{n}^{\dagger}\hat{b}_{n}^{}+J e^{i\delta}\,\hat{a}_{n+1}^{\dagger}\hat{b}_{n}^{}+\text{h.c.}\right),
\end{equation}

\begin{figure}[h]
  \centering
    \includegraphics[width=0.475\textwidth]{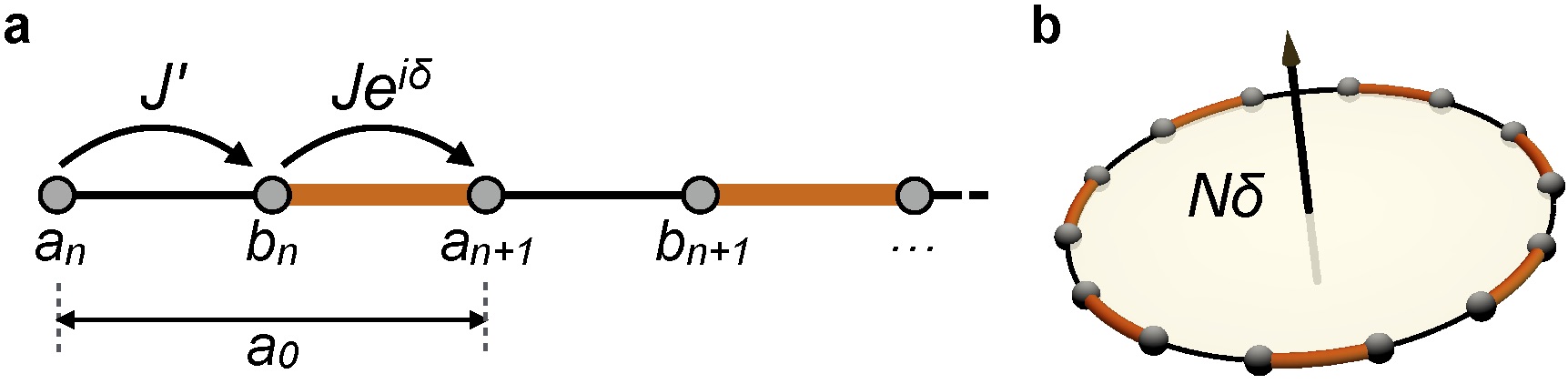}
    \caption{\textbf{The Model.} (\textbf{a}) Schematic illustration of the model for a topological insulator in the AIII class: a one-dimensional dimerized lattice model with two sites $a$ and $b$ per unit cell, and hopping amplitudes $J^\prime$ and $Je^{i\delta}$. The phase $\delta$  leads to breaking of time reversal symmetry. (\textbf{b}) For periodic boundary conditions and $N$ unit cells, a total phase $N\delta$ is acquired along the whole ring.}
    \label{fig:TheModel} 
\end{figure}

\begin{figure}[h]
  \centering
    \includegraphics[width=0.475\textwidth]{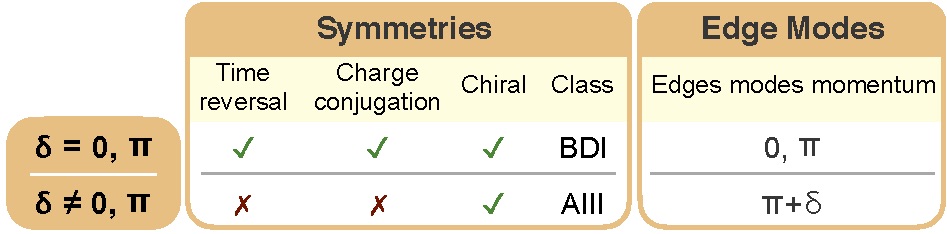}
    \caption{\textbf{Symmetries and edge modes: AIII class vs. BDI class.} The breaking (not-breaking) of time reversal symmetry is physically manifested in a non-zero (zero or $\pi$) momentum of the topological edge modes, characterizing the AIII (BDI) class.}
    \label{fig:Symmetries} 
\end{figure}

\noindent where $\hat{a}_{n}^{\dagger}(\hat{b}_{n}^{\dagger})$ are the particle creation
operators for a particle on the sublattice site $a_n(b_n)$ in
the $n$th lattice cell. For $\delta=0$ this Hamiltonian corresponds to the SSH model with tunneling amplitudes $J$ and $J^\prime$ \cite{Su1979, Delplace2011}. For $\delta \ne 0$ the model introduces a complex phase $e^{i\delta}$ that particles acquire when tunneling from one unit cell to the next [Fig.~\ref{fig:TheModel}]. In the presence of this extra phase the model breaks time reversal symmetry, entering the AIII symmetry class. This is better seen by writing the Hamiltonian (\ref{eq:hamiltoniano}) in momentum space. For periodic boundary conditions it takes the form 
$
 H=-J\sum_{k}\begin{pmatrix}
 \hat{a}_{k}^{\dagger} & \hat{b}_{k}^{\dagger}
 \end{pmatrix}
 M(k)\begin{pmatrix}
 \hat{a}_{k}\\
 \hat{b}_{k}
 \end{pmatrix},
$
with 
\begin{equation}
M(k)=\left[J'/J+\cos(k-\delta)\right]\sigma_{x}+\sin(k-\delta)\,\sigma_{y},
\end{equation}
$\sigma_{x(y,z)}$ being the Pauli matrices.  
The Hamiltonian exhibits chiral symmetry for any value of $\delta$, since $\sigma_{z}M(k)\sigma_{z}=-M(k)$. However, for $\delta\ne 0, \pi$, it is not time reversal symmetric (and thereby not charge-conjugation symmetric).
We have that:
\begin{eqnarray}
M^{*}(-k)\ne M(k),
\end{eqnarray}
since $e^{i(k+\delta)}\ne e^{i(k-\delta)}$.
Moreover, there is no $2 \times 2$ unitary transformation $U$ such that $U M^{*}(-k)U^{\dagger}=M^{}(k)$ (see Supplementary Material (SM)). 
For $\delta\ne0$ the model belongs to the AIII symmetry class [Fig.~\ref{fig:Symmetries}]. The symmetry properties of our model can be alternatively derived by realizing that it can be continuously transformed into a ladder Hamiltonian with a flux per plaquette equal to $\delta$ (see SM).
\begin{figure}[h]
  \centering
    \includegraphics[width=0.475\textwidth]{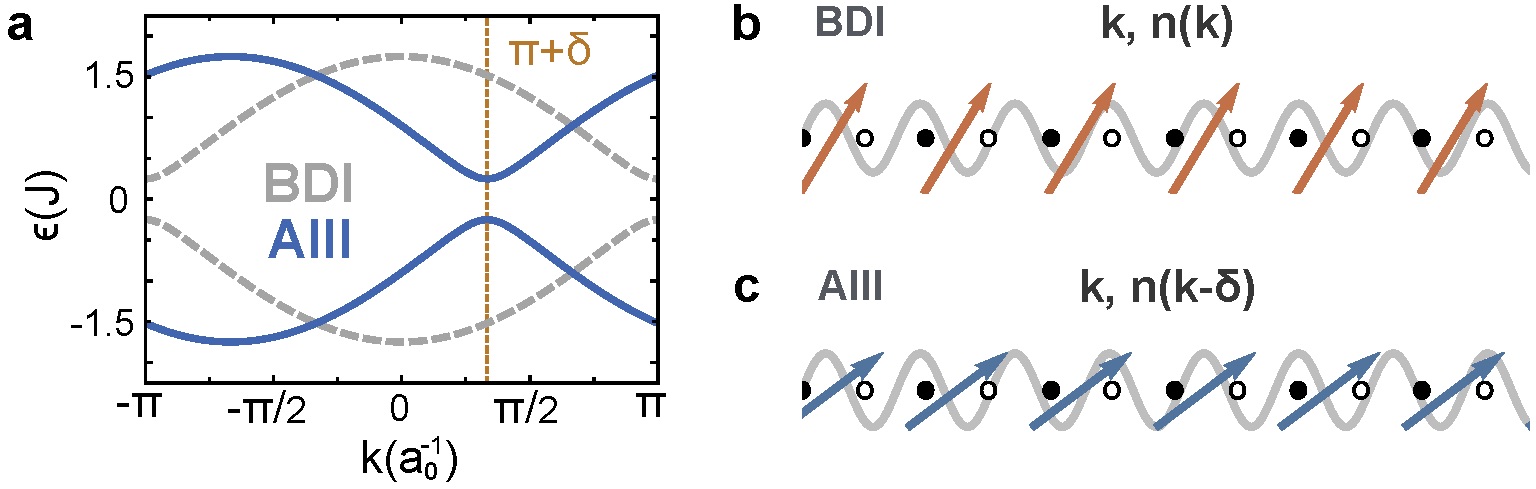}
    \caption{\textbf{Energy bands and bulk eigenstates. AIII class vs. BDI class.} (\textbf{a}) Energy bands for the AIII class ($\delta=-2\pi/3$, solid line) and the BDI class ($\delta=0$, dashed line), for $J^{\prime}/J=0.75$. Bulk eigenstates are fully characterized by the pair (quasimomentum $k$, isospin vector $\mathbf{n}$). (\textbf{b}) In the BDI class  the pair is given by $(k,\mathbf{n}(k))$. (\textbf{c}) In the AIII class, the correspondence between momentum and isospin is different: $(k,\mathbf{n}(k-\delta))$, resulting into a new set of eigenstates (see text and SM).}
    \label{fig:EnergyBands} 
\end{figure}

The different symmetry properties of our AIII model ($\delta\ne0$) are manifested in a distinctive set of eigenvectors, which are genuinely different from those of the SSH model. 
The modes are characterized by two quantities: the quasimomentum $k$ and the isospin vector $\mathbf{n}(k-\delta)$, which characterizes the superposition state between the modes 
$\hat{a}^{\dagger}_{k}$ and $\hat{b}^{\dagger}_{k}$ (see SM). In the AIII model the correspondence between momentum and isospin is different from the one of the SSH model [Fig.~\ref{fig:EnergyBands}]. 

\begin{figure}[t]
  \centering
    \includegraphics[width=0.475\textwidth]{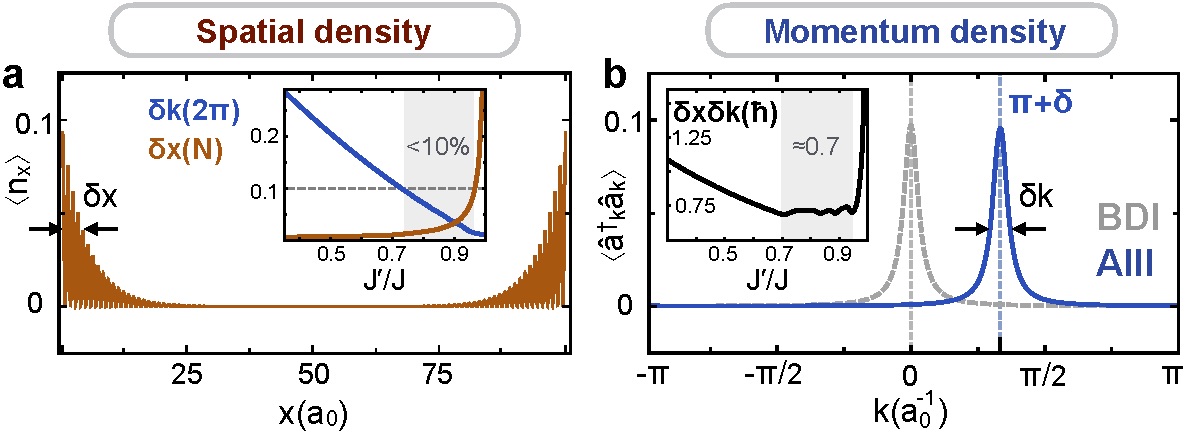}
    \caption{\textbf{Topological edge modes: AIII class vs BDI class} 
    (\textbf{a})  Average spatial density $\langle \hat{n}_x\rangle$ and (\textbf{b}) momentum density $\langle\hat{a}_{k}^{\dagger}\hat{a}^{}_{k}\rangle$ ($=\langle\hat{b}_{k}^{\dagger}\hat{b}^{}_{k}\rangle$) of edge modes for $N=100$ and $J'/J=0.9$.  (\textbf{a}) For both classes edge modes are spatially localized at the edges of the chain. (\textbf{b}) For the BDI class ($\delta=0$, dashed line) both edge modes are localized at momentum $k=\pi$ (for clarity, we plot them shifted to $k=0$). For the AIII class ($\delta=-2\pi/3$, solid line), both edge modes are localized at a non-zero momentum $k=\pi + \delta$. Insets show localization lengths in position and momentum as a function of $J'/J$. 
    }
    \label{fig:EdgeModesDistributions}
\end{figure}

{\em AIII edge modes}. The model exhibits a topological phase transition at $J'=J$, where (for an open chain) two topological edge modes arise with energies lying in the middle of the gap.
The symmetry properties of the AIII model manifest themselves in the properties of the edge modes [Fig.~\ref{fig:Symmetries},~Fig.~\ref{fig:EdgeModesDistributions}]. 
These have the form:
\begin{equation}
\hat{e}_{\pm}^{\dagger}=\frac{1}{\sqrt{2}}\left(\widetilde{a}_{n=1}^{\dagger}\pm\widetilde{b}_{n=N}^{\dagger}\right).
\end{equation}
Here, $\widetilde{a}_{n=1}^{\dagger}$ is a well localized mode at the left end of the chain:
\begin{equation}
\widetilde{a}_{n=1}^{\dagger}=\sum_{n}e^{i(\pi+\delta) n}\varphi(n)\,\hat{a}_{n}^{\dagger},
\label{Left_Edge_Mode}
\end{equation}
with $\varphi(n)\propto\sinh[\xi(N+1-n)]\approx e^{-\xi n}$ and $\xi\approx -\log{(J'/J)}$. Similarly, $\widetilde{b}_{n=N}^{\dagger}$ is a well localized mode at the right end of the chain.

The breaking of time reversal symmetry in the AIII model implies a non-zero average momentum $\pi +\delta$ for the edge modes [Fig.~\ref{fig:EdgeModesDistributions}\textbf{b}]. This chirality is the hallmark of the AIII class, since edge modes of the BDI class must have momentum zero or $\pi$. This statement can be proved as follows. 
In the presence of time reversal symmetry (BDI class), an edge mode $\ket{e}$ has to be equal (up to a global unitary $U$) to its time reversed: $\ket{e}=U\ket{e^*}$ (see SM). Since density operators are invariant under global unitaries, we have that 
\begin{equation}
\braket{e|\hat{n}_{k}|e}=\braket{e^*|\hat{n}_{-k}|e^*}=\braket{e|U\hat{n}_{-k}U^\dagger|e}=\braket{e|\hat{n}_{-k}|e}, \nonumber
\end{equation}
so that states with opposite momenta are on average equally occupied. This implies that the average momentum of an edge mode in the BDI class must be either $0$ or $\pi$.
Experimental observation of an edge mode with non-zero momentum is thus a direct evidence of having realized the AIII class.

Surprisingly, the edge modes are well localized around their average momentum [Fig.~\ref{fig:EdgeModesDistributions}\textbf{b}]. We have that $\widetilde{a}_{n=1}^{\dagger}\equiv \widetilde{a}_{k=\pi+\delta}^{\dagger}$, with $\widetilde{a}_{k=\pi+\delta}^{\dagger}=\sum_{k}F(k-\pi-\delta)\,\hat{a}_{k}^{\dagger} $, where $F(k -\pi-\delta)$ is a well localized function around $k=\pi+\delta$.
For a wide range of parameters in the topological phase, edge states are simultaneously localized both in position and momentum,  [see insets in Fig.~\ref{fig:EdgeModesDistributions}~\textbf{a},\textbf{b} and SM]. This simultaneous localization arises both in the AIII and BDI classes.
\nobalance

\begin{figure}[t]
  \centering
    \includegraphics[width=0.4\textwidth]{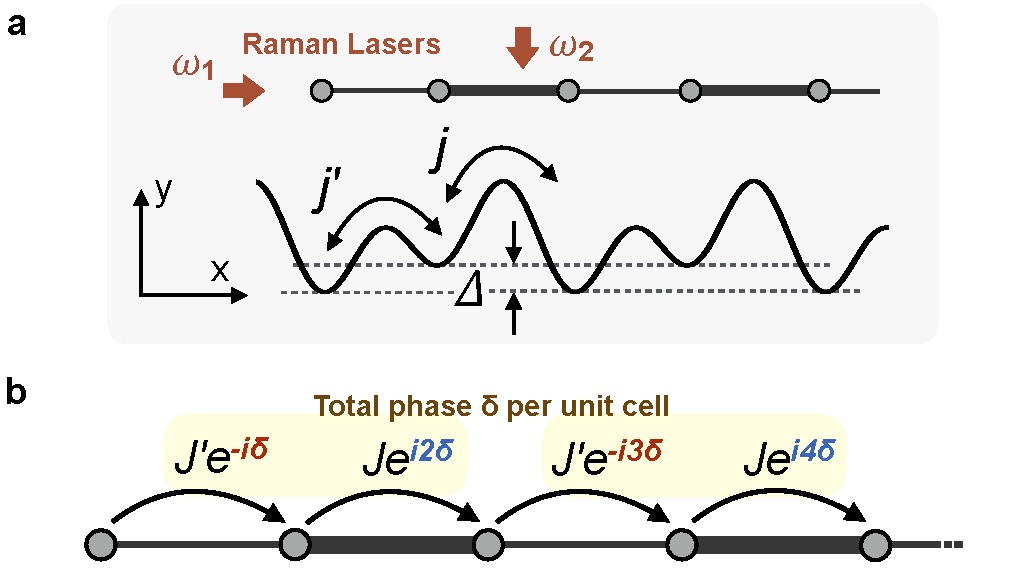}
\caption{\textbf{Experimental scheme.} (\textbf{a}) A double well potential with energy offset $\Delta$ is created from the superposition of two standing waves, forming a one-dimensional lattice with tunneling amplitudes $j$ and $j'$. A pair of Raman lasers is added along the $-x$ and $-y$ directions with frequency difference 
$\omega_{1}-\omega_{2}=\Delta/\hbar$. This leads to a Hamiltonian (\textbf{b}), which is gauge-equivalent to our model.}
 \label{fig:Experiment}
\end{figure}

\textit{Experimental realization.} 
We develop a scheme  [Fig.\ref{fig:Experiment}] for the realization of the model above.
The scheme combines a superlattice structure \citep{Atala2013} together with Raman assisted tunneling \cite{Aidelsburger2011, Aidelsburger2013} to realize
a Hamiltonian of the form:
\begin{equation}\label{eq:HamiltonianoExp}
H_{\text{exp}}=-\sum_{n}\left(J'e^{i(2n-1)\delta}\hat{a}_{n}^{\dagger}\hat{b}_{n}^{}+Je^{i2n\delta}\hat{a}_{n+1}^{\dagger}\hat{b}_{n}^{}+\text{h.c.}\right).\nonumber
\end{equation} 
This Hamiltonian is equivalent to our model (\ref{eq:hamiltoniano}) through the gauge transformation
$\hat{a}_{n}^{\dagger}\,(\hat{b}_{n}^{\dagger})\longrightarrow\hat{a}_{n}^{\dagger}\,(e^{i(2n-1)\delta}\hat{b}_{n}^{\dagger})$. This gives rise to a relative shift of the momentum modes $\hat{a}_{k}^{\dagger}\,(\hat{b}_{k}^{\dagger})\longrightarrow\hat{a}_{k}^{\dagger}\,(\hat{b}_{k-2\delta}^{\dagger})$, so that the edge modes are transformed as
\begin{equation}
\widetilde{a}_{k=\pi+\delta}^{\dagger}(\widetilde{b}_{k=\pi+\delta}^{\dagger})
\longrightarrow \widetilde{a}_{k=\pi+\delta}^{\dagger}\,(\widetilde{b}_{k=\pi-\delta}^{\dagger}).
\label{MomentumShift}
\end{equation}

\noindent For the BDI class ($\delta=0$) the two edge modes are localized at the same position in momentum space ($k=\pi$). In contrast, for the AIII class they are localized at opposite momenta ($k=\pi+\delta$ and $k=\pi-\delta$). This splitting is a distinctive feature of the AIII class. As we show below, it allows to directly observe fractionalization in momentum space.


\textit{Observing fractionalization in the AIII model.} 
We consider a chemical potential such that the lower band and both edge modes are occupied. The corresponding many-body state in the theoretical model is:

\begin{equation}
\ket{\Phi}=\hat{e}_{-}^{\dagger}\hat{e}_{+}^{\dagger}\prod_{q}\hat{c}_{+,q}^{\dagger}\ket{0},
\label{2-Quasiparticle-State}
\end{equation}
where $\hat{c}_{+,q}^{\dagger}$($\hat{c}_{-,q}^{\dagger}$) denotes a bulk mode with momentum $q$ in the lower (upper) band (see SM).
This state corresponds to a fermion added to a state with flat background density.
It can be written as $\ket{\Phi}=\hat{e}_{-}^{\dagger}\ket{\Phi_{+}}$,  where $\ket{\Phi_{\pm}}=\hat{e}_{\pm}^{\dagger}\prod_{q}\hat{c}_{\pm,q}^{\dagger}\ket{0}$. The state $\ket{\Phi_{+}}$, in which all bulk modes in the lower band plus one of the edge states are occupied, has a flat density profile, 
$\nu=\braket{\Phi_{+}|\hat{a}_{k}^{\dagger}\hat{a}_{k}^{}|\Phi_{+}}=\frac{1}{2}$.
This is easily derived by noticing that $\ket{\Phi_{+}}$ and $\ket{\Phi_{-}}$ have the same density profiles:
\begin{equation}
\braket{\Phi_{+}|\hat{a}_{k}^{\dagger}\hat{a}_{k}^{}|\Phi_{+}}=\braket{\Phi_{-}|C^{\dagger}\hat{a}_{k}^{\dagger}\hat{a}_{k}^{}C|\Phi_{-}}=\braket{\Phi_{-}|\hat{a}_{k}^{\dagger}\hat{a}_{k}^{}|\Phi_{-}}, \nonumber
\end{equation}
since they are obtained from each other through the unitary transformation $C$ that maps $\hat{a}_{n}^{\dagger}\rightarrow\hat{a}_{n}^{\dagger}$ and $\hat{b}_{n}^{\dagger}\rightarrow -\hat{b}_{n}^{\dagger}$.
Defining the projectors $P_{\pm}=\sum_{q}\ket{q_{\pm}}\bra{q_{\pm}}+\ket{e_{\pm}}\bra{e_{\pm}}$, with $\ket{q_{\pm}}=\hat{c}_{\pm,q}^{\dagger}\ket{0}$ and  $\ket{e_{\pm}}=\hat{e}_{\pm}^{\dagger}\ket{0}$ and taking into account that $P_{+}=\mathbb{I}-P_{-}$, we obtain:
\begin{align}
\nu&=
\text{tr}(\hat{a}_{k}^{\dagger}\hat{a}_{k}^{}P_{+})=\text{tr}(\hat{a}_{k}^{\dagger}\hat{a}_{k}^{})-\text{tr}(\hat{a}_{k}^{\dagger}\hat{a}_{k}^{}P_{-})=1-\nu
\end{align}
and thereby $\nu=1/2$.

In the state (\ref{2-Quasiparticle-State}) both edge states are occupied. Therefore the mode $\widetilde{a}_{n=1}^{\dagger}=\widetilde{a}_{k=\pi+\delta}^{\dagger}$, simultaneously localized in momentum and position is also occupied, and we have:
\begin{equation}
\left[\widetilde{a}_{k=\pi+\delta}^{\dagger}\widetilde{a}_{k=\pi+\delta}^{}-\nu\right]\ket{\Phi}=\frac{1}{2}\ket{\Phi}.
\end{equation}
The above expression states that the state $\ket{\Phi}$ (with one fermion on top of a flat background) is an exact eigenstate of the number operator $\widetilde{a}_{k=\pi+\delta}^{\dagger}\widetilde{a}_{k=\pi+\delta}^{}-\nu$ with eigenvalue $1/2$. Thus, a sharp $1/2$ fraction of particles is localized at momentum $k=\pi+\delta$.
The same holds for the number operator $\widetilde{b}_{k=\pi+\delta}^{\dagger}\widetilde{b}_{k=\pi+\delta}^{}-\nu$.
\begin{figure}[t]
  \centering
    \includegraphics[width=0.475\textwidth]{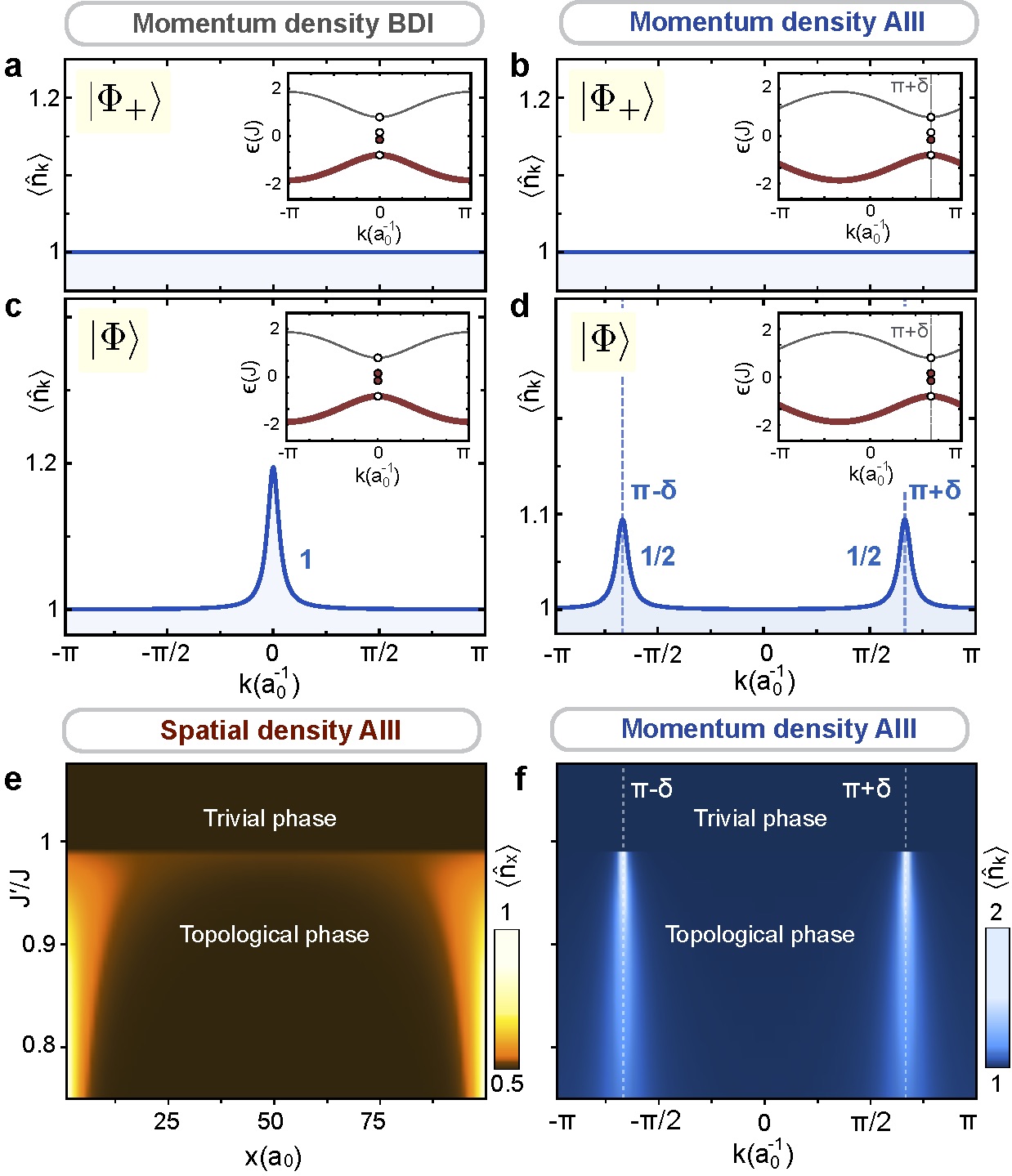}
    \caption{\textbf{Direct observation of fractionalization in the AIII class.} 
    (\textbf{a}-\textbf{d}). Momentum density profiles (\textbf{a},\textbf{c}) for the BDI class ($\delta=0$), and (\textbf{b},\textbf{d}) the AIII class ($\delta=-\pi/3$), for $N=100$ and $J^{\prime}/J=0.9$. For both classes the many body state $\ket{\Phi_+}$ (see text), 
   shows a flat density (\textbf{a},\textbf{b}).  
   For the state $\ket{\Phi}$, in which a fermion is added to the flat background, the momentum density in the BDI class (\textbf{c}) shows a single peak of charge 1 at momentum $k=\pi$ (plotted shifted at $k=0$). In contrast, in the AIII class (\textbf{d}), the momentum density shows two peaks of charge $1/2$ localized at momenta $\pi\pm\delta$. 
(\textbf{e},\textbf{f}) Simultaneous fractionalization in position and momentum space in the AIII class. 
(\textbf{e}) Average spatial density and (\textbf{f}) momentum density for the many body state such that the upper band is empty. In the topological phase localized peaks of charge $1/2$ (at the edges of the chain, and at momenta $\pi\pm\delta$) arise. 
    }
    \label{fig:DensityProfiles}
\end{figure}

What are the implications for the experiment? In the experimental gauge, the edge modes $\widetilde{a}$ and $\widetilde{b}$  are shifted according to (\ref{MomentumShift}). For the SSH model ($\delta=0$) this means that a particle added to a uniform background will consist of two quasiparticles bound together at the same momentum position. The momentum distribution will show a single peak at momentum $\pi$ enclosing a total charge $1$ [Fig.~\ref{fig:DensityProfiles}\textbf{c}]. For the AIII case, in contrast, a particle added to a uniform background will split into two halves located at opposite momenta ($\pi-\delta,\pi+\delta$). The momentum distribution will show two peaks, each enclosing a $1/2$ charge [Fig.~\ref{fig:DensityProfiles}\textbf{d}].
In the AIII class splitting of the fermion occurs therefore both in position and momentum space [Fig.~\ref{fig:DensityProfiles}\textbf{e},\textbf{f}]. The splitting in momentum is a direct consequence of the non-zero momentum of the edge modes, which is in turn a direct manifestation of the breaking of time reversal symmetry characterizing the AIII class.

We have presented a one-dimensional model realizing the AIII symmetry class which can be readily realized in experiments. 
We have identified distinctive features characterizing the AIII class, which can serve as experimental signatures discriminating it with respect to the BDI class. For instance, the characteristic non-zero momentum of AIII edge modes could be observed by realizing our model for a system of bosons. By making the bosons condense in a topological edge mode, a time of flight experiment will show a macroscopic signal at a non-zero momentum, evidencing the realization of the AIII class.

Our findings open a path for the detection of fractionalization in the AIII model by directly imaging the momentum distribution of the particles. 
Our protocol relies on the preparation of the many-body state (\ref{2-Quasiparticle-State}) and on the observation of 1/2 fractions on top of a flat background. These requirements are common to previous protocols to observe fractionalization in position space. In contrast to those, which require in situ imaging of the system, our scheme benefits from the magnification of the cloud after time of flight. Analysis of temperature and soft edge effects in previous protocols show that edge states are robust against them \cite{Goldman2012, Goldman2013, Goldman2013b, Buchhold2012}. We believe that this also holds for our momentum protocol. Moreover, localization in momentum might be even less susceptible to soft edges. 

Finally, it is interesting to investigate how the simultaneous localization of edge modes in momentum and position can be exploited to prove the topological protection of fractionalized quasiparticles.
It is a challenge to explore variations of our AIII model in ladder architectures and to investigate whether they can serve as physical pathways to Riemann's conjecture \cite{Sierra2014}.

This research was funded by the Deutsche Forschungsgemeinschaft (DFG, German Research Foundation) via Research Unit FOR 2414 under project number 277974659.

The work of C. G. V. was supported through the grant BES-2013-064443 of the Spanish MINECO and by the Deutsche Forschungsgemeinschaft (DFG, German Research Foundation) via Research Unit FOR 2414 under project number 277974659.

\appendix

\section{A model for an AIII topological insulator}

\subsection{Hamiltonian of the model}

\begin{figure}[H]
  \centering
    \includegraphics[width=0.45\textwidth]{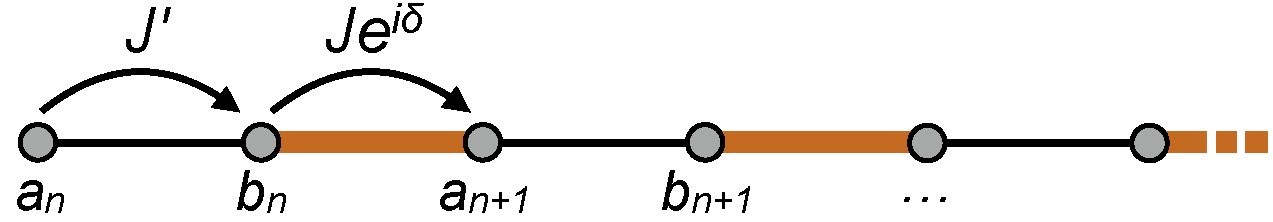}
    \caption[The model.]{\textbf{The model.} Schematic illustration of our model for a topological insulator in the AIII class: a one-dimensional dimerized lattice with two different sites per unit cell, $a$ and $b$, and two tunneling amplitudes $J^{\prime}$ and $Je^{i\delta}$. Particles pick up a phase $\delta$ when travelling from one sublattice site to the same sublattice site in the following lattice cell.}
    \label{fig:FiguraAIII01} 
\end{figure}

The model conceived by B. Paredes for a topological insulator in the AIII symmetry class consists of a dimerized one-dimensional lattice with two different sites per unit cell and two distinct coupling terms. The two hopping amplitudes are $J^{\prime}$ and $Je^{i\delta}$, being one of them complex so that particles pick up a phase $\delta$ when traveling from one unit cell to the next one (Fig.~\ref{fig:FiguraAIII01}).
This model is described by the following tight-binding Hamiltonian:
\begin{equation}\label{eq:HamiltonianPosition}
H_{\delta}=-\sum_{n}^{N}\big(J^{\prime}\,\hat{a}_{n}^{\dagger}\hat{b}_{n}^{}+Je^{i\delta}\,\hat{a}_{n+1}^{\dagger}\hat{b}_{n}^{}+\text{H.c.}\big),
\end{equation}
being $\hat{a}_{n}^{\dagger}$ and $\hat{b}_{n}^{\dagger}$ the creation operators of a particle at the sublattice site $a_{n}$ and $b_{n}$, respectively, in the n-th unit cell, and $N$ the total number of unit cells in the lattice. 
This model is a generalization of the SSH model \cite{Su1979,Delplace2011}. In the particular case in which $\delta=0$ our model becomes the SSH model. As we show throughout this work, the presence of the phase $\delta$ is sufficient to:\\

\textit{i)} break time reversal and charge conjugation symmetries and, therefore, make the model be a realization of the AIII symmetry class;\\

\textit{ii)} make the symmetry protected zero modes that the model exhibits in its topological phase under open boundary conditions have different properties from those present in the case in which $\delta=0$;\\

\textit{iii)} allow the development of an alternative way of observing the phenomenon of fractionalization, in momentum space ,which would not be possible in the BDI class.

\subsection{Symmetries of the model}

In order to know what are the symmetry properties of the Hamiltonian, we exploit its translational invariance and write it in the momentum representation. The Bloch modes $\hat{a}_{k}^{\dagger}$ and $\hat{b}_{k}^{\dagger}$, which constitute the momentum basis, are defined from the position basis as:
\begin{align}
&\hat{a}_{k}^{\dagger}=\frac{1}{\sqrt{N}}\sum_{n=1}^{N}e^{ikn}\,\hat{a}_{n}^{\dagger}\\
&\hat{b}_{k}^{\dagger}=\frac{1}{\sqrt{N}}\sum_{n=1}^{N}e^{ikn}\,\hat{b}_{n}^{\dagger},
\end{align}
being the momentum $k$ quantized as $k=2\pi m/N$, with $m\in\mathbb{Z}$, and having set the lattice constant to the unit. The Hamiltonian of the system under periodic boundary conditions can be written as:
\begin{equation}
H_{\delta}=-\sum_{k}\hat{\psi}_{k}^{\dagger}\,M_{\delta}(k)\,\hat{\psi}_{k},
\end{equation}
where $\hat{\psi}_{k}^{\dagger}=\begin{pmatrix}\hat{a}_{k}^{\dagger} & \hat{b}_{k}^{\dagger}\end{pmatrix}$ and being $M_{\delta}(k)$ the so-called Hamiltonian matrix, which depends on the momentum and takes the form:
\begin{equation}\label{eq:AIIIHamiltonianMatrix}
M_{\delta}(k)=\big[J^{\prime}+J\cos(k-\delta)\big]\,\sigma_{x}+J\sin(k-\delta)\,\sigma_{y}.
\end{equation}
Once we know the Hamiltonian matrix of the model, it is straightforward to obtain the symmetry properties of the system. These are the following:\\

\textit{i) Chiral symmetry.}\\

The condition that the Hamiltonian matrix needs to fulfil for the model to present chiral symmetry is that $U_{S}M_{\delta}(k)U_{S}^{\dagger}=-M_{\delta}(k)$ for some unitary transformation $U_{S}$, called the chiral operator, which does not depend on the momentum (this transformation acts on the space spanned by the two sublattice sites and consists of the same transformation on every cell in the lattice). By looking at the Hamiltonian matrix, Eq.~(\ref{eq:AIIIHamiltonianMatrix}), we conclude that the model has always chiral symmetry, for every possible parameter configuration, as it consists of a linear superposition of the two Pauli matrices $\sigma_{x}$ and $\sigma_{y}$ and thus:
\begin{equation}
\sigma_{z}M_{\delta}(k)\sigma_{z}=-M_{\delta}(k),
\end{equation}
so that the chiral operator is $U_{S}=\sigma_{z}$.\\

\begin{figure}[t]
  \centering
    \includegraphics[width=0.475\textwidth]{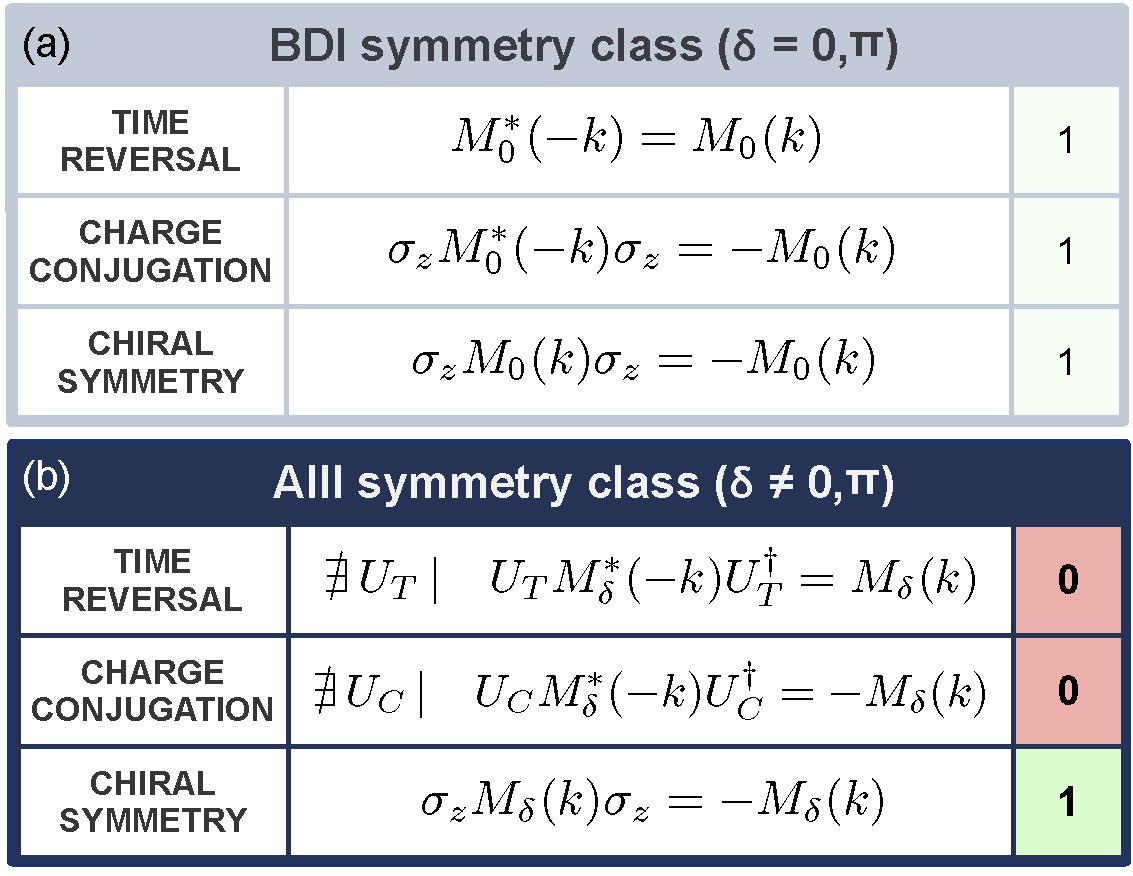}
    \caption[Symmetry class of the model.]{\textbf{Symmetry class of the model.} Presence (1) or absence (0) of time reversal, charge conjugation and chiral symmetries in our model. (a) On one hand, for $\delta=0$ or $\delta=\pi$ (cases in which our model becomes the SSH model), the system has chiral, time reversal and charge conjugation symmetries and thus belongs to the BDI class. (b) On the other hand, for $\delta\neq0,\pi$, the model breaks both time reversal and charge conjugation symmetries while it remains chiral symmetric. Consequently, it is a realization of the AIII class.}
    \label{fig:FiguraAIII02}
\end{figure}

\textit{ii) Time reversal and charge conjugation symmetries.}\\

Chiral symmetry is the composition of time reversal and charge conjugation symmetries and our model is always chiral symmetric. Therefore, the condition for time reversal symmetry is all we need to analyse in order to know weather the Hamiltonian exhibits both time reversal and charge conjugation symmetries or none of them. That condition is $U_{T}M_{\delta}^{*}(-k)U_{T}^{\dagger}=M_{\delta}(k)$ for some unitary operator $U_{T}$ with no dependence on the momentum.
In the particular case in which $\delta=0$ or $\delta=\pi$ the Hamiltonian matrix is such that:
\begin{equation}
M_{0}^{*}(-k)=M_{0}(k),
\end{equation}
and therefore the model is both time reversal and charge conjugation symmetric. These two parameter configurations, $\delta=0$ and $\delta=\pi$, belong then to the BDI symmetry class [see Fig.~\ref{fig:FiguraAIII02}(a)].

On the contrary, in the more general situation in which $\delta\neq0,\pi$, the model presents neither time reversal symmetry nor charge conjugation symmetry. We have that:
\begin{align}
M_{\delta}^{*}&(-k)=\begin{pmatrix}
\quad0 & J^{\prime}+Je^{-i(k+\delta)}\quad\\
\quad J^{\prime}+Je^{i(k+\delta)} & 0\quad\end{pmatrix}\neq\nonumber\\
&\begin{pmatrix}
\quad0 & J^{\prime}+Je^{-i(k-\delta)}\quad\\
\quad J^{\prime}+Je^{i(k-\delta)} & 0\quad\end{pmatrix}=M_{\delta}(k),
\end{align}
and there is no unitary matrix $U_{T}$ such that:
\begin{equation}
U_{T}M_{\delta}^{*}(-k)U_{T}^{\dagger}=M_{\delta}(k).
\end{equation}
In case such unitary transformation existed, it would be necessarily a rotation around the $z$-axis, as any other unitary operator would add a component proportional to $\sigma_{z}$ to the Hamiltonian matrix. A rotation $R_{z}(\theta)$ of an arbitrary angle $\theta$ around the $z$-axis transforms $M_{\delta}^{*}(-k)$ in the following way:
\begin{align}
&R_{z}(\theta)M_{\delta}^{*}(-k)R_{z}^{\dagger}(\theta)=\nonumber\\
&\begin{pmatrix}
\quad0 & e^{i\theta}\left[J^{\prime}+Je^{-i(k+\delta)}\right]\quad\\
\quad e^{-i\theta}\left[J^{\prime}+Je^{i(k+\delta)}\right]  0\quad\end{pmatrix}\neq\nonumber\\
&\begin{pmatrix}
\quad0 & J^{\prime}+Je^{-i(k-\delta)}\quad\\
\quad J^{\prime}+Je^{i(k-\delta)} & 0\quad\end{pmatrix}=M_{\delta}(k).
\end{align}
As we see, there is no unitary matrix $U_{T}$ able to fulfil the time reversal symmetry condition. As a result, the presence of a phase $\delta\neq0,\pi$ breaks both time reversal and charge conjugation symmetries, while keeping chiral symmetry, and thus makes the model belong to the AIII class [see Fig.~\ref{fig:FiguraAIII02}(b)].

\subsection{Topological nature of the model}

\begin{figure}[t]
  \centering
    \includegraphics[width=0.485\textwidth]{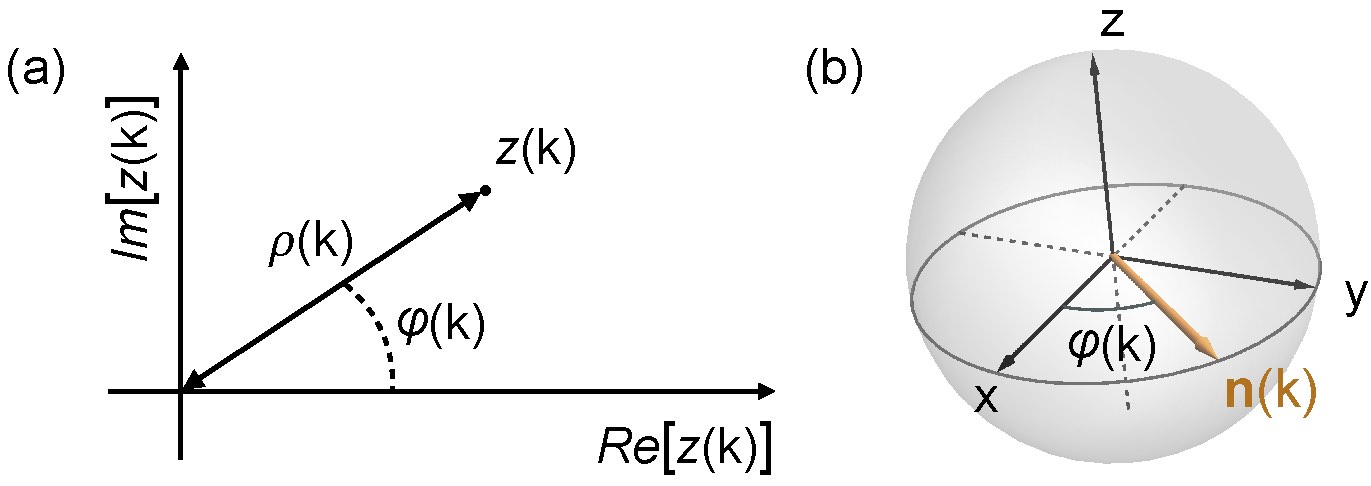}
    \caption[Hamiltonian matrix and isospin.]{\textbf{Hamiltonian matrix and isospin.} (a) The Hamiltonian matrix is completely characterized by two functions: $\rho(k)$ and $\varphi(k)$, which are the modulus and argument of the complex function $z(k)$. On one hand, the function $\rho(k)$ determines the two energy bands of the system as $E_{\pm}(k)=\mp\rho(k-\delta)$. On the other hand, the function $\varphi(k)$ defines a unit vector $\bm{n}(k)$ that characterizes, through the Bloch sphere representation, the superposition state between the modes $\hat{a}_{k}^{\dagger}$ and $\hat{b}_{k}^{\dagger}$ that constitutes an eigenstate of the Hamiltonian.}
    \label{fig:FiguraAIII03}
\end{figure}

The topological nature of the model is characterized through the Zak phase, which is the Berry phase that particles pick up when they complete an adiabatic path across the first Brillouin zone \cite{Zak1989}. The Zak phase can be computed as:
\begin{equation}\label{eq:DefZakPhase}
\mathcal{Z}=-i\oint_{1^{st}\mathcal{BZ}}dk\, \langle k,\delta|_{\pm}\,\partial_{k}\,|k,\delta\rangle_{\pm},
\end{equation}
where $\ket{k,\delta}_{\pm}$ are the Bloch eigenstates of the Hamiltonian, that is, the eigenstates for periodic boundary conditions.
Therefore, in order to calculate the Zak phase, we first need to obtain the Bloch eigenstates of the Hamiltonian. For that, we consider the Hamiltonian matrix, Eq.~(\ref{eq:AIIIHamiltonianMatrix}), and write it in the following form:
\begin{equation}
M_{\delta}(k)=M_{0}(k-\delta),
\end{equation}
with:
\begin{equation}
M_{0}(k)=\rho(k)\,\bm{n}(k)\cdot\bm{\sigma},
\end{equation}
being $\bm{\sigma}$ a vector containing the three Pauli matrices, and:
\begin{align}
&\rho(k)=\sqrt{J^{2}+J^{\prime 2}+2JJ^{\prime}\cos k},\label{eq:FuncionRhoDimerizedLattice}\\
&\bm{n}(k)=\cos\varphi(k)\,\hat{x}+\sin\varphi(k)\,\hat{y},
\end{align}
where $\varphi(k)=\arg(J^{\prime}+Je^{ik})$. That is, the Hamiltonian matrix is characterized by the two functions $\rho(k)$ and $\varphi(k)$, which are the modulus and argument of the complex number $z(k)=J^{\prime}+Je^{ik}$ [see Fig.~\ref{fig:FiguraAIII03}(a)]. The angle $\varphi(k)$ corresponds to the azimuthal angle in spherical coordinates of the unit vector $\bm{n}(k)$ in the three-dimensional space [see Fig.~\ref{fig:FiguraAIII03}(b)].

Decomposing the Hamiltonian matrix in this way is really convenient, as its eigenvectors can be easily obtained from the vector $\bm{n}(k)$ through the Bloch sphere representation. That is, for a particular value of the momentum $k$ the Hamiltonian matrix has two eigenvectors $\hat{n}_{\pm}(k)$, with eigenvalues $\pm\rho(k)$, which correspond to the vectors $\pm\bm{n}(k)$ in the Bloch sphere representation:
\begin{align}
&\pm\bm{n}(k)=\cos\varphi(k)\,\hat{x}+\sin\varphi(k)\,\hat{y}\,\longrightarrow\nonumber\\
&\longrightarrow\,\hat{n}_{\pm}(k)=\frac{1}{\sqrt{2}}\begin{pmatrix}1\\ \pm e^{i\varphi(k)}\end{pmatrix}.
\end{align} 
The eigenstates of the Hamiltonian are then obtained from the eigenvectors of the Hamiltonian matrix and the two-component momentum creation operator:
\begin{equation}
\ket{k,\delta}_{\pm}=\hat{\psi}_{k}^{\dagger}\,\hat{n}_{\pm}(k-\delta)\ket{0}=\frac{1}{\sqrt{2}}\left(\,\hat{a}_{k}^{\dagger}\pm e^{i\varphi(k-\delta)}\,\hat{b}_{k}^{\dagger}\,\right)\ket{0},
\end{equation}\label{eq:AIIIBlochEigenModes}
with energies $E_{\pm}(k)=\mp\rho(k-\delta)$.
Substituting these eigenstates in Eq.~(\ref{eq:DefZakPhase}) we can compute the Zak phase:
\begin{equation}
\mathcal{Z}=\frac{1}{2}\oint_{1^{st}\mathcal{BZ}}\,dk\,\frac{d\varphi}{dk}=\frac{\Delta\varphi}{2}=\begin{cases}\,0\,(\text{winding number }0)\\ \,\pi\,(\text{winding number }1).\end{cases}
\end{equation}
\begin{figure}[t]
  \centering
    \includegraphics[width=0.485\textwidth]{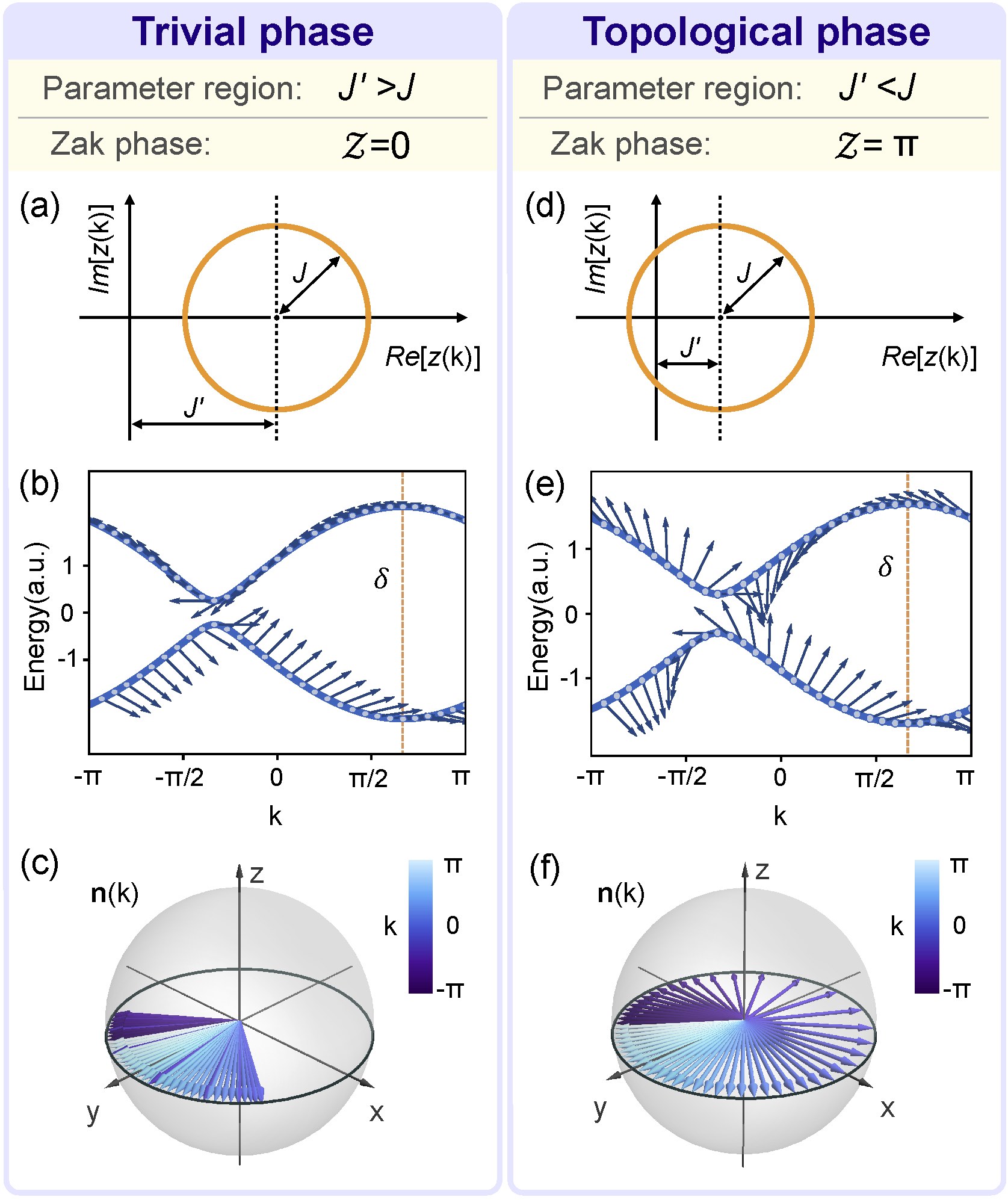}
    \caption[Trivial and topological phases.]{\textbf{Trivial and topological phases.} Plot of $z(k)$, the energy bands and the isospin vector $\bm{n}(k)$ in the trivial phase, (a), (b) and (c), and in the topological phase (d), (e) and (f). In the trivial phase, when $J^{\prime}>J$, the circle that describes the function $z(k)$ does not enclose the origin and, thus, the winding number of the vector $\bm{n}(k)$ on the $xy$-plane is $0$. On the contrary, in the topological phase, when $J^{\prime}<J$, the circle described by $z(k)$ encloses the origin and, equivalently, the isospin vector has winding number $1$.}
    \label{fig:FiguraAIII04}
\end{figure}
As we see, the Zak phase is quantized according to the winding number of the vector $\bm{n}(k)$ in the $xy$-plane as the momentum runs over the first Brillouin zone or, equivalently, according to the winding number of $z(k)$ in the complex plane. In this way, the system can be found to be in two distinct topological phases: on one hand, a topologically trivial phase characterized by a Zak phase $\mathcal{Z}=0$, and on the other hand, a topologically nontrivial phase characterized by a Zak phase $\mathcal{Z}=\pi$. The complex function $z(k)$ describes a circle of radius $J$ centred at the point $(J^{\prime},0)$. Therefore, the system is in its topological phase for $J^{\prime}<J$, whereas the trivial phase corresponds to $J^{\prime}>J$, being both phases separated by the critical point $J^{\prime}=J$ (see Fig.~\ref{fig:FiguraAIII04}). To be more precise, the exact values of the Zak phase depends on our definition of the unit cell \cite{Atala2013}. The important aspect about the Zak phase is the fact that it is quantized, corresponding each of the two possible values to a distinct topological class distinguished by a different winding number of $z(k)$ and $\bm{n}(k)$. It is the difference between the Zak phase in the trivial phase and the Zak phase in the topological phase what constitutes an invariant quantity that characterizes the topological nature of the system. Nevertheless, for the sake of simplicity, we use in this work the values $\mathcal{Z}=0$ and $\mathcal{Z}=\pi$ for the Zak phase in the trivial and topological phases, respectively.

\section{Analysis of the phase $\delta$}

As we have already proved, the presence of a phase $\delta$ in one of the two hopping terms in a dimerized lattice model is sufficient to produce the breaking of time reversal and charge conjugation symmetries and, thus, make the model belong to the AIII symmetry class. We obtained this result after following the simplest and more direct procedure, which consists of analysing the condition that the Hamiltonian matrix needs to fulfil for the Hamiltonian to present chiral, time reversal and charge conjugation symmetries.

Nevertheless, it might seem surprising that just a simple phase is all we need in order to get a realization of a distinct class of topological insulators, characterized by different symmetry properties, as we are dealing with a one-dimensional system and one could think that any phase can be removed by means of gauge transformations and, therefore, conclude that all possible phases are irrelevant.

For this reason, we dedicate this section to show a detailed and profound analysis of the role that such phase plays in our model. We provide several ways to realize how relevant  the phase $\delta$ is and why it is directly related to the breaking of time reversal and charge conjugation symmetries.

\subsection{Momentum-isospin correspondence}

The phase $\delta$ has direct implications in the correspondence, established by the Hamiltonian matrix, between the two characteristic physical magnitudes associated to each eigenmode of the system for periodic boundary conditions. These magnitudes are, on one hand, the momentum and, on the other, the isospin vector corresponding to the two-level quantum system formed by the two different sublattice sites.

\subsubsection{Momentum shift in a simple lattice}

In order to understand the relation between these two physical quantities and the phase $\delta$, we first analyse a more simple model. We consider a one-dimensional lattice with just one single site per unit cell, that is, the simplest lattice model we can imagine. The Hamiltonian for such lattice model is:
\begin{equation}
H_{\text{s}}=-\sum_{n=1}^{N}\left( t\,\hat{a}_{n+1}^{\dagger}\hat{a}_{n}^{}+\text{H.c.}\right),
\end{equation}
where $t$ is a real first-neighbour hopping amplitude. If we write the Hamiltonian using the momentum representation we obtain directly its spectral decomposition:
\begin{equation}
H_{\text{s}}=-\sum_{k}2t\cos k\, \hat{a}_{k}^{\dagger}\hat{a}_{k}^{},
\end{equation}
from which we can identify the eigenstates of the Hamiltonian: $\ket{k}=\hat{a}_{k}^{\dagger}\ket{0}$. They fulfil the eigenvalue equation:
\begin{equation}
H_{\text{s}}\ket{k}=-2t\cos k\ket{k},
\end{equation}
so that the energy associated to the eigenstate $\ket{k}$ is given by the function $E_{\text{s}}(k)=-2t\cos k$.

We know consider another Hamiltonian, which is the result of adding a phase $\alpha$ to every hopping term in the previous model. That is:
\begin{equation}
H_{\text{s}}(\alpha)=-\sum_{n=1}^{N}\left( te^{i\alpha}\,\hat{a}_{n+1}^{\dagger}\hat{a}_{n}^{}+\text{H.c.}\right).
\end{equation}
In the momentum basis this new Hamiltonian takes the following form:
\begin{equation}
H_{\text{s}}(\alpha)=-\sum_{k}2t\cos (k-\alpha)\, \hat{a}_{k}^{\dagger}\hat{a}_{k}^{},
\end{equation}
and, therefore, the states $\ket{k}$ ,which are the eigenstates of $H_{\text{s}}(\alpha=0)$, are also eigenstates of the new Hamiltonian for any value that the phase $\alpha$ can take. However, the energy associated to the eigenstate $\ket{k}$ is given by $E_{\text{s}}(k-\alpha)$, as we have that:
\begin{equation}
H_{\text{s}}(\alpha)\ket{k}=E_{\text{s}}(k-\alpha)\ket{k}.
\end{equation}
In this way, we can say that a shift in the momentum has been introduced in the Hamiltonian by adding a phase to the hopping terms, that is, by making them complex. It is important to remark that the eigenstates of the Hamiltonian are the same for any value of $\alpha$. The only thing that depends on such phase is the energy associated to each eigenstate, in such a way that the dispersion relation is the one corresponding to the Hamiltonian with no phase $\alpha$, but shifted with respect to the momentum by $\alpha$.\\

Let us consider the eigenstates $\ket{k}$ of the Hamiltonian with no phase $\alpha$ and perform a momentum relabelling so that we obtain the states $\ket{k-\alpha}$. In general these new states are not eigenstates of the Hamiltonian for any value of $\alpha$, because the value of the shifted momentum $k-\alpha$ will not coincide with any of the quantized values of the momentum that form the orthogonal basis $\{\ket{k_{n}}\}_{n=1}^{N}$, with $k_{n}=2\pi n/N$ and $n\in\mathbb{Z}$. Only when $\alpha=2\pi m/N$, with $m\in\mathbb{Z}$, will the new states still be eigenstates of the Hamiltonian; which is due to the fact that, in that case, the momentum relabelling produces just a rearrangement of the set of eigenstates.

As we see, relabelling the eigenstates is not equivalent to making a shift in the momentum. A genuine shift in the momentum has to be implemented by adding the same phase to all coupling amplitudes in the model. Why is that? What do we mean when we say that a particle occupies a wave function with a certain well-defined linear momentum? We mean that, when we perform a translation in space along a certain direction, this wave function transforms into itself times a complex phase proportional to the translated distance and being the proportionality constant precisely the value of its linear momentum along that particular direction. In other words, the momentum is the generator of the group of translations in space and, thus, a well defined momentum is a consequence of having translational symmetry in space. In the particular case of a system in the lattice, only those translations given by a lattice vector leave the system invariant and, thus, are symmetries. Therefore, the momentum of a particle is the phase that its wave function picks up when a translation of one unit cell is performed, since the momentum, called in this context quaismomentum, is the generator of this discrete group of translations. That is the reason why adding a phase $\alpha$ to all coupling terms in the simple lattice model $H_{text{s}}$ produces a shift in the momentum.

\subsubsection{Momentum shift in a dimerized lattice}

On the contrary, in the case of our AIII model things are different due to the fact that it consists of a dimerized lattice and not a simple one.

We start by considering the model with no phase $\delta$, that is the SSH model, whose Hamiltonian is:
\begin{equation}
H_{\delta=0}=-\sum_{k}\hat{\psi}_{k}^{\dagger}\,M_{0}(k)\,\hat{\psi}_{k}.
\end{equation}
The eigenstates of this Hamiltonian are made from two quantities. On one hand $\hat{\psi}_{k}^{\dagger}$, a two-component creation operator with a particular momentum $k$. On the other hand $\hat{n}_{\pm}(k)$, a two-component complex vector that depends on the momentum and determines the superposition state between the modes $\hat{a}_{k}^{\dagger}$ and $\hat{b}_{k}^{\dagger}$ that constitutes an eigenstate of the Hamiltonian. This vector corresponds, through the Bloch sphere representation, to a unit vector $\pm\bm{n}(k)$ in the three-dimensional real space, which is determined by the Hamiltonian matrix when we write it as:
\begin{equation}\label{eq:HamiltonianMatrixVectorAIII}
M_{0}(k)=\rho(k)\,\bm{n}(k)\cdot{\sigma},
\end{equation}
The two eigenvectors of the Hamiltonian matrix, for a particular momentum value $k$, are precisely $\hat{n}_{\pm}(k)$ with energies $\pm\rho(k)$, that is:
\begin{equation}
M_{0}(k)\,\hat{n}_{\pm}(k)=\pm\rho(k)\,\hat{n}_{\pm}(k).
\end{equation}
Therefore, the eigenstates of the Hamiltonian $H_{\delta=0}$ for periodic boundary conditions are:
\begin{equation}
\ket{k,\delta=0}_{\pm}=\hat{\psi}_{k}^{\dagger}\,\hat{n}_{\pm}(k)\ket{0},
\end{equation}
with energies $\mp\rho(k)$. 

What happens if we introduce a non-zero phase $\delta$ to the model? In that case the Hamiltonian written in the momentum representation would be:
\begin{equation}\label{eq:HamiltonianDeltaMatrixShift}
H_{\delta}=-\sum_{k}\hat{\psi}_{k}^{\dagger}\,M_{\delta}(k)\,\hat{\psi}_{k}=-\sum_{k}\hat{\psi}_{k}^{\dagger}\,M_{0}(k-\delta)\,\hat{\psi}_{k}.
\end{equation}
If we apply it onto an eigenstate of the former Hamiltonian, $H_{\delta=0}$, we get:
\begin{align}
H_{\delta}\ket{k,0}_{\pm}=&-\sum_{q}\hat{\psi}_{q}^{\dagger}\,M_{0}(q-\delta)\,\hat{\psi}_{q}\,\hat{\psi}_{k}^{\dagger}\,\hat{n}_{\pm}(k)\ket{0}=\nonumber\\
&-\hat{\psi}_{k}^{\dagger}\,M_{0}(k-\delta)\,\hat{n}_{\pm}(k)\ket{0}.
\end{align}
As we see, the Hamiltonian matrix is evaluated at the momentum value $k-\delta$, whereas the isospin vector is evaluated at $k$. In order to satisfy the eigenvector equation, both need to be evaluated at the same momentum value. Therefore, the states $\ket{k,0}_{\pm}$, which are the eigenstates of the Hamiltonian with no phase $\delta$, are no longer eigenstates of the Hamiltonian once the phase $\delta$ has been added. This is due to the fact that the Hamiltonian has an internal structure corresponding to the two-level system that constitute the two sites within each unit cell.

In the case of a simple chain, the momentum is the only magnitud that constitutes each eigenstate. Therefore, if a certain state has the appropriate momentum value, i.e. one of the momentum values that form the momentum basis, it will always be an eigenstate of the system, being its corresponding energy the only quantity that depends on the phase $\alpha$.

\begin{figure}[t]
  \centering
    \includegraphics[width=0.485\textwidth]{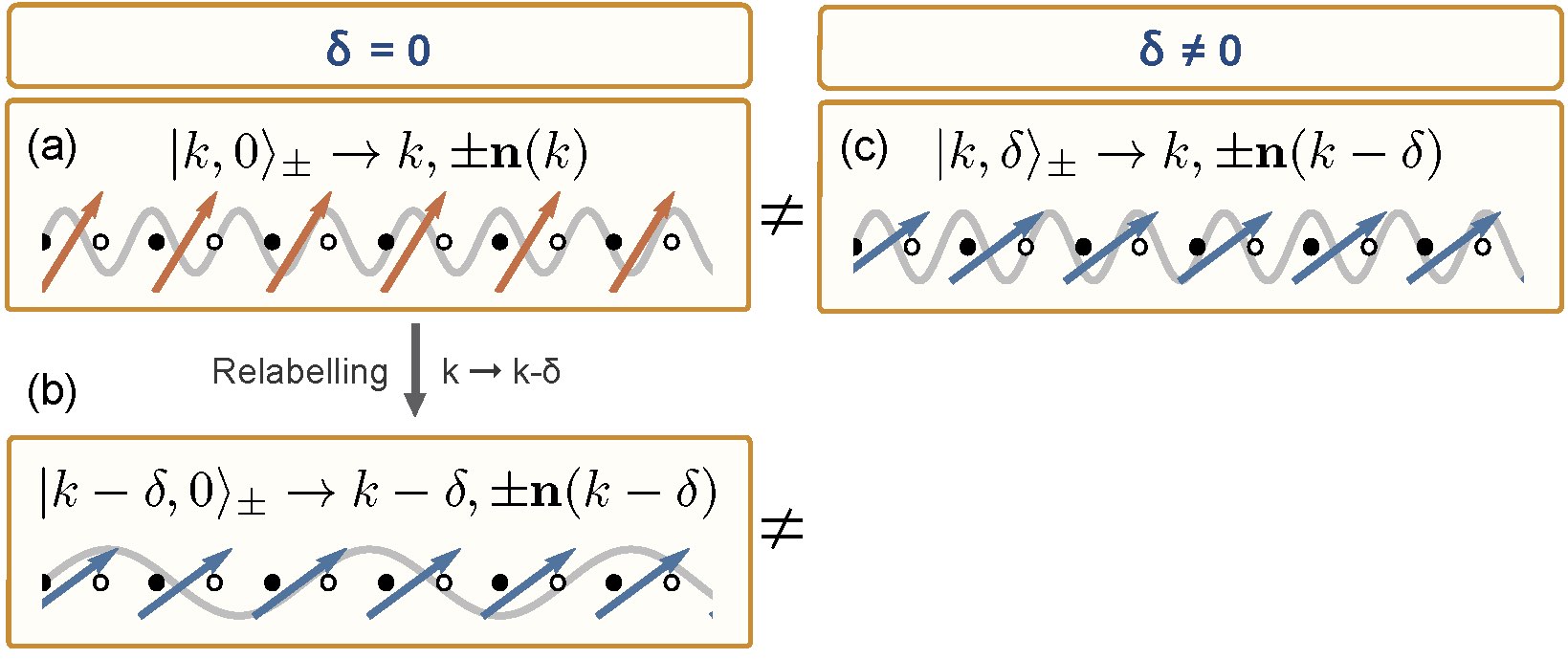}
    \caption[Momentum-isospin correspondence.]{\textbf{Momentum-isospin correspondence.} (a) The eigenstates of the Hamiltonian $H_{\delta=0}$ are characterized by a momentum $k$ and an isospin vector $\bm{n}(k)$. (b) Relabelling the momentum of such states gives us a new set of states, with momentum $k-\delta$ and isospin vector $\bm{n}(k-\delta)$, however the relation between these two magnitudes remains the same. (c) The eigenstates of the Hamiltonian $H_{\delta}$ are characterized by a momentum $k$ and an isospin vector $\bm{n}(k-\delta)$, what makes them be genuinely different from those eigenstates of $H_{\delta=0}$ (and any relabelling of such states), as they show a distinct momentum-isospin correspondence.}
    \label{fig:FiguraAIII05}
\end{figure}

On the contrary, in the case of a dimerized lattice there are two physical magnitudes associated to each eigenstate: the momentum and the isospin vector. Having the appropriate momentum value is not enough for a state to be an eigenstate of the Hamiltonian. In addition, the correspondence between its momentum and its isospin vector has to be the same as the one in the Hamiltonian. That is, if there is a phase $\delta$ in the model, the Hamiltonian shows a shift in the momentum-isospin correspondence, as the Hamiltonian matrix is evaluated at the momentum value $k-\delta$, whereas the creation and annihilation operators correspond to a momentum value $k$ [see Eq.~(\ref{eq:HamiltonianDeltaMatrixShift})]. As a consequence, the eigenstates of such Hamiltonian must have the same shift between its momentum and isospin in order to be truly eigenstates. The eigenstates of $H_{\delta}$ are then:
\begin{equation}
\ket{k,\delta}_{\pm}=\hat{\psi}_{k}^{\dagger}\,\hat{n}_{\pm}(k-\delta)\ket{0},
\end{equation}
and are such that: $H_{\delta}\ket{k,\delta}_{\pm}=\mp\rho(k-\delta)\ket{k,\delta}_{\pm}$. These states are genuinely different from the eigenstates of $H_{\delta=0}$, and cannot be obtained from them by making a relabelling of the momentum. If we consider an eigenstate of $H_{\delta=0}$, $\ket{k,\delta=0}_{\pm}$, and just evaluate it at a different momentum $k-\delta$, we will get $\ket{k-\delta,\delta=0}_{\pm}$. This new state will not be an eigenstate of $H_{\delta}$, as they do not share the same momentum-isospin correspondence (see Fig.~\ref{fig:FiguraAIII05}).

What is the reason for this behavior? It is due to the fact that in a dimerized lattice a shift in the momentum is nothing trivial. In order to perform a shift in the momentum, as in any lattice with a discrete translational symmetry, we need to implement a phase that particles pick up when traveling from one unit cell to the next one. In the simple lattice, this means the same phase for all couplings in the model, whereas for a dimerized lattice this introduces a modification in the isospin vector that determines the Hamiltonian matrix [see Eq.~(\ref{eq:HamiltonianMatrixVectorAIII})].

\subsection{Equivalence to a ladder Hamiltonian with flux}

\begin{figure}[t]
  \centering
    \includegraphics[width=0.485\textwidth]{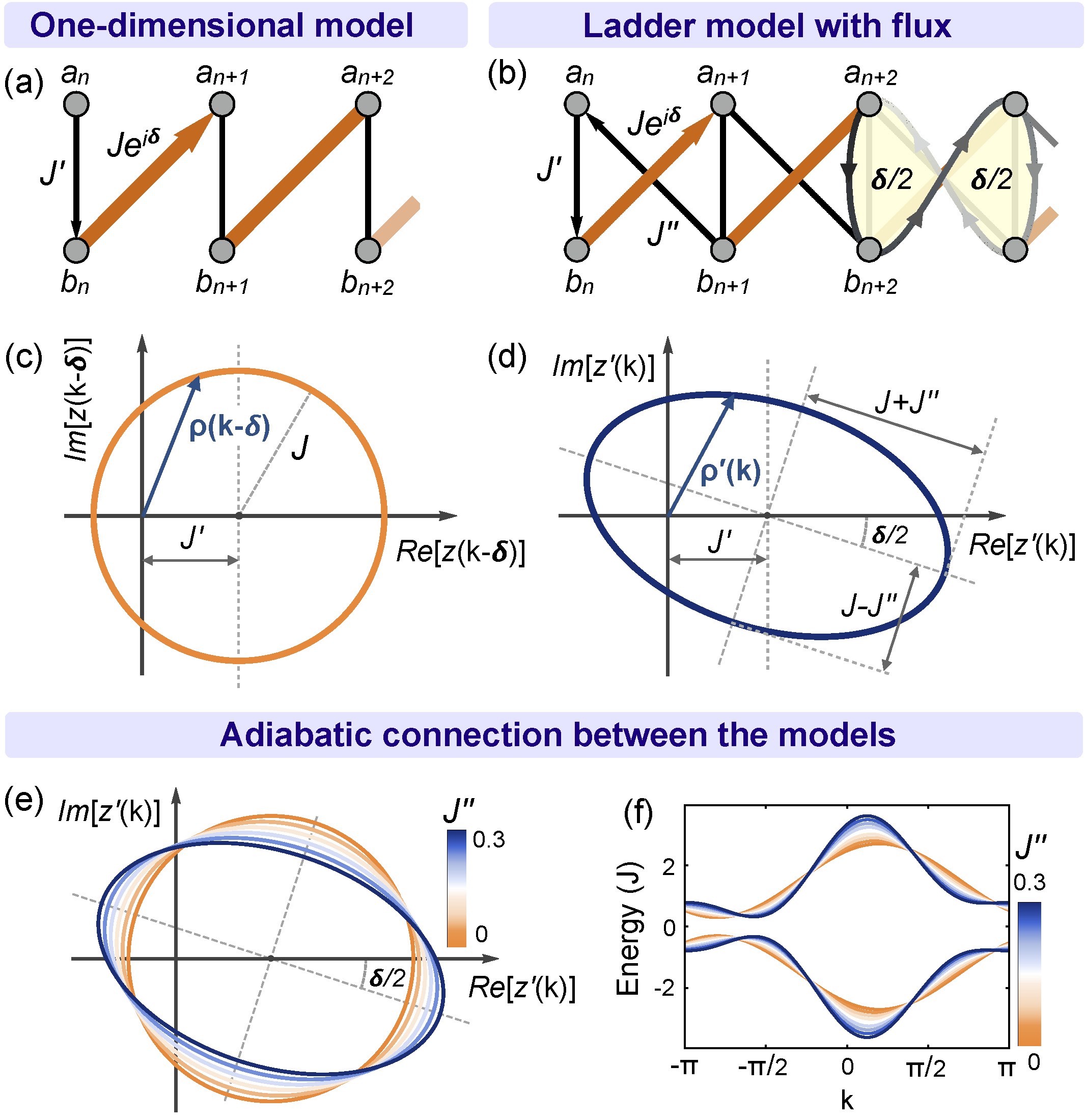}
    \caption[Equivalence to a ladder model with flux.]{\textbf{Equivalence to a ladder model with flux.} (a) Schematic illustration of our model for a topological insulator in the AIII class seen as a ladder. It can be deformed into a ladder Hamiltonian (b) in the same symmetry class. Here, the phase $\delta$ corresponds to the total phase accumulated along a closed path, and thereby to an effective magnetic flux. For the one-dimensional model ($J=1.2$, $J^{\prime}=0.8$ and $\delta=\pi/5$) the two components of the Hamiltonian matrix describe a circle (c), whereas they describe an ellipse (d) for the ladder model  ($J=1.2$, $J^{\prime}=0.8$, $J^{\prime\prime}=0.3$ and $\delta=\pi/5$). (e) Both models can be smoothly connected without crossing the origin in the complex plane. Therefore, their corresponding energy bands can be continuous deformed without closing the gap (f).}
    \label{fig:LadderModel} 
\end{figure}

There is and alternative way of seeing how the phase $\delta$ determines the symmetry properties of our one-dimensional Hamiltonian and thus its symmetry class. It consists of continuously deforming our one-dimensional Hamiltonian, Eq.~(\ref{eq:HamiltonianPosition}), into the following ladder Hamiltonian:
\begin{equation}\label{eq:LadderHamiltonian}
H_{\text{Ladder}}=H_{\delta}-J''\sum_{n}\left(\hat{a}_{n}^{\dagger}\hat{b}_{n+1}^{}+\text{h.c.}\right).
\end{equation}
When transforming the one-dimensional model into the ladder model, the phase $\delta$ is transformed into a non-zero magnetic flux per plaquette, which is precisely equal to  $\delta$
[Fig.~\ref{fig:LadderModel}(b)].
In the ladder model, it is clear that for $\delta=0$ the magnetic flux vanishes and the system is time reversal symmetric, whereas for $\delta \ne 0$ the magnetic flux leads to a time reversal breaking model in the AIII symmetry class. The particular case
\noindent in which $\delta=\pi$ is an exception, as, although constituting a non-vanishing effective magnetic flux per plaquette, it just introduces a negative sign in one of the tunneling amplitudes. The corresponding Hamiltonian is real so that time reversal and charge conjugation symmetries are preserved and therefore the Hamiltonian belongs to the BDI class.
Both Hamiltonians, the one-dimensional one and the ladder one, are continuously connected without closing any energy gap, therefore they share the same symmetry properties. As a result, the one-dimensional model is also in the AIII symmetry class for $\delta \ne 0,\pi$.\\

In order to see how the two Hamiltonians are continuously connected we need to analyse the Hamiltonian matrices and the energy bands. The Hamiltonian matrix of our one-dimensional model, Eq.~(\ref{eq:AIIIHamiltonianMatrix}), can be written as:
\begin{equation}
M_{\delta}(k)=\text{Re}\left[z(k-\delta)\right]\sigma_{x}+\text{Im}\left[z(k-\delta)\right]\sigma_{y},
\end{equation}
with:
\begin{equation}
z(k)=J^{\prime}+Je^{ik}.
\end{equation}
If we draw the function $z(k-\delta)$ on the complex plane, we obtain a circle parametrized by the momentum $k$ of radius $J$ centred at the point $(J^{\prime},0)$ [see Fig.~\ref{fig:LadderModel}(c)]. The model has two energy bands with energies $E_{\pm}(k)=\mp\rho(k-\delta)$, where $\rho(k-\delta)$ is given by the distance between the origin and the curve point $z(k-\delta)$ [see Fig.~\ref{fig:LadderModel}(c)]. Hence, the energy gap closes only if the curve crosses the origin, something that only happens at the topological transition point $J^{\prime}=J$.

On the other side, the ladder model $H_{\text{Ladder}}$ in Eq.~(\ref{eq:LadderHamiltonian}) corresponds to a Hamiltonian matrix:
\begin{equation}
M_{\text{Ladder}}(k)=\text{Re}\left[z^{\prime}(k)\right]\sigma_{x}+\text{Im}\left[z^{\prime}(k)\right]\sigma_{y},
\end{equation}
with
\begin{equation}
z^{\prime}(k)=z(k-\delta)+J^{\prime\prime}e^{-ik}
\end{equation}

The new function $z^{\prime}(k)$ that characterizes the Hamiltonian matrix describes an ellipse with axes $J+J^{\prime\prime}$ and $|J-J^{\prime\prime}|$, rotated an angle $\delta/2$ [Fig.~\ref{fig:LadderModel}(d)], and the energy bands of the model are given by $E^{\prime}_{\pm}(k)=\mp\rho^{\prime}(k)$, where $\rho^{\prime}(k)$ corresponds to the distance between the origin and the curve point $z^{\prime}(k)$ [see Fig.~\ref{fig:LadderModel}(d)]

As the tunneling amplitude $J^{\prime\prime}$ is increased from zero to a given value, the circle is continuously deformed into the ellipse [Fig.~\ref{fig:LadderModel}(e)]. Since this occurs without the curves crossing the origin in the complex plane, the corresponding energy bands are continuously deformed without closing the energy gap [Fig.~\ref{fig:LadderModel}(f)]. As a conclusion, both Hamiltonians share the same symmetry properties and thus belong to the same symmetry class. Therefore, our one-dimensional model belongs to the BDI symmetry class only when $\delta=0,\pi$, being in the AIII symmetry class for $\delta\neq0,\pi$. 

\subsection{Non-zero average momentum edge modes}

There exists a relation between the symmetry class of the Hamiltonian and the momentum of the symmetry protected edge modes that the system exhibits when it is found to be in its topologically non-trivial phase. This correspondence serves to make evident the relevance of the phase $\delta$ and its direct relation with the breaking of time reversal and charge conjugation symmetries.

In this context, if a Hamiltonian $H$ presents timer reversal symmetry there is a global unitary transformation $U_{T}$ such that:
\begin{equation}
U_{T}\,H\,U^{\dagger}_{T}=H.
\end{equation}
Therefore, if a certain state $\ket{e}$ is an eigenstate of the Hamiltonian with some energy $E$, then the transformed state $U_{T}\ket{e}^{*}$ is also an eigenstate of the Hamiltonian with the same energy. In case of a non degenerate state, as the edge states, this means that this transformation leaves the state invariant up to a phase, which can be absorbed in the unitary operator $U_{T}$. That is:
\begin{equation}
U_{T}\ket{e}^{*}=\ket{e},
\end{equation}
and consequently, opposite momentum modes are on average equally occupied:
\begin{align}
&\langle e|\,\hat{n}_{k}\,|e\rangle=\langle e^{*}|\,\hat{n}_{-k}\,|e^{*}\rangle=\nonumber\\
&\langle e^{*}|\,U^{\dagger}_{T}\,\hat{n}_{-k}\,U_{T}\,|e^{*}\rangle=\langle e|\,\hat{n}_{-k}\,|e\rangle.
\end{align}
As a result, the momentum density distribution of the edge modes of a time reversal symmetric Hamiltonian is an even function of the momentum and, thus, such states have a zero average momentum.

\subsection{Gauge transformations, topology and boundary conditions}

Despite everything we have already explained in this section about the role that the phase $\delta$ plays in our one-dimensional model, there is a gauge transformation that removes the phase $\delta$ form the system. This unitary is given by:
\begin{equation}\label{eq:GaugeTransformationDelta}
V:\begin{cases}
\hat{a}_{n}^{\dagger}\longrightarrow e^{i\delta n}\hat{a}_{n}^{\dagger}\\
\hat{b}_{n}^{\dagger}\longrightarrow e^{i\delta n}\hat{b}_{n}^{\dagger},\end{cases}
\end{equation}
so that, after all, the Hamiltonian with a phase $\delta$ and the one with no phase, can be connected through a gauge transformation:
\begin{equation}
V\,H_{\delta}\,V^{\dagger}=H_{\delta=0}.
\end{equation}
However, this does not mean that both models are equivalent, nor they share the same symmetry properties.

Why is that? The fundamental reason is that not every unitary operator that can be applied to the Hamiltonian are allowed in the definition of the different symmetries that serve to classify topological insulators. The model can exhibit topological features because of having two sites per unit cell. This makes the Hamiltonian be able to be written using a spinor-like discrete field $\hat{\psi}^{\dagger}_{k}$, and it is in this internal space where the topological properties are found. As a two level-quantum system, this internal space can be described by using the Bloch sphere representation. Chiral symmetry implies that, instead of being able to reach any point in the sphere, the vectors that characterize the system are constrained to a particular plane crossing the center of the sphere. Therefore there are two possible situations that correspond to two distinct topological phases: enclosing the origin forming a complete circle (winding number $1$) or forming just an arc that does not enclose the origin and can be continuously deformed into a point (winding number $0$). When using the symmetry properties in order to classify distinct topological insulators we need to consider those unitary transformations that preserve the topological nature of the system. In other words, they need to consist of the same unitary transformation applied to every vector in this internal space.

We can give an example that serves to easily visualize this. Let us consider a tight-binding model with two sites per unit cell and translational invariance, so that its corresponding Hamiltonian can be written in momentum space as:
\begin{equation}
H=\sum_{k}\hat{\psi}^{\dagger}_{k}\,\bm{v}(k)\cdot\bm{\sigma}\,\hat{\psi}^{\dagger}_{k},
\end{equation}
being $\bm{v}(k)$ a vector that depends on the momentum. In this way, the system is totally characterized by a finite set of vectors, one for each quantized momentum value. If we were allowed to consider any unitary transformation that can be applied to the whole Hamiltonian in order to define the symmetry transformations, we could define the following unitary:
\begin{equation}
U=\sum_{k}\hat{\psi}^{\dagger}_{k}\,\bm{u}(k)\cdot\bm{\sigma}\,\hat{\psi}^{\dagger}_{k},
\end{equation}
choosing each vector $\bm{u}(k)$ to be orthogonal to each vector $\bm{v}(k)$, that is: $\bm{u}(k)\cdot\bm{v}(k)=0\quad\forall k$. Consequently, the unitary operator $U$ is such that $UHU^{\dagger}=-H$ and thus the Hamiltonian would be chiral symmetric. This means that every Hamiltonian of that kind, which is quite general, would be topologically non-trivial, would represent a topological insulator and would exhibit symmetry protected edge modes, which is clearly not true.\\
\begin{figure}[t]
    \includegraphics[width=0.475\textwidth]{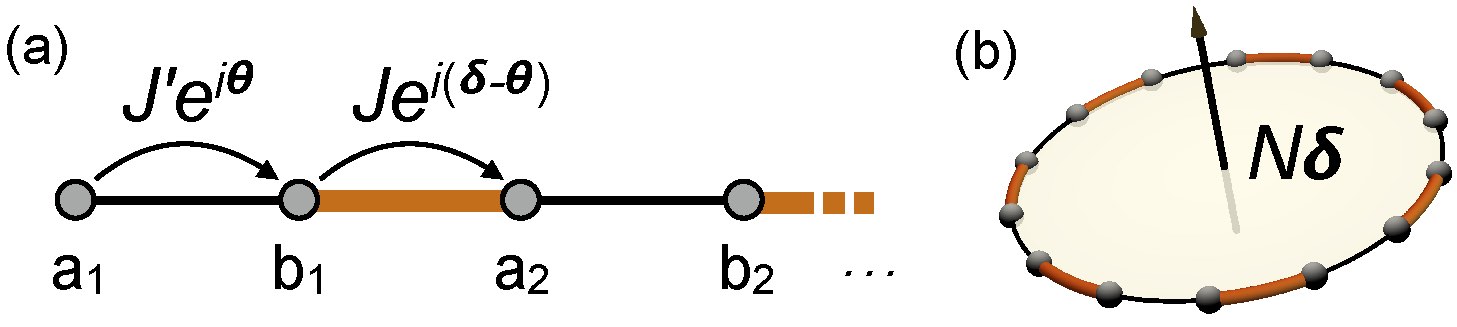}
    \caption[Phase $\delta$ as an effective magnetic flux.]{\textbf{Phase $\delta$ as an effective magnetic flux.} (a) In a dimerized one-dimensional lattice with two different complex hopping amplitudes only the sum of the two phases is relevant, i.e. the total phase accumulated when a particle travels from one site to the same type of site in the following cell. In this way, each possible value of $\delta$ constitutes a different model that belongs to a continuous set of equivalent models parametrised by the phase $\theta$. All equivalent models are connected by global rotations around the $z$-axis (see text). (b) When joining the two ends of the lattice particles pick up a phase $N\delta$ when completing a closed path along the whole ring. Therefore, the phase $\delta$ is directly related to an effective magnetic flux penetrating the system.}
    \label{fig:FiguraAIII07} 
\end{figure}
We can consider a more general Hamiltonian than our one-dimensional Hamiltonian in Eq.~(\ref{eq:HamiltonianPosition}), which consist of a dimerized lattice with two different phases in the hopping amplitudes, $\delta$ and $\theta$ [see Fig.~\ref{fig:FiguraAIII07}(a)]. The Hamiltonian of such model is:
\begin{equation}
H_{\delta,\theta}=-\sum_{n}^{N}\big(J^{\prime}e^{-i\theta}\,\hat{a}_{n}^{\dagger}\hat{b}_{n}^{}+Je^{i(\delta-\theta)}\,\hat{a}_{n+1}^{\dagger}\hat{b}_{n}^{}+\text{H.c.}\big).
\end{equation}
In the momentum representation this Hamiltonian looks like:
\begin{equation}
H_{\delta,\theta}=-\sum_{k}\hat{\psi}^{\dagger}_{k}\,R_{z}(\theta)\,M_{\delta}(k)\,R^{\dagger}_{z}(\theta)\hat{\psi}_{k},
\end{equation}
being $M_{\delta}(k)$ the same Hamiltonian matrix as the one corresponding to our on-dimensional model, Eq.~(\ref{eq:AIIIHamiltonianMatrix}), and $R_{z}(\theta)=e^{i\theta\sigma_{z}/2}$ a rotation around the $z$-axis of an angle $\theta$.

As we see, the phase $\theta$ is an irrelevant quantity, as it can be removed by performing a global unitary transformation, whereas $\delta$ is not. This is due to the fact that $\delta$ is the total phase accumulated by a particle when travelling from any site in the lattice to the same site in the following lattice cell. As we explained before in this section, this phase constitutes a shift in the momentum and changes the momentum-isospin correspondence. Furthermore, if we consider periodic boundary conditions, it turns out that there is an effective magnetic flux penetrating the ring formed by the whole lattice which is precisely equal to $N\delta$ [see Fig.~\ref{fig:FiguraAIII07}(b)]. It is clear then that the phase $\delta$ can be removed by a gauge transformation only for open boundary conditions, being a gauge invariant quantity when the system is subjected to periodic boundary conditions.

A topological insulator like our model exhibits a bulk-edge correspondence, that is a direct relation between two ways in which its topological nature can be manifested. On one hand, for periodic boundary conditions, it is manifested in the bulk through the non-trivial Zak phase. On the other hand, for open boundary conditions, through the existence of symmetry protected zero modes localized at the edges of the system. Following this bulk-edge correspondence, if the phase $\delta$ is relevant and determines the symmetry properties and topological class of the model in the bulk, it has to be the same when we look at the edges of the system. 

In conclusion, not every gauge transformation is allowed when considering the symmetry transformations that classify topological insulators. They do not only need to be unitary, but also preserve the topology of the system. In other words, only those that act equally on every vector in the internal space are permitted. Alternatively, we can say that only those unitary transformations that act on the system independently of having periodic or open boundary conditions can be considered to constitute symmetry transformations. In fact, the gauge transformation in Eq.~(\ref{eq:GaugeTransformationDelta}), which is not a global transformation, does not commute with a change of the boundary conditions.


\section{Edge modes in the AIII class}

\subsection{Emergence of edge modes in the topological phase}

When the edges of the one-dimensional lattice are not connected with each other, and thus the system consists of an open chain, the Bloch modes in Eq.~(\ref{eq:AIIIBlochEigenModes}) are no longer the proper eigenstates of the Hamiltonian, as they do not fulfil the open boundary conditions. Nevertheless, they constitute an orthonormal basis of the space of states and are really convenient for obtaining the right eigenstates of the Hamiltonian. For that, we need to consider linear superpositions of two Bloch modes with arbitrary momenta such that their corresponding energy, according to the dispersion relation, are the same. That is, we need to combine the Bloch mode with momentum $q$ with the one with momentum $q^{\prime}=-q+2\delta$. Then, we impose the boundary conditions, what gives us, on one hand, the relation between the two coefficients in such linear superposition and, on the other hand, the quantization condition for $q$.

The eigenmodes of the model for open boundary conditions are:
\begin{equation}
\hat{c}_{\pm,q}^{\dagger}\propto\sum_{n}e^{i\delta n}\left[ \sin\big(qn-\varphi(q)\big)\hat{a}_{n}^{\dagger}\pm\sin qn\,\hat{b}_{n}^{\dagger} \right],
\end{equation}\label{eq:AIIIOpenBulkModes}
\begin{figure}[H]
  \centering
    \includegraphics[width=0.375\textwidth]{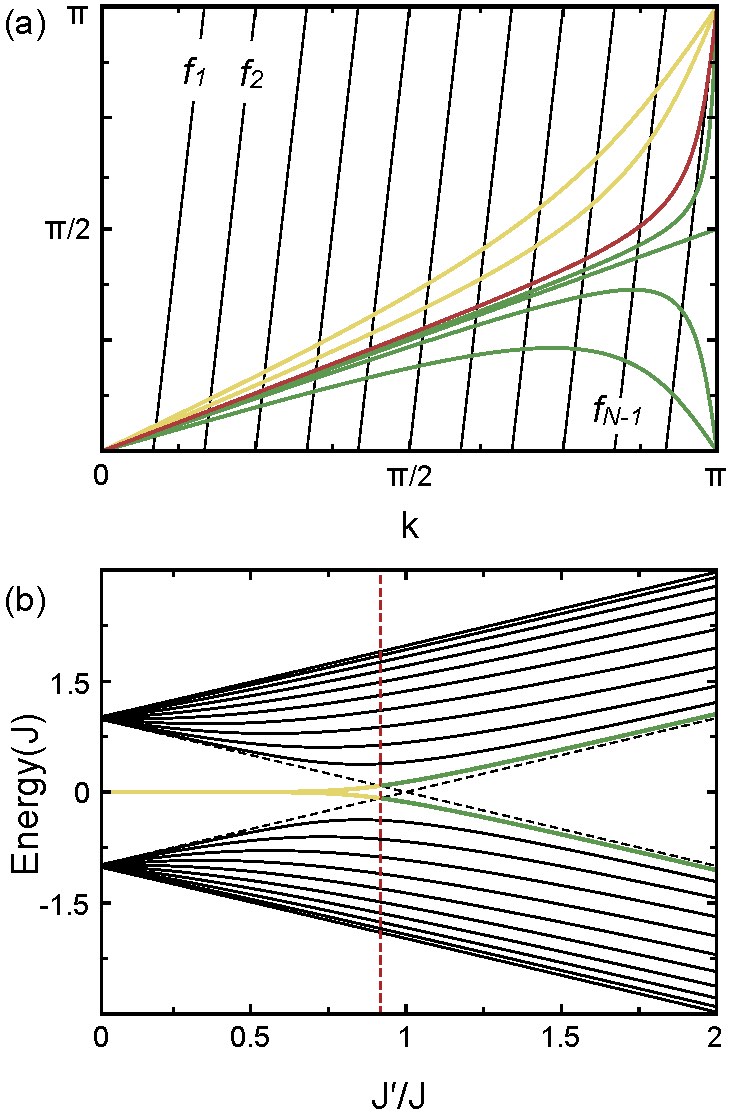}
    \caption[Emergence of edge modes in the topological phase]{\textbf{Emergence of edge modes in the topological phase.} (a) Graphic representation of the momentum quantization condition in Eq.~(\ref{eq:AIIIOpenBulkModes}) for a total number of unit cells $N=11$. The black lines correspond to the functions $f_{m}(q)=q(N+1)-m\pi$ and the coloured curves to the function $\varphi(q)$ for different values of the ratio between the two hopping amplitudes, $J^{\prime}/J$. The crossing points between $\varphi(q)$ and each line $f_{m}(q)$ correspond to each quantized value of $q$ and, thus, to pairs of bulk eigenstates of the Hamiltonian, one for each energy band. For $J^{\prime}/J>1-1/(N+1)$, the function $\varphi(q)$ crosses every line $f_{m}(q)$ and, therefore, there exist $2N$ bulk eigenstates (green curves, corresponding to $J^{\prime}/J=1.5,\,1.1,\,1$ and $0.967$ starting from the bottom). For $J^{\prime}/J<1-1/(N+1)$, the function $\varphi(q)$ crosses just the first $N-1$ lines $f_{m}(q)$, so that the Hamiltonian exhibits only $2N-2$ bulk eigenstates, being the two missing states the edge modes (yellow curves, corresponding to $J^{\prime}/J=0.5$ and $0.7$ starting from above). The transition point between the absence and presence of edge modes, $J^{\prime}/J=1-1/(N+1)$, corresponds to the case in which the slope of the tangent line to the function $\varphi(q)$ (red curve) at $q=\pi$ is equal to the slope of the line $f_{N}(q)$. (b) Energy diagram for a total number of unit cells $N=11$. The black curves correspond to the energy of the bulk eigenstates of the Hamiltonian as a function of the ratio $J^{\prime}/J$. The green curves are the energy of the lowest energy bulk eigenstate in the upper band and the highest energy bulk eigenstate in the lower band. At the transition point $J^{\prime}/J=1-1/(N+1)$ (red dashed line), these two eigenstates are no longer in the bulk and they become the two edge states, whose energy is exponentially close to zero (yellow curves).}
    \label{fig:FiguraAIII08} 
\end{figure}
being the quantum number $q$ quantized according to the following equation:
\begin{equation}
q(N+1)-m\pi=\varphi(q),
\end{equation}
where $\varphi(k)=\arg(J^{\prime}+Je^{ik})$ and being $m$ and integer running from $1$ to $N$. In this way, for each value of $m$ we would obtain a different value of $q$ that satisfies the quantization condition and corresponds to a pair of eigenstates, one for each energy band, with energies $E_{\pm}(q)=\mp\rho(q)$, being $\rho(q)=|J^{\prime}+Je^{ik}|$. It is important to notice that the quantum number $q$, that labels all bulk eigenstates of the Hamiltonian for open boundary conditions, does not coincide with the momentum of such states. A bulk mode of the form in Eq.~(\ref{eq:AIIIOpenBulkModes}) has two different momentum components, one with momentum $q+\delta$ and another one with momentum $-q+\delta$.

There is a region in the parameter space in which the quantization condition has $N$ different solutions, leading to $2N$ bulk eigenstates, $N$ for each energy band. However, when $J^{\prime}/J<1-1/(N+1)$ there are just $N-1$ solutions and therefore only $2N-2$ bulk eigenstates of the form in (Eq.~\ref{eq:AIIIOpenBulkModes}). In that case the two missing eigenstates will be the so-called edge modes or zero modes [see Fig.~\ref{fig:FiguraAIII08}(a)].

In this way, if we consider the system to be somewhere in its topologically trivial phase and then we continuously modify the ratio between the two hopping amplitudes $J^{\prime}/J$ towards the topological phase, the edge modes will not appear exactly at the topological transition point $J^{\prime}=J$ but slightly later on, when the two tunneling amplitudes are such that $J^{\prime}/J=1-1/(N+1)$ [see Fig.~\ref{fig:FiguraAIII08}(b)]. Consequently, we can say that there are two critical points, one that indicates the topological phase transition between the two topologically distinct phases and another one that separates the  absence and presence of edge modes in the system. Both are quite close to each other and, when considering a large number of unit cells in the lattice, $N>>1$, they coincide, what stands for a bulk-edge correspondence. In other words, the presence of symmetry protected zero modes at the edges of the system is directly related to a non-trivial topology characterised by the whole set of bulk eigenstates through the Zak phase.\\

In order to obtain the wave function of the edge modes we need to consider Bloch modes with complex momentum, as there exist no more real solutions to the quantization condition other that the $N-1$ that correspond to the $2N-2$ bulk eigenstates. Combining a mode with momentum $k=\pi+\delta+i\xi$ and another one with momentum $k^{\prime}=-\pi+\delta-i\xi$, and imposing the boundary conditions we get the two edge modes of the Hamiltonian:
\begin{equation}
\hat{e}_{\pm}^{\dagger}=\frac{1}{\sqrt{2}}\Big(\tilde{a}_{n=1}^{\dagger}\pm\tilde{b}_{n=N}^{\dagger}\Big),
\end{equation}
where:
\begin{align}
&\tilde{a}_{n=1}^{\dagger}=\frac{1}{\sqrt{c}}\sum_{n}e^{i(\pi+\delta)n}\,\sinh\xi(N+1-n)\,\hat{a}_{n}^{\dagger},\label{eq:EdgeModeWaveFunctionLeft}\\
&\tilde{b}_{N=1}^{\dagger}=\frac{1}{\sqrt{c}}\sum_{n}e^{i(\pi+\delta)n}\,\sinh\xi n\,\hat{b}_{n}^{\dagger},\label{eq:EdgeModeWaveFunctionRight}
\end{align}
being $c=\sum_{n}\sinh^{2}\xi n$ a normalization constant and being $\xi$ determined by the following equation:
\begin{equation}
J^{\prime}\sinh\xi(N+1)=J\sinh\xi N.
\end{equation}
That is, the two edge modes of the Hamiltonian, $\hat{e}_{\pm}^{\dagger}$, are the symmetric and antisymmetric linear combinations of the modes $\tilde{a}_{n=1}^{\dagger}$ and $\tilde{b}_{n=N}^{\dagger}$, which are localized at the left and right edges of the system, respectively. Furthermore, they also exhibit a polarization property, as the mode at the left edge of the lattice occupies only $a$-modes, whereas the one at the right occupies only $b$- modes. In other words, the modes localized at opposite ends of the lattice correspond to orthogonal isospin states.
\begin{figure}[t]
  \centering
    \includegraphics[width=0.35\textwidth]{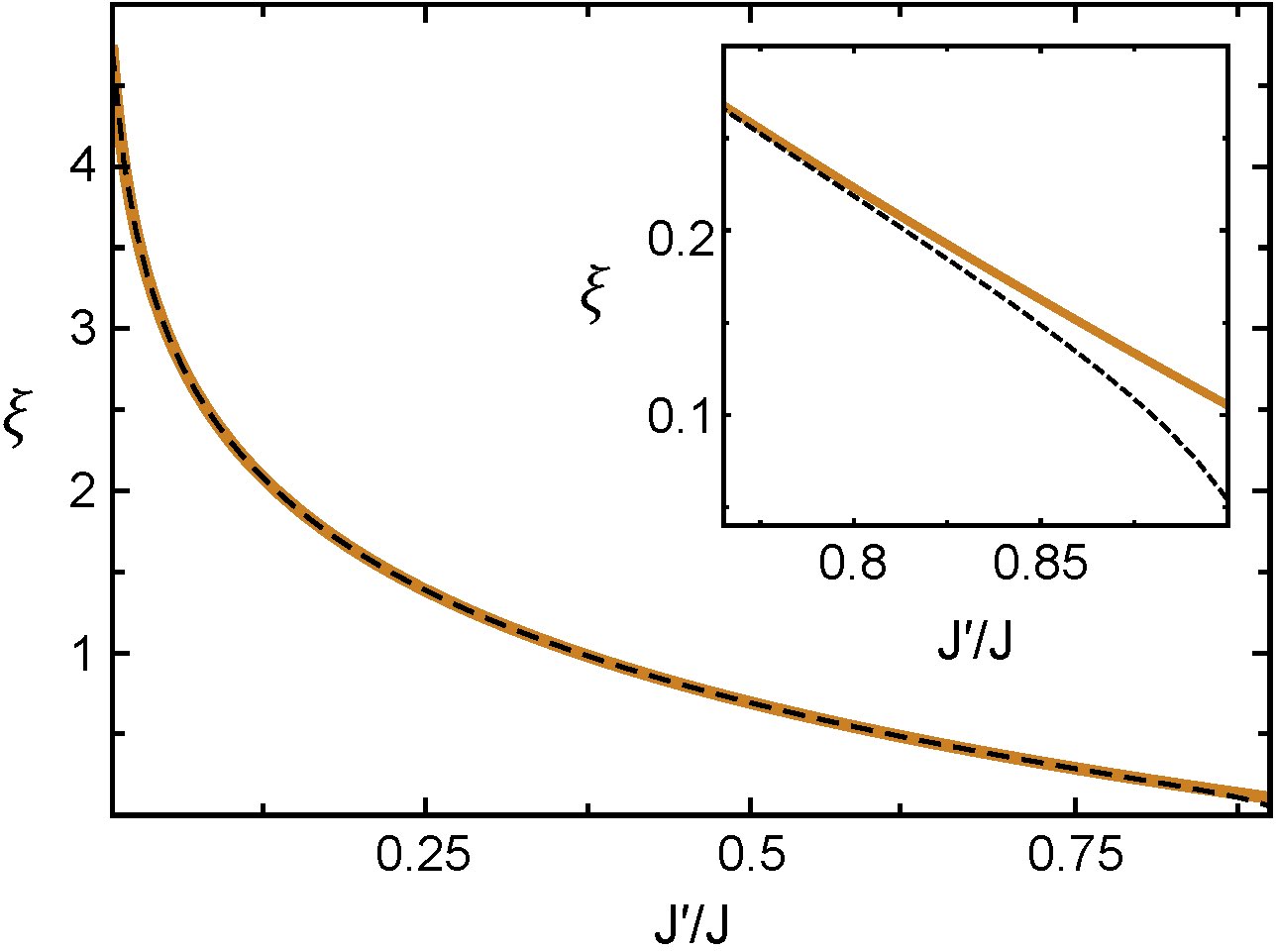}
    \caption[Zero energy approximation.]{\textbf{Zero energy approximation.} Plot of the exact value of $\xi$ (black dashed curve) and its approximated value according to the expression in Eq.~(\ref{eq:ApproximationXi}) (orange curve) as a function of the ratio between the two hopping amplitudes, $J^{\prime}/J$. The approximation fits quite well in most of the topological phase, being not as good in the vicinity of the transition point. As a result, considering the edge modes to have zero energy is a quite good approximation.}
    \label{fig:FiguraAIII09} 
\end{figure}

The quantity $\xi$ is the imaginary part of the edge modes momentum, so that its inverse is a length and, as we can see from the edge modes wave functions, it determines the localization length associated to the edge modes spatial density distribution. In this way, if we consider the edge modes to be spatially concentrated in a region much smaller than the size of the whole system, we can take the limit $\xi N>>1$, what allows us to obtain a approximated expression for $\xi$:
\begin{equation}\label{eq:ApproximationXi}
\xi\approx-\log(J^{\prime}/J).
\end{equation}
As a consequence of this approximated expression for the quantity $\xi$, the energies of the edge modes, which in general are $\mp\sqrt{J^{\prime2}+J^{2}-2J^{\prime}J\cosh\xi}$, can also be approximated and become zero. This approximation holds quite well in most of the topologically non-trivial phase, being not as good in a vicinity of the topological phase transition (see Fig.~\ref{fig:FiguraAIII09}).
The fact that the energy of the edge modes is almost zero implies that they can be consider to constitute a two-dimensional degenerate eigensubspace and, thus, the two modes from which they are formed, $\tilde{a}_{n=1}^{\dagger}$ and $\tilde{b}_{n=N}^{\dagger}$, are also zero energy eigenstates of the Hamiltonian. This is of course not exactly true, just an approximation. If we prepared a particle in the mode localized at one particular edge of the lattice and let it evolve in time under the Hamiltonian of the system, it would not remain the same as a truly eigenstate would do. It would oscillate between itself and the mode localized at the opposite edge. Nevertheless, the time that we would need to wait until it becomes the other mode is really large as it is proportional to the inverse of the absolute value of the edge modes energy.

\subsection{Edge modes momentum and symmetry class}

As we have anticipated in Sec.~3.4.3., there is a relation between the topological symmetry class of the Hamiltonian and the edge modes that it exhibits. In other words, the edge modes that correspond to the BDI symmetry class, for $\delta=0,\pi$ (SSH model), and the ones in the AIII class, for $\delta\neq0,\pi$, are different and this difference can be measured. As a result, the symmetry class of the system is characterised by the properties of the edge modes.
\begin{figure}[t]
  \centering
    \includegraphics[width=0.475\textwidth]{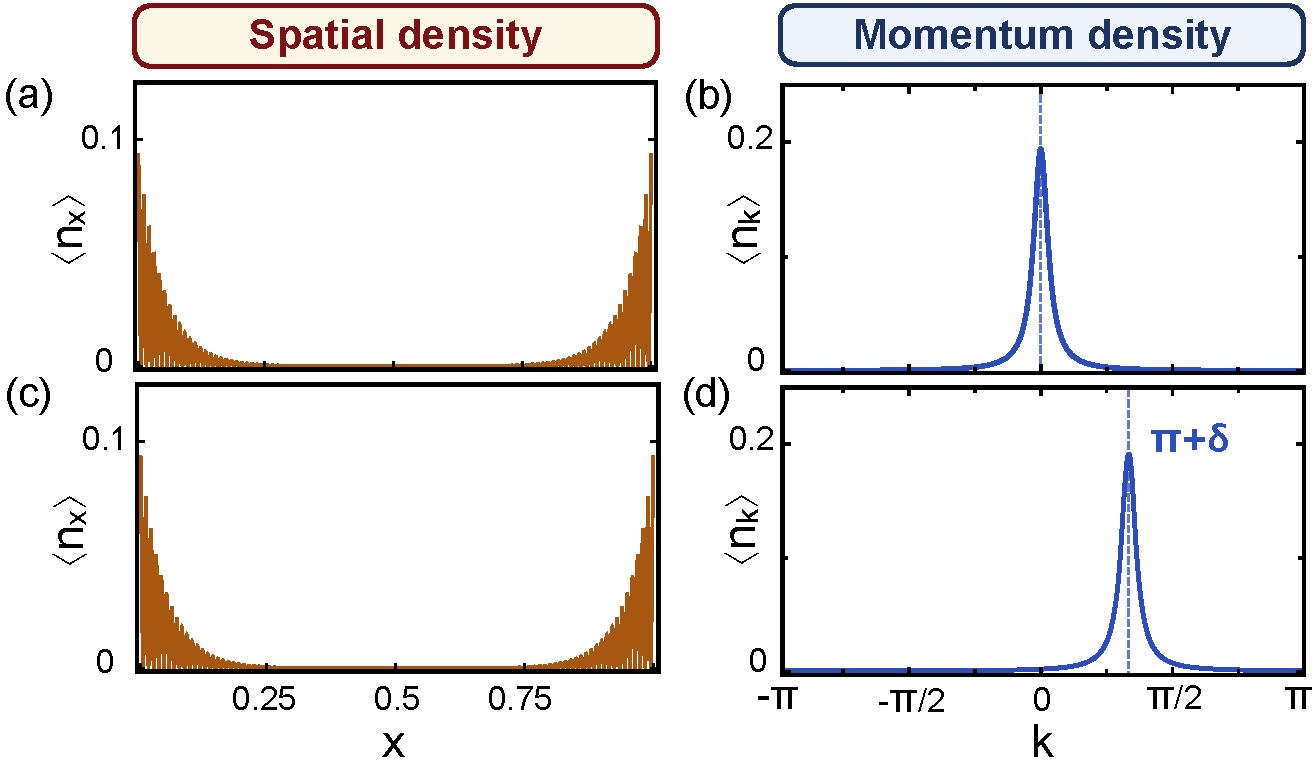}
    \caption[Edge modes momentum and symmetry class]{\textbf{Edge modes momentum and symmetry class.} (a) Spatial density distribution $\langle\hat{n}_{x}\rangle$ and (b) momentum density distribution $\langle\hat{a}_{k}^{\dagger}\hat{a}_{k}\rangle=\langle\hat{b}_{k}^{\dagger}\hat{b}_{k}\rangle$ of the edge modes for $N=100$, $J^{\prime}/J=0.9$ and $\delta=0$ (BDI symmetry class). (c) Spatial and (d) momentum density distributions of the edge modes for $N=100$, $J^{\prime}/J=0.9$ and $\delta=-2\pi/3$ (AIII symmetry class). While the edge modes spatial density distributions are the same in both symmetry classes, it is through their momentum distributions how the Hamiltonian symmetry class is manifested. Edge modes in the BDI class have a zero average momentum, imposed by the presence of time reversal symmetry. On the contrary, the AIII symmetry class is characterised by non-zero average momentum edge modes. All data represented in this figure has been obtained by exact numerical diagonalization of the Hamiltonian.}
    \label{fig:FiguraAIII10} 
\end{figure}
When we look at the edge modes spatial density distribution, we find no difference at all between the BDI class and the AIII class. In both symmetry classes the edge modes are concentrated at the edges of the system [see Fig.~\ref{fig:FiguraAIII10}(a) and (c)]. On the contrary, it is their momentum density distribution what has the information about the Hamiltonian symmetry class. If we look at the edge modes wave functions, Eq.~(\ref{eq:EdgeModeWaveFunctionLeft}) and Eq.~(\ref{eq:EdgeModeWaveFunctionRight}), we can identify a momentum term $e^{i(\pi+\delta)n}$, what indicates that their momentum density distribution is shifted by $\pi+\delta$. In the BDI symmetry class, $\delta=0,\pi$, the edge modes momentum distribution consists of an even function centred at $k=0$ or $k=\pi$ [see Fig.\ref{fig:FiguraAIII10}(b)], so that edge modes in the BDI class have a zero average momentum. In contrast, in the AIII symmetry class, $\delta\neq0,\pi$, the edge modes density distribution is centred at $k=\pi+\delta$ [see Fig.\ref{fig:FiguraAIII10}(d)], so that the edge modes average momentum is different from zero.

This difference between the momentum of the edge modes in the BDI class and in the AIII class is nothing casual. As we explained in Sec.~3.4.3., the presence of time reversal symmetry forces the edge modes to have a zero average momentum. Therefore, edge modes with a non-zero average momentum are a direct consequence of the breaking of time reversal symmetry and constitute a hallmark of the AIII symmetry class.

\subsection{Edge modes spatial and momentum localizations}
\begin{figure}[t]
  \centering
    \includegraphics[width=0.485\textwidth]{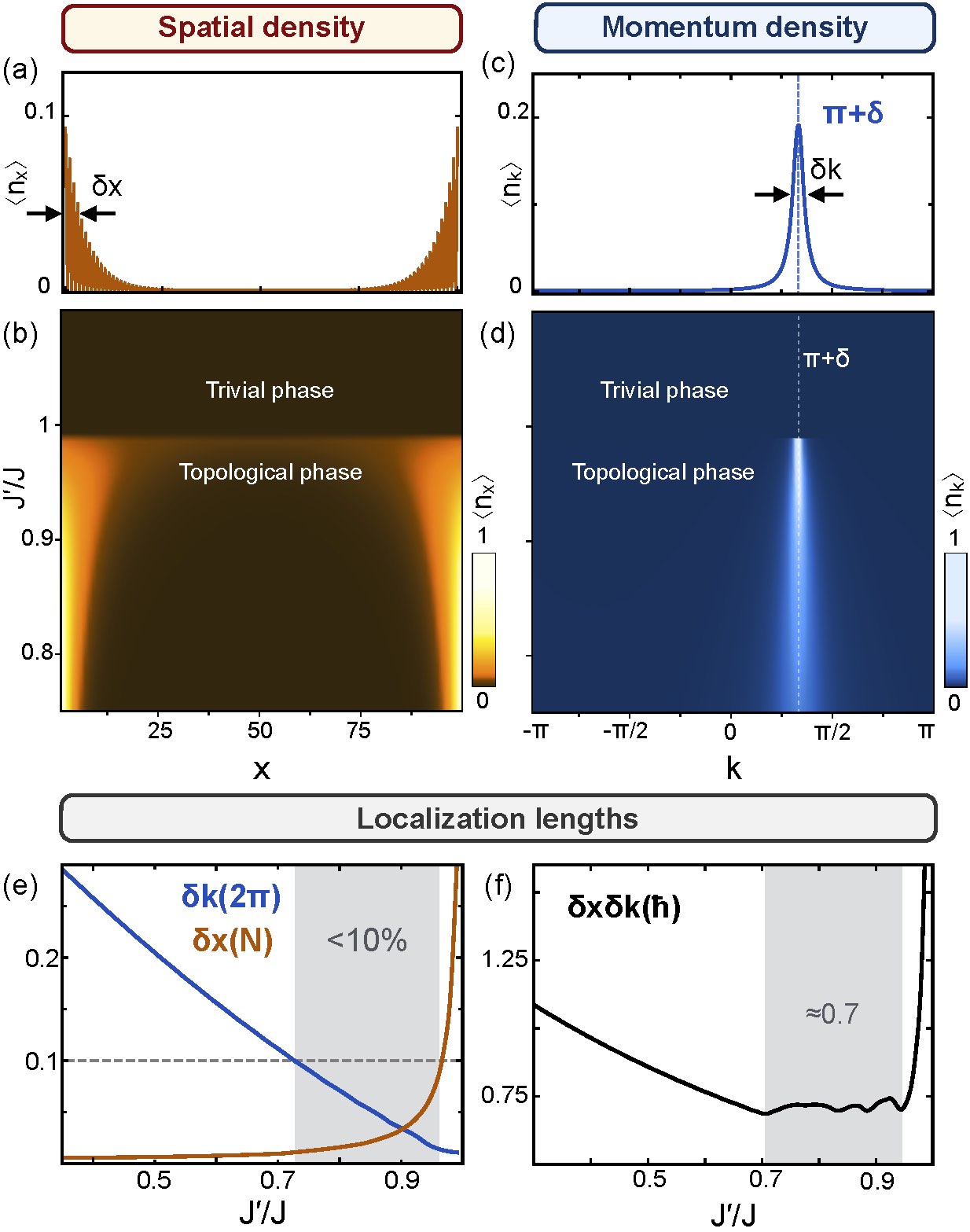}
    \caption[Edge modes spatial and momentum localizations.]{\textbf{Edge modes spatial and momentum localizations.} (a) Spatial density distribution $\langle\hat{n}_{x}\rangle$ and (c) momentum density distribution $\langle\hat{a}_{k}^{\dagger}\hat{a}_{k}\rangle=\langle\hat{b}_{k}^{\dagger}\hat{b}_{k}\rangle$ of the edge modes for $N=100$, $J^{\prime}/J=0.9$ and $\delta=-2\pi/3$. (b) Spatial and (d) momentum density distributions of the edge modes as a function of the ratio $J^{\prime}/J$ for $N=100$ and $\delta=-2\pi/3$. (e) Edge modes spatial and momentum localization lengths and (f) combined position-momentum uncertainty as a function of the ratio $J^{\prime}/J$ for $N=100$. The spatial localization of the edge modes is better the farther to the critical point the system is, whereas their momentum localization behaves in the opposite way. Nevertheless, both magnitudes can be simultaneously well defined, reaching a minimum position-momentum uncertainty of $0.7\hbar$. All data represented in this figure has been obtained by exact numerical diagonalization of the Hamiltonian.}
    \label{fig:FiguraAIII11} 
\end{figure}

Besides the manifestation of the Hamiltonian symmetry class through the edge modes momentum distribution, studying the edge modes in momentum space has allowed us to find out another interesting feature of such states. Surprisingly, the edge modes are well localized around their average momentum, in addition to their good spatial localization at the edges of the system.

From the edge modes wave functions, Eq.~(\ref{eq:EdgeModeWaveFunctionLeft}) and Eq.~(\ref{eq:EdgeModeWaveFunctionRight}), we can compute their spatial localization length, denoted as $\delta x$, as the FWHM of their spatial density distribution [see Fig.~\ref{fig:FiguraAIII11}(a)]. This localization length can also be approximated by taking the limit $\xi N>>1$ and is given by: 
\begin{equation}\label{eq:Deltax}
\delta x=N-\frac{1}{\xi}\text{arsinh}\left(\frac{1}{\sqrt{2}}\sinh\xi N\right)\simeq\frac{\log2}{2\xi},
\end{equation}
expressed in units of the total size of the system.

In order to get the edge modes momentum localization length, we first need to write them using the momentum basis. In momentum representation the edge modes are:
\begin{equation}\label{eq:EdgeModesMomentumRepresentation}
\hat{e}_{\pm}^{\dagger}=\frac{1}{\sqrt{2}}\sum_{k}\left[ F(\tilde{k})\,\hat{a}_{k}^{\dagger} \pm e^{i\tilde{k}(N+1)}F(-\tilde{k})\,\hat{b}_{k}^{\dagger}\right],
\end{equation}
with $F(k)=\frac{1}{\sqrt{Nc}}\sum_{n}e^{-ikn}\sinh\xi(N+1-n)$ and $\tilde{k}=k-\pi-\delta$. By making an integral instead of a discrete sum we can obtain the following approximation of the function $F(k)$:
\begin{equation}\label{eq:EdgeModeFourierTransformed}
F(k)=\frac{f(k)}{1+(k/\xi)^2},
\end{equation}
with:
\begin{equation}
f(k)=\cosh\xi N-i\frac{k}{\xi}\sinh\xi N-e^{-ikN}.
\end{equation}
The real and imaginary parts of $f(k)$ have a well defined parity, being its real part even and its imaginary part odd. Therefore, the modulus square of $f(k)$ is an even function of the momentum. In consequence, $|F(k)|^2=|F(-k)|^2$ and, thus, the two edge modes components have the same momentum distribution [see Eq.~(\ref{eq:EdgeModesMomentumRepresentation})]. That is:
\begin{equation}
\langle0|\,\hat{e}_{\pm}\,\hat{a}_{k}^{\dagger}\,\hat{a}_{k}\,\hat{e}_{\pm}^{\dagger}\,|0\rangle=\langle0|\,\hat{e}_{\pm}\,\hat{b}_{k}^{\dagger}\,\hat{b}_{k}\,\hat{e}_{\pm}^{\dagger}\,|0\rangle=\frac{1}{2}|F(k-\pi-\delta)|^{2}.
\end{equation} 

The modulus square $|F(k)|^2$ consists of the product between the square of a Cauchy distribution centred at $k=\pi+\delta$ with scale parameter $\xi$ and the even function of the momentum $|f(k)|^{2}$. The first factor tends to highly localize the distribution around its center. Therefore, in order to get a simpler expression for the edge modes momentum density distribution that allows us to obtain its localization length, we can approximate the second factor by its second order Taylor expansion:
\begin{equation}
|F(k)|^2\simeq\left[\frac{1}{1+(k/\xi)^2}\right]^2\left[A_{0}+A_{2}(k/\xi)^2\right],
\end{equation}
with $A_{0}$ and $A_{2}$ being constant coefficients. The edge modes localization length in momentum space, defined as the FWHM of their momentum density distribution [see Fig.~\ref{fig:FiguraAIII11}(c)] is then given by:
\begin{equation}\label{eq:DeltaK}
\delta k\simeq\xi(1+A_{2}/A_{0})\simeq2\xi(1+\xi^2 N^2e^{-\xi N}),
\end{equation}
expressed in terms of the total size of the first Brillouin zone. These two expressions for the spatial and momentum localization lengths of the edge modes, Eq.~(\ref{eq:Deltax}) and Eq.~(\ref{eq:DeltaK}), hold for a wide range of parameters.

The bigger the ratio between the two hopping amplitudes $J^{\prime}/J$ is, the smaller the quantity $\xi$ is [see Eq.~(\ref{eq:ApproximationXi}) and Fig.~\ref{fig:FiguraAIII09}]. Therefore, by analysing the expressions for the edge modes spatial and momentum localizations, Eq.~(\ref{eq:Deltax}) and Eq.~(\ref{eq:DeltaK}), we can conclude that the spatial localization of the edge modes gets better as the system gets farther from the topological transition point, whereas their momentum localization gets better as the system gets closer to the topological transition point. This behaviour can be checked and visualized in Fig.~\ref{fig:FiguraAIII11}(b) and (d), where we show the edge modes spatial and momentum density distributions as a function of the ratio $J^{\prime}/J$, obtained by exact numerical diagonalization of the Hamiltonian.

Interestingly, there is a region in the topological phase where both the position and momentum of the edge modes are well defined and correspond to localization lengths lower than $0.1$, in terms of the total size of the lattice and total size of the first Brillouin zone, respectively [see Fig.~\ref{fig:FiguraAIII11}(e)]. In that region of the parameter space the product between the spatial and momentum localization lengths reaches its minimum value of $0.7\hbar$, quite close to the minimum possible value it could take according to the Heisenberg uncertainty principle [see Fig.~\ref{fig:FiguraAIII11}(f)].

\begin{figure}[t]
  \centering
    \includegraphics[width=0.45\textwidth]{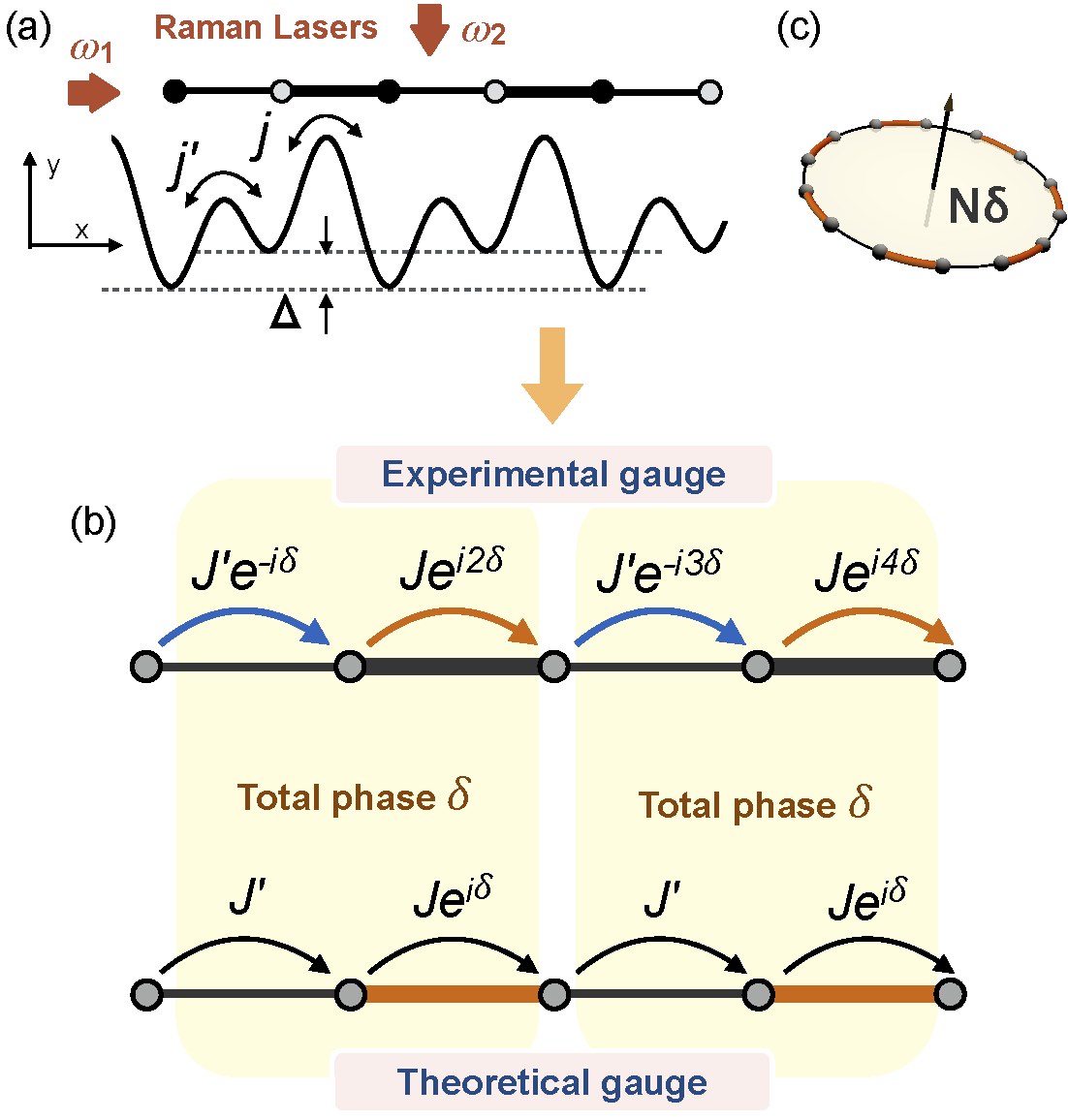}
    \caption[Experimental scheme]{\textbf{Experimental scheme.} (a) A doble well potential with energy offset $\Delta$ is created from the superposition of two standing waves, forming a one-dimensional dimerized lattice with tunneling amplitudes $j$ and $j^{\prime}$. A pair of Raman lasers is added along the $-x$ and $-y$ directions with frequencies $\omega_{1}$ and $\omega_{2}$, such that $\omega_{1}-\omega_{2}=\Delta/\hbar$. This produces a laser assisted tunneling that leads to the Hamiltonian in Eq.~(\ref{eq:HamiltonianExperimentPosition}). (b) Schematic representation of the experimental Hamiltonian and the theoretical one. Both Hamiltonians are equivalent, as the total phase accumulated by a particle when travelling from a site in the lattice to the same site in the next lattice cell is always $\delta$. (c) In both cases the effective magnetic flux that penetrates the ring formed by the lattice under periodic boundary conditions is $N\delta$.}
    \label{fig:FiguraAIII12} 
\end{figure}

\section{Experimental realization and observing fractionalization}

\subsection{Experimental realization}

We have developed an experimental protocol for the realization of our one-dimensional model for a topological insulator in the AIII symmetry class (see Fig.~\ref{fig:FiguraAIII12}). For that, we need to combine a superlattice structure \cite{Atala2013}, which serves as a dimerized one-dimensional lattice with two different hopping amplitudes, together with Raman assisted tunneling \cite{Aidelsburger2011,Aidelsburger2013}, which allows the implementation of complex tunneling amplitudes. As a result, the following experimental Hamiltonian can be realize:
\begin{equation}\label{eq:HamiltonianExperimentPosition}
H_{\text{exp}}=-\sum_{n}^{N}\big(J^{\prime}\,e^{i(2n-1)\delta}\,\hat{a}_{n}^{\dagger}\hat{b}_{n}^{}+Je^{i2n\delta}\,\hat{a}_{n+1}^{\dagger}\hat{b}_{n}^{}+\text{H.c.}\big).
\end{equation}
It is not exactly the same Hamiltonian as the one in Eq.~(\ref{eq:HamiltonianPosition}) that corresponds to our model. However, both Hamiltonians are equivalent and are connected through the following gauge transformation:
\begin{equation}
\begin{cases}
\hat{a}_{n}^{\dagger}\,\rightarrow\,\hat{a}_{n}^{\dagger},\\
\hat{b}_{n}^{\dagger}\,\rightarrow\, e^{i(2n-1)\delta}\,\hat{b}_{n}^{\dagger}.
\end{cases}
\end{equation}
The reason why the experimental Hamiltonian and the theoretical one are equivalent to each other is the fact that the total phase accumulated by a particle after travelling from a particular site in the lattice to the same site in the following lattice cell is always $\delta$ [see Fig.~\ref{fig:FiguraAIII12}(b)]. Therefore, the two Hamiltonians can be connected by means of this gauge transformation both for periodic and for open boundary conditions, as the effective magnetic flux that penetrates the sample for periodic boundary conditions is in both cases $N\delta$ [see Fig.~\ref{fig:FiguraAIII12}(c)].

The gauge transformation that connects the experimental and the theoretical Hamiltonians can be written in the momentum representation as $\hat{a}_{k}^{\dagger}(\hat{b}_{k}^{\dagger})\,\rightarrow\,\hat{a}_{k}^{\dagger}(\hat{b}_{k-2\delta}^{\dagger})$. In consequence, the edge modes in the experimental gauge have the following form:
\begin{align}\label{eq:EdgeModesExperimentalGauge}
\hat{e}_{\text{exp},\pm}^{\dagger}=\frac{1}{\sqrt{2}}\sum_{n}\bigg(e^{i(\pi+\delta)n}\sinh\xi(N+1-n)\,\hat{a}_{n}^{\dagger}\pm\nonumber\\
e^{i(\pi-\delta)n}\sinh\xi n\,\hat{b}_{n}^{\dagger} \bigg),
\end{align}
so that the mode located at the left edge of the system has momentum $\pi+\delta$, whereas the one at the right edge has momentum $\pi-\delta$. In this way, for the BDI symmetry class ($\delta=0,\pi$) the two edge modes are localized at the same position in momentum space ($k=\pi$ or $k=0$). In contrast, for the AIII symmetry class ($\delta\neq0,\pi$) they are localized at opposite momenta ($k=\pi+\delta$ and $k=\pi-\delta$). This splitting is a distinctive feature of the AIII symmetry class and is not possible in the BDI symmetry class, as the presence of time reversal symmetry forces the edge modes to be at the same momentum position. As we show in the following, this feature of the AIII symmetry class allows us to directly observe fractionalization in momentum space.

\subsection{Observing fractionalization in the AIII model}

The phenomenon of fractionalization was first studied in the context of quantum field theory and consist of the appearance of excitations carrying a fractionalized particle number in some soliton quantum field theories \cite{Jackiw1976,Goldstone1981,Stone1985,Niemi1986}. Such field theories are not present in the Standard Model of particle physics and, thus, such fractionalized excitations do not exist as conventional free particles in nature. However, some condensed matter systems can serve as physical realizations of such soliton field theories and, therefore, give rise to fractionalized particles, which in this context are called \textit{quasiparticles}.

That is the case of our one-dimensional model for a topological insulator. At a filling factor such that all states of negative energy are occupied, this many body state can be interpreted as a Dirac sea on top of which particles and antiparticles (holes) can be created, by occupying positive energy states and emptying negative energy states, respectively. In that situation, and when the system is in its topological phase, the lowest energy excitation that can be created, which corresponds to one of the edge modes, is described by a discretized version of a soliton quantum field theory with fractionalized excitations. Therefore, it inherits the property of fractionalization.

What does exactly the phenomenon of fractionalization mean in the context of a condensed matter system and not a soliton field theory? In order to understand the correspondence between these two levels, the soliton quantum field theory and the topological insulator that realizes it at a very specific filling factor, we need to compare two situations. On one hand, a fermion occupying an edge state with no more occupied states; and, on the other hand, the situation in which a fermion occupies the same edge state and all negative energy states are also occupied, which represent the vacuum of the field theory.\\

\subsubsection{Fractionalization in position space}

We start by considering the many body state in which all bulk states in the lower energy band plus the negative energy edge state are occupied. This many body state is:
\begin{equation}\label{eq:DefFlatDensityState}
|\Phi_{+}\rangle=\hat{e}_{+}^{\dagger}\prod_{q}\hat{c}_{+,q}^{\dagger}|0\rangle,
\end{equation}
being $\hat{c}_{+,q}^{\dagger}$ the bulk modes in Eq.~(\ref{eq:AIIIOpenBulkModes}). We want to know what is the spatial density distribution, denoted as $\nu_{x}$, for such many body state. That is, the expected value of the particle number operator $\hat{n}_{x}$; being $\hat{n}_{x=2n-1}=\hat{a}_{n}^{\dagger}\hat{a}_{n}$ and $\hat{n}_{x=2n}=\hat{b}_{n}^{\dagger}\hat{b}_{n}$, runing $x$ from $1$ to $2N$ and $n$ from $1$ to $N$.

First, it is convenient to define the two following projectors:
\begin{equation}\label{eq:DefProjectorsBandas}
P_{\pm}=\hat{e}_{\pm}^{\dagger}\,|0\rangle\langle0|\,\hat{e}_{\pm}+\sum_{q}\hat{c}_{\pm,q}^{\dagger}\,|0\rangle\langle0|\,\hat{c}_{\pm,q},
\end{equation}
which correspond to the subspaces generated by all positive and negative energy states, respectively. Now, we can write an expression for the spatial density distribution corresponding to the many body state $|\Phi_{+}\rangle$ using such projectors:
\begin{align}\label{eq:SpatialDensityDistribution}
\nu_{x}=\langle0|\,\hat{e}_{+}\,\hat{n}_{x}\,\hat{e}_{+}^{\dagger}\,|0\rangle+\sum_{q}\langle0|\,\hat{c}_{+,q}\,\hat{n}_{x}\,\hat{c}_{+,q}^{\dagger}\,|0\rangle=\nonumber\\
\text{tr}\big(\hat{n}_{x}\,P_{+}\big)=1-\text{tr}\big(\hat{n}_{x}\,P_{-}\big),
\end{align}
where we have used the fact that $P_{+}+P_{-}=\mathbb{I}$ and $\text{tr}\big(\hat{n}_{x}\big)=1$.

Due to the presence of chiral symmetry, the chiral operator $U_{S}:\hat{a}_{n}^{\dagger}(\hat{b}_{n}^{\dagger})\rightarrow\hat{a}_{n}^{\dagger}(-\hat{b}_{n}^{\dagger})$ transforms every positive energy eigenstate into a negative energy eigenstate and vice versa, so that $U_{S}P_{\pm}U_{S}^{\dagger}=P_{\mp}$. For this reason, together with the fact that the chiral operator leaves the density operator $\hat{n}_{x}$ invariant, we have that $\text{tr}\big(\hat{n}_{x}\,P_{+}\big)=\text{tr}\big(\hat{n}_{x}\,P_{-}\big)$. Therefore, substituting in Eq.~(\ref{eq:SpatialDensityDistribution}), we get:
\begin{equation}
\nu_{x}=1-\text{tr}\big(\hat{n}_{x}\,P_{-}\big)=1-\text{tr}\big(\hat{n}_{x}\,P_{+}\big)=1-\nu_{x}\implies\nu_{x}=\frac{1}{2}.
\end{equation}
In this way, the many body state $|\Phi_{+}\rangle$, that consists of having all negative energy states occupied, corresponds to a flat spatial density distribution.

Now we consider the resulting state of adding a fermion, occupying the positive energy edge mode, to that flat density profile:
\begin{equation}\label{eq:QuasiparticleManyBodyState}
|\Phi_{\text{qp}}\rangle=\hat{e}_{-}^{\dagger}|\Phi_{+}\rangle=\hat{e}_{-}^{\dagger}\hat{e}_{+}^{\dagger}\prod_{q}\hat{c}_{+,q}^{\dagger}|0\rangle.
\end{equation}
In this many-body state both edge modes are occupied. Therefore the modes $\tilde{a}_{n=1}^{\dagger}$ and $\tilde{b}_{n=N}^{\dagger}$ are also occupied and we have:
\begin{align}
&\Big(\tilde{a}_{n=1}^{\dagger}\tilde{a}_{n=1}-\nu_{x}\Big)|\Phi_{\text{qp}}\rangle=\frac{1}{2}\,|\Phi_{\text{qp}}\rangle,\label{eq:QuasiparticleFractionalizationEquationA}\\
&\Big(\tilde{b}_{n=N}^{\dagger}\tilde{b}_{n=N}-\nu_{x}\Big)|\Phi_{\text{qp}}\rangle=\frac{1}{2}\,|\Phi_{\text{qp}}\rangle.\label{eq:QuasiparticleFractionalizationEquationB}
\end{align}
This result states that the many-body state $|\Phi_{\text{qp}}\rangle$, with a fermion on top of a flat background, is an exact eigenstate of the number operators $\tilde{a}_{n=1}^{\dagger}\tilde{a}_{n=1}-\nu_{x}$ and $\tilde{b}_{n=N}^{\dagger}\tilde{b}_{n=N}-\nu_{x}$. Therefore, we can say that two fractionalized quasiparticles with particle number $1/2$ exist at the edges of the system. Analogously, we can define the quasihole sate:
\begin{equation}
|\Phi_{\text{qh}}\rangle=\hat{e}_{+}|\Phi_{+}\rangle=\prod_{q}\hat{c}_{+,q}^{\dagger}|0\rangle,
\end{equation}
which is the result of removing a fermion from the flat density many-body state in Eq.~(\ref{eq:DefFlatDensityState}) and satisfies:
\begin{align}
&\Big(\tilde{a}_{n=1}^{\dagger}\tilde{a}_{n=1}-\nu_{x}\Big)|\Phi_{\text{qh}}\rangle=-\frac{1}{2}\,|\Phi_{\text{qh}}\rangle,\\
&\Big(\tilde{b}_{n=N}^{\dagger}\tilde{b}_{n=N}-\nu_{x}\Big)|\Phi_{\text{qh}}\rangle=-\frac{1}{2}\,|\Phi_{\text{qh}}\rangle.
\end{align}
\begin{figure}[t]
  \centering
    \includegraphics[width=0.49\textwidth]{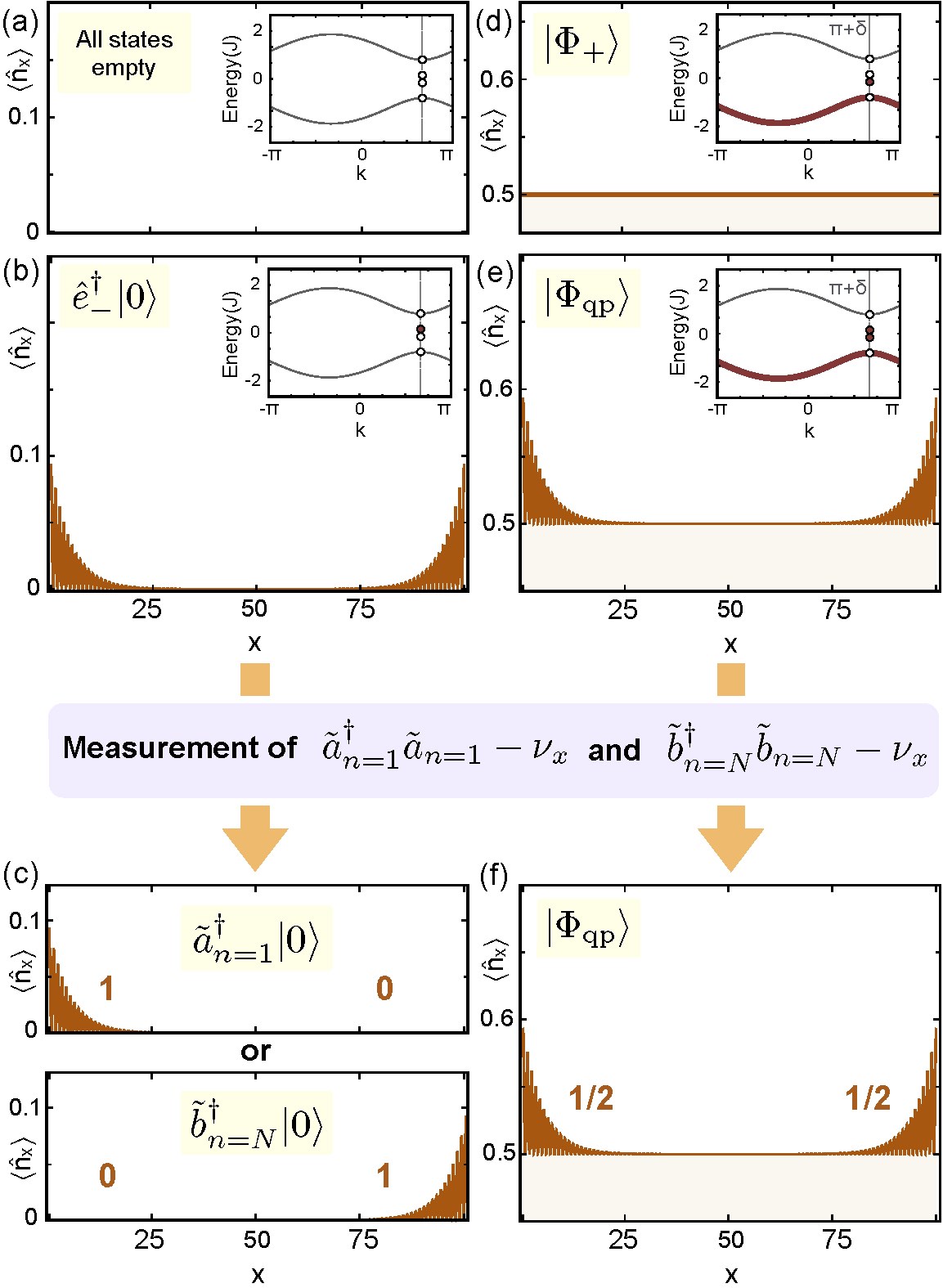}
    \caption[Fractionalization in a one-dimensional topological insulator.]{\textbf{Fractionalization in a one-dimensional topological insulator.} (b) Spatial density distribution of a fermion occupying the edge mode $\hat{e}_{-}^{\dagger}$, being all other eigenstates of the Hamiltonian empty. (c) After measuring the number operators $\tilde{a}_{n=1}^{\dagger}\tilde{a}_{n=1}-\nu_{x}$ and $\tilde{b}_{n=N}^{\dagger}\tilde{b}_{n=N}-\nu_{x}$ the state of the fermion will be $\tilde{a}_{n=1}^{\dagger}|0\rangle$ or $\tilde{b}_{n=N}^{\dagger}|0\rangle$, being both possibilities equally probable. In this case the fermion is found to be at one of the two edges of the system. (d) Spatial density distribution of the many-body state $|\Phi_{+}\rangle$, in which all bulk modes in the lower energy band plus one edge mode are occupied, see Eq.~(\ref{eq:DefFlatDensityState}). (e) Spatial density distribution of the many-body state $|\Phi_{\text{qp}}\rangle$, consisting of a fermion added to a flat background density, see Eq.~(\ref{eq:QuasiparticleManyBodyState}). (f) The measurement of $\tilde{a}_{n=1}^{\dagger}\tilde{a}_{n=1}-\nu_{x}$ and $\tilde{b}_{n=N}^{\dagger}\tilde{b}_{n=N}-\nu_{x}$ leaves the sate of the system invariant and gives the outcomes $1/2$ and $1/2$, what indicates that two quasiparticles of fractional charge $1/2$ exist at the edges of the chain on top of a flat background density. All data in this figure has been obtained by exact numerical diagonalization of our Hamiltonian $H_{\delta}$ for $J^{\prime}/J=0.9$, $\delta=-2\pi/3$ and $N=100$.}
    \label{fig:FiguraAIII13} 
\end{figure}
As we have already mentioned, in order to better understand the phenomenon of fractionalization in our one-dimensional model, we can compare two different situations: a fermion occupying one edge state wave function and a fermion occupying the same state but on top of the flat density distribution that corresponds to the many-body state in Eq.~(\ref{eq:DefFlatDensityState}).

In the first case [see Fig.~\ref{fig:FiguraAIII13}(a) and (b)] we have a particle whose wave function is the superposition between two states, each of them localized at each edge of the system. Thereby, if we measure the number operators $\tilde{a}_{n=1}^{\dagger}\tilde{a}_{n=1}-\nu_{x}$ and $\tilde{b}_{n=N}^{\dagger}\tilde{b}_{n=N}-\nu_{x}$ we will obtain the outcome $1$ for one of them and the outcome $0$ for the other, being the two alternatives equally probable. In other words, the particle will be found at one edge of the chain and its state after the measurement will be $\tilde{a}_{n=1}^{\dagger}|0\rangle$ or $\tilde{b}_{n=N}^{\dagger}|0\rangle$ [see Fig.~\ref{fig:FiguraAIII13}(c)].

In contrast, in the second situation the system is in a many-body state consisting of a fermion on top of a flat background density [see Fig.~\ref{fig:FiguraAIII13}(d) and (e)]. As we have explained, see Eq.~(\ref{eq:QuasiparticleFractionalizationEquationA}) and Eq.~(\ref{eq:QuasiparticleFractionalizationEquationB}), this state is an eigenstate of the number operators $\tilde{a}_{n=1}^{\dagger}\tilde{a}_{n=1}-\nu_{x}$ and $\tilde{b}_{n=N}^{\dagger}\tilde{b}_{n=N}-\nu_{x}$. Therefore, if we measure such number operators we will always obtain the outcome $1/2$ for both of them and the state of the system after the measurement will remain the same [see Fig.~\ref{fig:FiguraAIII13}(f)]. That is, a fermion occupying an edge mode behaves in a different way, depending on whether it has been added to a flat background many-body state or not. And the reason behind this difference is the fact that the flat background many-body state represents the vacuum of a soliton field theory in which excitations with fractional particle number arise.

\subsubsection{Fractionalization in momentum space}

Up to now, we have discussed the phenomenon of fractionalization in position space. However, the well definition of the edge modes momentum leads to an interesting result: fractionalization also occurs in momentum space.

In the following we show how our analysis of the edge modes momentum makes possible the observation of the phenomenon of fractionalization in momentum space, which represents a new alternative way in addition to the position space. Furthermore, the fact that the edge modes in the AIII symmetry class have a momentum different from $0$ and $\pi$ facilitates the observation of fractionalization, as we explain later on in this section.

The momentum representation of the edge modes, see Eq.~(\ref{eq:EdgeModesMomentumRepresentation}) and Eq.~(\ref{eq:EdgeModeFourierTransformed}), allows as to decompose them as:
\begin{equation}
\hat{e}_{\pm}^{\dagger}=\frac{1}{\sqrt{2}}\Big(\tilde{a}_{k=\pi+\delta}^{\dagger}\pm\tilde{b}_{k=\pi+\delta}^{\dagger}\Big),
\end{equation}
with:
\begin{align}
&\tilde{a}_{k=\pi+\delta}^{\dagger}=\sum_{k}F(k-\pi-\delta)\,\hat{a}_{k}^{\dagger},\\
&\tilde{b}_{k=\pi+\delta}^{\dagger}=\sum_{k}e^{i(k-\pi-\delta)(N+1)}\,F(-k+\pi+\delta)\,\hat{b}_{k}^{\dagger},
\end{align}
being $\tilde{a}_{k=\pi+\delta}^{\dagger}$ and $\tilde{b}_{k=\pi+\delta}^{\dagger}$ two momentum components highly localized in momentum space at $k=\pi+\delta$.

In order to show the phenomenon of fractionalization in momentum space we proceed in the same way as we did in the previous section for the position space. We start by considering the many-body state $|\Phi_{+}\rangle$, in which all negative energy eigenstates of the Hamiltonian are occupied, see Eq.~(\ref{eq:DefFlatDensityState}). We denote by $\nu_{k}$ its corresponding momentum density distribution, which consists of the expected value of $\hat{a}_{k}^{\dagger}\hat{a}_{k}$. Using the projectors defined in Eq.~(\ref{eq:DefProjectorsBandas}) and following the same reasoning as before we can easily prove that this momentum density distribution is also flat:
\begin{align}
&\nu_{k}=\text{tr}\Big(\hat{a}_{k}^{\dagger}\hat{a}_{k}P_{+}\Big)=1-\text{tr}\Big(\hat{a}_{k}^{\dagger}\hat{a}_{k}P_{-}\Big)=\nonumber\\
&1-\nu_{k}\implies\nu_{k}=\frac{1}{2}.
\end{align}
In the same way, the momentum density distribution for $\hat{b}_{k}^{\dagger}\hat{b}_{k}$ is also uniform and equal to $1/2$. Adding a fermion to this flat background density leads to the many-body state $|\Phi_{\text{qp}}\rangle$, in which both edge modes are occupied. Therefore, the two momentum modes $\tilde{a}_{k=\pi+\delta}^{\dagger}$ and $\tilde{b}_{k=\pi+\delta}^{\dagger}$ are also occupied and we have:
\begin{align}
&\Big(\tilde{a}_{k=\pi+\delta}^{\dagger}\tilde{a}_{k=\pi+\delta}-\nu_{k}\Big)|\Phi_{\text{qp}}\rangle=\frac{1}{2}\,|\Phi_{\text{qp}}\rangle,\\
&\Big(\tilde{b}_{k=\pi+\delta}^{\dagger}\tilde{b}_{k=\pi+\delta}-\nu_{k}\Big)|\Phi_{\text{qp}}\rangle=\frac{1}{2}\,|\Phi_{\text{qp}}\rangle.
\end{align}
This result states that two quasiparticles of fractional charge $1/2$ exist in momentum space at the position $k+\pi+\delta$.
\begin{figure}[t]
  \centering
    \includegraphics[width=0.49\textwidth]{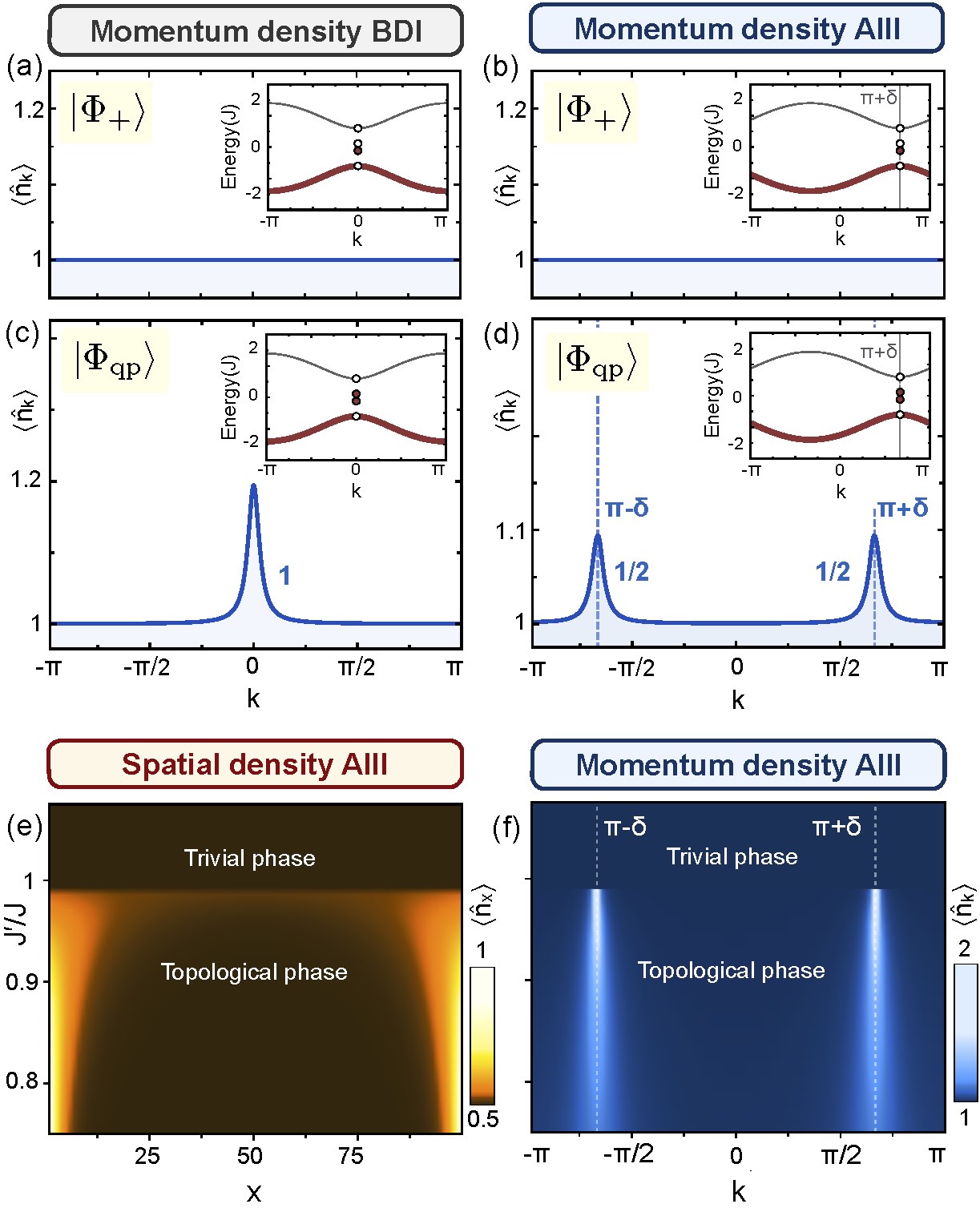}
    \caption[Direct observation of fractionalization in the AIII class.]{\textbf{Direct observation of fractionalization in the AIII class.} (a)-(d). Momentum density distributions (a), (c) for the BDI symmetry class ($\delta=0$), and (b), (d) the AIII symmetry class ($\delta=-2\pi/3$), for $N=100$ and $J^{\prime}/J=0.9$. For both symmetry classes the many body state $|\Phi_{+}\rangle$, see Eq.~(\ref{eq:DefFlatDensityState}), corresponds to a flat momentum density distribution (a), (b). For the state $|\Phi_{\text{qp}}\rangle$, in which a fermion is added to the flat background, the momentum density distribution in the BDI symmetry class (c) shows a single peak of charge $1$ at momentum $k=\pi$ (plotted shifted at $k=0$). In contrast, in the AIII symmetry class (d), the momentum density distribution shows two peaks of charge $1/2$ localized at momenta $\pi\pm\delta$. (e), (f) Simultaneous fractionalization in position and momentum space in the AIII symmetry class. (e) Spatial density distribution and (f) momentum density distribution for the many body state $|\Phi_{\text{qp}}\rangle$, such that the upper band is empty, as a function of the ratio between the two hopping amplitudes $J^{\prime}/J$ for $N=100$ and $\delta=-2\pi/3$. In the topological phase two peaks of charge $1/2$ arise at the edge of the chain and also at momenta $\pi\pm\delta$ in momentum space.}
    \label{fig:FiguraAIII14} 
\end{figure}

What are the implications for the experiment? In the experimental gauge, the edge modes are [see Eq.~(\ref{eq:EdgeModesExperimentalGauge})]:
\begin{equation}
\hat{e}^{\dagger}_{\text{exp},\pm}=\frac{1}{\sqrt{2}}\Big(\tilde{a}_{\text{exp},k=\pi+\delta}^{\dagger}\pm\tilde{b}_{\text{exp},k=\pi-\delta}^{\dagger}\Big),
\end{equation}
with:
\begin{align}
&\tilde{a}_{\text{exp},k=\pi+\delta}^{\dagger}=\sum_{k}F(k-\pi-\delta)\,\hat{a}_{k}^{\dagger},\\
&\tilde{b}_{\text{exp},k=\pi+\delta}^{\dagger}=\sum_{k}e^{i(k-\pi+\delta)(N+1)}\,F(-k+\pi-\delta)\,\hat{b}_{k}^{\dagger}.
\end{align}

That is, the two momentum modes that form the edge states are localized at different positions in the momentum space. The mode $\tilde{a}$ is localized in position space at the left edge of the chain and at momentum $k=\pi+\delta$ in momentum space, whereas the mode $\tilde{b}$ is localized in position space at the right edge of the system and at momentum $k=\pi-\delta$ in momentum space.

This means a significant difference between the BDI symmetry class and the AIII symmetry class.
For the BDI class ($\delta=0,\pi$) a particle added to a uniform background will consist of two quasiparticles bound together at the same momentum position. The momentum distribution will show a single peak at momentum $k=\pi$ or $k=0$ enclosing a total charge $1$ [Fig.~\ref{fig:FiguraAIII14}(c)]. For the AIII case, in contrast, a particle added to a uniform background will split into two halves located at opposite momenta ($\pi-\delta,\pi+\delta$). The momentum distribution will show two peaks, each enclosing a $1/2$ charge [Fig.~\ref{fig:FiguraAIII14}(d)].
In the AIII class splitting of the fermion occurs therefore both in position and momentum spaces [Fig.~\ref{fig:FiguraAIII14}(e),(f)]. The splitting in momentum is a direct consequence of the non-zero momentum of the edge modes, which is in turn a direct manifestation of the breaking of time reversal symmetry characterizing the AIII class.

\section{Interference of fractionalized quasiparticles}

\subsection{Quench to the critical point}

We have designed a quench protocol for the one-dimensional model $H_{\delta}$ in which two fractionalized quasiparticles move along the system at constant velocity and almost without dissipation. They interfere with each other and keep on moving until they reach the edges of the system. Afterwards, they bounce and repeat the same process over and over.

Such dynamics emerges after quenching the ratio between the two hopping amplitudes from the topological phase to the critical point:
\begin{equation}
H_{\delta}(J^{\prime}<J)\longrightarrow H_{\delta}(J^{\prime}=J),
\end{equation}
and is the consequence of two ingredients:\\

\textit{i)} The uniform background spatial density of the many-body state $\ket{\Phi_{+}}$ [Eq.~(\ref{eq:DefFlatDensityState})], in which all bulk eigenstates in the lower energy band plus one edge mode are occupied, does not evolve in time. Thereby, quasiparticles can be defined on top on that flat background density at any time.\\

\begin{figure}[t]
  \centering
    \includegraphics[width=0.485\textwidth]{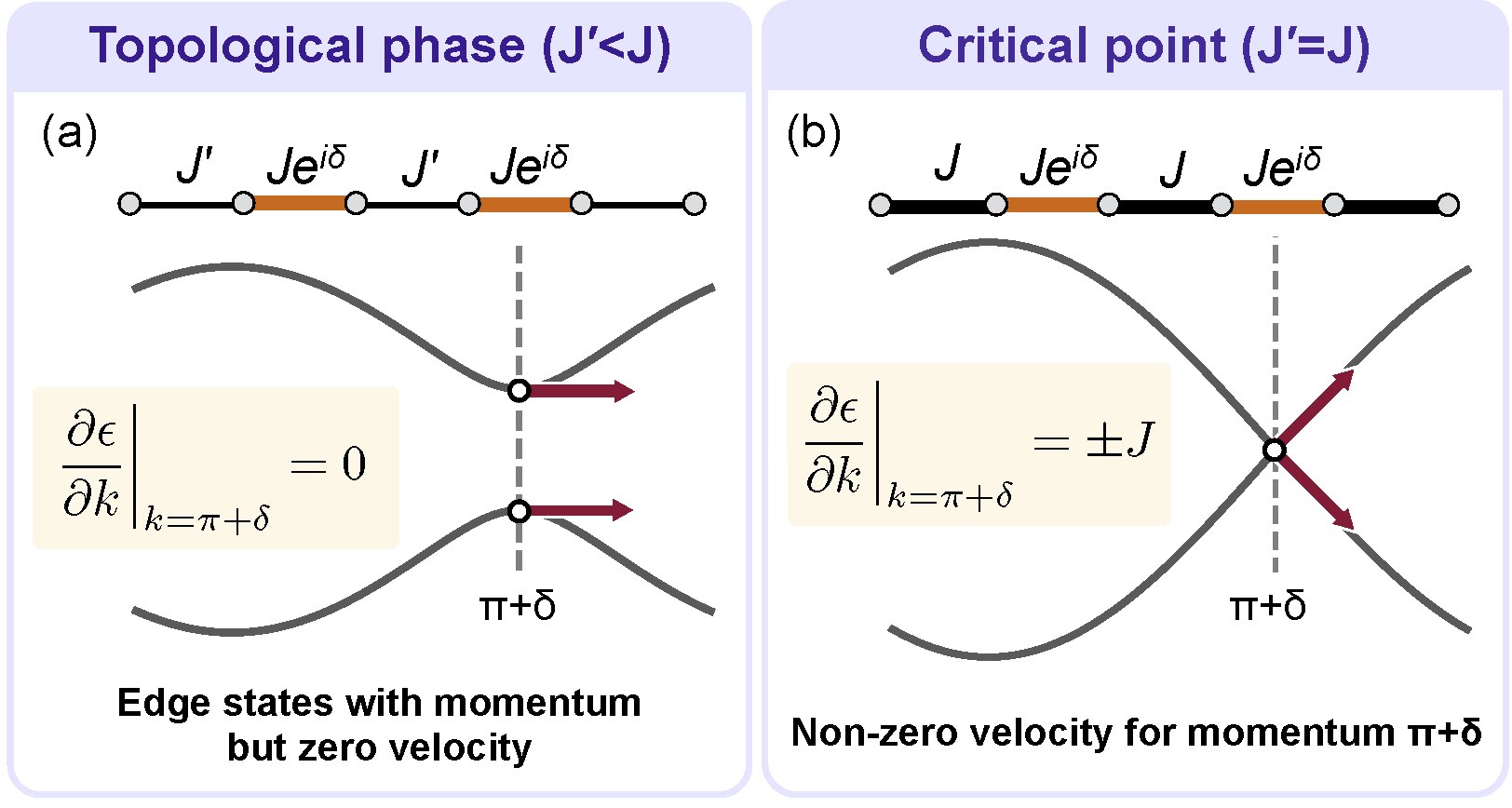}
    \caption[Quench to the critical point.]{\textbf{Quench to the critical point.} The system is prepared somewhere in its topologically non-trivial phase, so that two quasiparticles exist at the edges of the chain. For that, we need $J^{\prime}<J$. In that situation, (a), the two edge modes have momentum $\pi+\delta$ and no velocity. After quenching the ratio between the two hopping amplitudes to $J^{\prime}/J=1$, the edge modes are localized in momentum space where the gap has been closed and the dispersion relation is almost linear (b). In consequence, they have a non-zero velocity and propagate along the system.}
    \label{fig:FiguraInt01} 
\end{figure}

\textit{ii)} At the critical point, the gap between two energy bands of the system closes at $k=\pi+\delta$ [see Fig.\ref{fig:FiguraInt01}(b)]. Therefore, the dispersion relation is almost linear around that momentum value, which is precisely the momentum at which the edge modes are localized in momentum space. The simultaneous good definition of both the position and the momentum of the edge modes, together with an almost linear dispersion relation, makes them propagate at constant velocity along the system while preserving their good spatial and momentum localizations.\\

The combination of \textit{i)} and \textit{ii)} leads to quench dynamics in which two quasiparticles move along the system on top of a uniform background density and interfere with each other. In the following sections we explain in detail such process.

\subsection{Absence of evolution of flat background}

We want to compute the spatial density distribution of the many-body state $|\Phi_{+}\rangle$ as a function of time, that is:
\begin{equation}
\nu_{x}(t)=\langle\Phi_{+}|\,U^{\dagger}(t)\,\hat{n}_{x}\,U(t)\,|\Phi_{+}\rangle,
\end{equation} 
where $U(t)=\exp(-itH_{\delta}/\hbar)$ is the time evolution operator.

In order to compute this spatial density evolution we start by exploiting the chiral symmetry of the Hamiltonian. First, the chiral operator $U_{S}$ transforms every positive energy eigenstate into a negative energy eigenstate and vice versa, and thus $U_{S}\ket{\Phi_{-}}=\ket{\Phi_{+}}$, where $\ket{\Phi_{-}}$ is the many-body state in which all positive energy eigenstates are occupied. Second, the chiral operator is such that $U_{S}^{\dagger}\,H\,U_{S}=-H$, so that $U_{S}^{\dagger}\,U(t)\,U_{S}=U(-t)$. Furthermore, it leaves the number operator $\hat{n}_{x}$ invariant.
The combination of these three properties implies that the spatial density distribution of the many-body state $|\Phi_{+}\rangle$ at positive time $t$ is equal to the spatial density distribution of the many-body state $|\Phi_{-}\rangle$ at negative time $-t$. That is:
\begin{align}\label{eq:BackgroundTimeEvol1}
\nu_{x}(t)=&\langle\Phi_{+}|\,U^{\dagger}(t)\,\hat{n}_{x}\,U(t)\,|\Phi_{+}\rangle=\nonumber\\
&\langle\Phi_{-}|\,U^{\dagger}_{S}\,U^{\dagger}(t)\,\hat{n}_{x}\,U(t)\,U_{S}\,|\Phi_{-}\rangle=\nonumber\\
&\langle\Phi_{-}|\,U^{\dagger}(-t)\,U^{\dagger}_{S}\,\hat{n}_{x}\,U_{S}\,U(-t)\,|\Phi_{-}\rangle=\nonumber\\
&\langle\Phi_{-}|\,U^{\dagger}(-t)\,\hat{n}_{x}\,U(-t)\,|\Phi_{-}\rangle
\end{align} 
We can simplify this computation by using the unitary transformation $V$ that connects the Hamiltonian with a phase $\delta$ to the one with no phase at all, that is: $V\,H_{\delta}\,V^{\dagger}=H_{\delta=0}$ [see Eq.~(\ref{eq:GaugeTransformationDelta})]. We have:
\begin{align}
&\langle\Phi_{-}|\,U^{\dagger}(-t)\,\hat{n}_{x}\,U(-t)\,|\Phi_{-}\rangle\overset{a}{=}\nonumber\\
&\langle\Phi_{-}^{\delta=0}|\,V\,U^{\dagger}(-t)\,\hat{n}_{x}\,U(-t)\,V^{\dagger}\,|\Phi_{-}^{\delta=0}\rangle\overset{b}{=}\nonumber\\
&\langle\Phi_{-}^{\delta=0}|\,U^{\dagger}_{\delta=0}(-t)\,V\,\hat{n}_{x}\,V^{\dagger}\,U_{\delta=0}(-t)\,|\Phi_{-}^{\delta=0}\rangle\overset{c}{=}\nonumber\\
&\langle\Phi_{-}^{\delta=0}|\,U^{\dagger}_{\delta=0}(-t)\,\hat{n}_{x}\,U_{\delta=0}(-t)\,|\Phi_{-}^{\delta=0}\rangle\overset{d}{=}\nonumber\\
&\langle\Phi_{-}|\,U^{\dagger}(t)\,\hat{n}_{x}\,U(t)\,|\Phi_{-}\rangle.\label{eq:BackgroundTimeEvol2}
\end{align}
Here, we have used that (a) $|\Phi_{\pm}\rangle=V^{\dagger}\,|\Phi_{\pm}^{\delta=0}\rangle$. We have also used that (b) the quench does not change the value of $\delta$ and thereby the time evolution operator for arbitrary $\delta$ is connected to the one for $\delta=0$ through the same unitary $V$:
\begin{equation}
U(t)=V^{}U_{\delta=0}(t)V^{\dagger}.
\end{equation}
Moreover, we have used that (c) the density operator $\hat{n}_{x}$ is invariant under $V$, since $V$ only introduces a local phase shift. Finally we have that (d) for $\delta=0$ the system has time reversal symmetry and thus we can invert the time evolution. The result can be generalized for any $\delta$ following the same reasoning and using again (a), (b) and (c).

From Eq.~(\ref{eq:BackgroundTimeEvol1}) and Eq.~(\ref{eq:BackgroundTimeEvol2}) we conclude that the two many-body states $|\Phi_{+}\rangle$ and $|\Phi_{-}\rangle$ show the same spatial density distribution at all times. In addition, the evolved projectors $P_{+}(t)$ and $P_{-}(t)$, which correspond to the subspaces generated by all negative and positive eigenstates of the Hamiltonian, respectively [see Eq.~(\ref{eq:DefProjectorsBandas})], sum to identity at any time. In consequence, we have:
\begin{align}
\nu_{x}(t)=&\text{tr}\Big(\hat{n}_{x}P_{+}(t)\Big)=1-\text{tr}\Big(\hat{n}_{x}P_{-}(t)\Big)=\nonumber\\
&1-\text{tr}\Big(\hat{n}_{x}P_{+}(t)\Big)=1-\nu_{x}(t),
\end{align}
what leads us to the final result:
\begin{equation}
\nu_{x}(t)=\frac{1}{2}.
\end{equation}
In conclusion, the flat background density of the many-body state $|\Phi_{+}\rangle$ remains constant in time.

\subsection{Free propagation of edge modes}

The spatial density distribution of the edge modes at time $t$ is given by:
\begin{equation}
\langle e_{\pm}|\,U^{\dagger}(t)\,\hat{n}_{x}\,U(t)\,|e_{\pm}\rangle,
\end{equation}
being $|e_{\pm}\rangle\equiv\hat{e}_{\pm}^{\dagger}|0\rangle$ and $U(t)$ the time evolution operator.

As we did before, we can use the unitary transformation $V$ that connects the Hamiltonian with a phase $\delta$ to the one with no phase to simplify this computation. We have:
\begin{align}
&\langle e_{\pm}|\,U^{\dagger}(t)\,\hat{n}_{x}\,U(t)\,|e_{\pm}\rangle\overset{a}{=}\nonumber\\
&\langle e_{\pm,\delta=0}|\,V\,U^{\dagger}(t)\,\hat{n}_{x}\,U(t)\,V^{\dagger}\,|e_{\pm,\delta=0}\rangle\overset{b}{=}\nonumber\\
&\langle e_{\pm,\delta=0}|\,U_{\delta=0}^{\dagger}(t)\,V\,\hat{n}_{x}\,V^{\dagger}\,U_{\delta=0}(t)\,|e_{\pm,\delta=0}\rangle\overset{c}{=}\nonumber\\
&\langle e_{\pm,\delta=0}|\,U_{\delta=0}^{\dagger}(t)\,\hat{n}_{x}\,U_{\delta=0}(t)\,|e_{\pm,\delta=0}\rangle.
\end{align}
Where we have used that (a) $|e_{\pm}\rangle=V^{\dagger}\,|e_{\pm,\delta=0}\rangle$, (b) $U(t)=V^{}U_{\delta=0}(t)V^{\dagger}$, and (c) the density operator $\hat{n}$ is invariant under $V$. In conclusion, we can compute the edge states spatial density evolution for $\delta=0$, as we know that the result holds for arbitrary $\delta$.\\

\begin{figure}[t]
  \centering
    \includegraphics[width=0.4\textwidth]{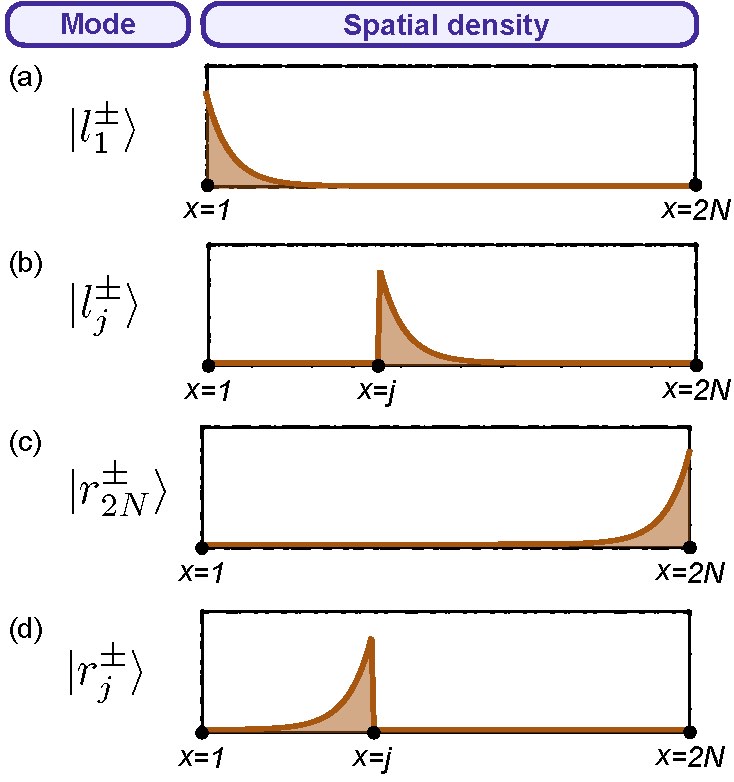}
    \caption[Edge modes decomposition and shifted modes.]{\textbf{Edge modes decomposition and shifted modes.} The edge modes of $H_{\delta=0}$, as well as their time evolution after the quench $H_{\delta=0}(J^{\prime}<J)\longrightarrow H_{\delta=0}(J^{\prime}=J)$, can be written using the states $\ket{l^{\pm}_{j}}$ and $\ket{r^{\pm}_{j}}$. (a) The states $\ket{l^{\pm}_{1}}$ are localized at the left edge of the system and form the left edge eigenstate of $H_{\delta=0}$, see Eq.~(\ref{eq:IntLeftEdgeStateDecomp}). (b) For an arbitrary $j$, the states $\ket{l^{\pm}_{j}}$ consist of the the same spatial density distribution, but shifted so that its maximum value corresponds to the position $x=j$. (c) The states $\ket{r^{\pm}_{2N}}$ are localized at the right edge of the system and form the right edge eigenstate of $H_{\delta=0}$, see Eq.~(\ref{eq:IntRightEdgeStateDecomp}). (d) For arbitrary $j$, the states $\ket{r^{\pm}_{j}}$ show a shifted spatial density distribution with its maximum value at position $x=j$. The superscript $\pm$ in all states indicates a factor $e^{\pm i\pi x/2}$ in the wave function, so that the state has momentum $\pm\pi/2$.}
    \label{fig:FiguraInt02} 
\end{figure}

For $\delta=0$ the time evolution Hamiltonian is:
\begin{equation}
H_{\text{Quench}}=H_{\delta=0}(J=J^{\prime})=-\sum_{x=1}^{2N}\hat{d}_{x}^{\dagger}\hat{d}_{x+1}^{}+\text{h.c.},
\end{equation}
where $\hat{d}_{2n-1}^{\dagger}\equiv\hat{a}_{n}^{\dagger}$ and $\hat{d}_{2n}^{\dagger}\equiv\hat{b}_{n}^{\dagger}$, with $n=1,...,N$. The eigenstates of this time evolution Hamiltonian are:
\begin{equation}\label{eq:QuenchHamiltonianEigenstates}
|q\rangle=\sum_{x=1}^{2N}\sin qx\,\hat{d}_{x}^{\dagger},
\end{equation}
with eigenvalues $\epsilon(q)=-2J\cos q$, where $q=m\pi/2L$, $m=1,...,2N$ and $L=N+1/2$.
The edge eigenstates of $H_{\delta=0}$ (before the quench, with $J^{\prime}<J$), are:
\begin{equation}
|e_{\pm,\delta=0}\rangle=\frac{1}{\sqrt{2}}\big(\ket{l}\pm\ket{r}\big),
\end{equation}
being $\ket{l}=\tilde{a}^{\dagger}_{n=1}\ket{0}$ and $\ket{r}=\tilde{b}^{\dagger}_{n=N}\ket{0}$ the edge modes located at the let and right ends of the system, respectively [see Eq.~(\ref{eq:EdgeModeWaveFunctionLeft}) and Eq.~(\ref{eq:EdgeModeWaveFunctionRight})]. They are written in terms of the $\hat{d}_{x}^{\dagger}$ modes as:
\begin{align}
&\ket{l}=\sum_{x=1}^{2N}\sin\frac{\pi}{2}x\,\psi(x)\,\hat{d}_{x}^{\dagger}\ket{0},\label{eq:IntEdgeModeL}\\
&\ket{r}=\sum_{x=1}^{2N}\cos\frac{\pi}{2}x\,\psi(2L-x)\,\hat{d}_{x}^{\dagger}\ket{0},\label{eq:IntEdgeModeR}
\end{align}
with:
\begin{equation}
\psi(x)=\begin{cases}
0,\quad x<1,\\
\frac{1}{\sqrt{\kappa}}\sinh\frac{\xi}{2}(2N+1-x),\quad 1\leq x\leq 2N,\\
0,\quad x>2N,
\end{cases}
\end{equation}
being $\kappa=\sum_{x=1}^{2N}\sinh^{2}(\xi x/2)$ a normalization constant. It is convenient to define the following states: 
\begin{align}
&\ket{l^{\pm}_{j}}=\sum_{x=1}^{2N}e^{\pm i\frac{\pi}{2}x}\,\psi(x+1-j)\,\hat{d}_{x}^{\dagger}\ket{0},\\
&\ket{r^{\pm}_{j}}=\sum_{x=1}^{2N}e^{\pm i\frac{\pi}{2}x}\,\psi(j+1-x)\,\hat{d}_{x}^{\dagger}\ket{0}.
\end{align}
On one hand, $\ket{l^{\pm}_{j}}$ is a state of momentum $\pm\pi/2$ with a spatial density distribution equal to the spatial density distribution of the state $\ket{l}$ but shifted a distance $j+1$, so that it reaches its maximum value at position $x=j$ [see Fig.~\ref{fig:FiguraInt02}(a) and (b)]. On the other hand, $\ket{r^{\pm}_{j}}$ is a state of momentum $\pm\pi/2$ with a spatial density distribution equal to the spatial density distribution of the state $\ket{r}$ but shifted a distance $2N-j$, so that it reaches its maximum value at position $x=j$ [see Fig.~\ref{fig:FiguraInt02}(c) and (d)].

Using these states, the edge states in Eq.~(\ref{eq:IntEdgeModeL}) and Eq.~(\ref{eq:IntEdgeModeR}) can be decomposed into two components of opposite momenta:
\begin{align}
&\ket{l}=\frac{1}{2i}\Big(\ket{l^{+}_{1}}-\ket{l^{-}_{1}}\Big),\label{eq:IntLeftEdgeStateDecomp}\\
&\ket{r}=\frac{1}{2}\Big(\ket{r^{+}_{2N}}+\ket{r^{-}_{2N}}\Big).\label{eq:IntRightEdgeStateDecomp}
\end{align}

\begin{figure*}
  \centering
    \includegraphics[width=0.75\textwidth]{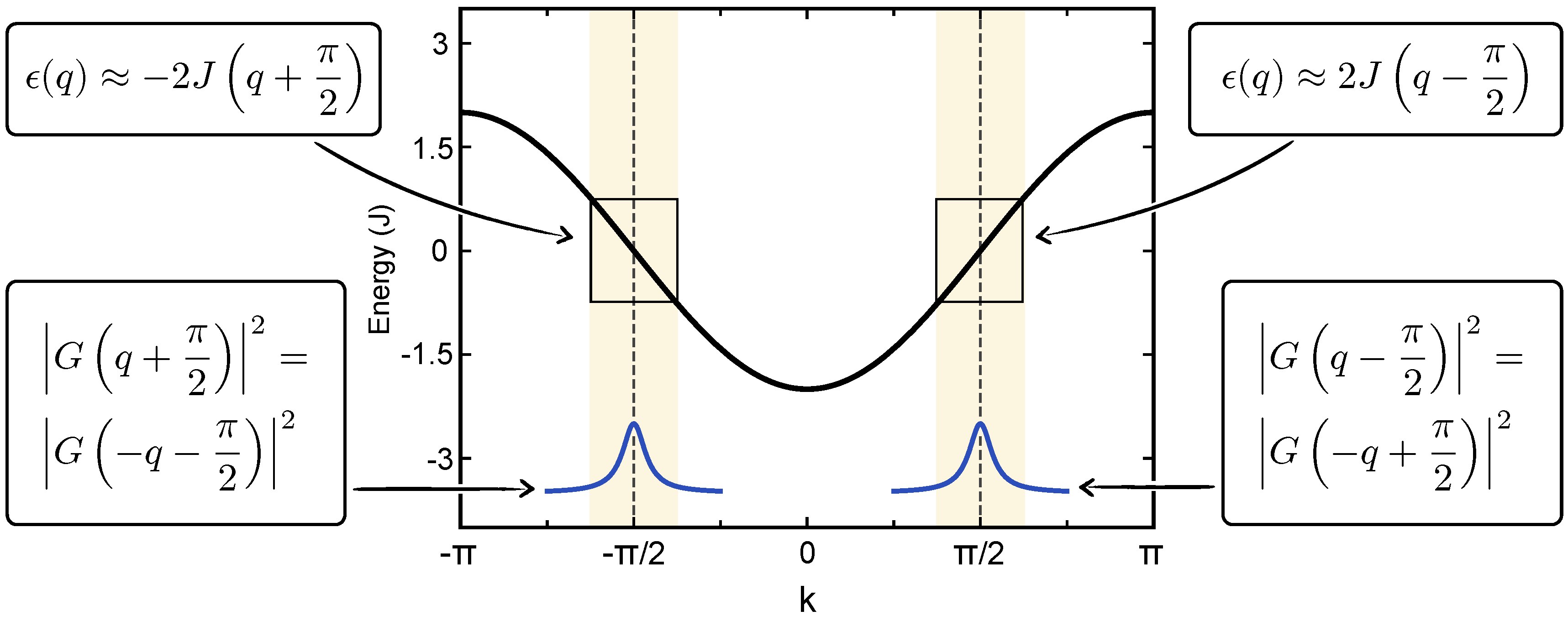}
    \caption[Linear dispersion relation approximation.]{\textbf{Linear dispersion relation approximation.} Dispersion relation of the quench Hamiltonian $H_{\delta=0}(J^{\prime}=J)$ (black line) and detail of the edge modes momentum distribution (blue lines). The dispersion relation is almost linear for momenta close to the values $\pm\pi/2$. This property, together with the well localization of the edge modes in momentum space at precisely those particular values, allows us to make the approximation $\epsilon(q)\approx\pm2J(q\mp\pi/2)$.}
    \label{fig:FiguraInt03} 
\end{figure*}
In order to compute the time evolution of the edge states, we first write them using the basis $\{\ket{q}\}$ of eigenstates of the quench Hamiltonian, see Eq.~(\ref{eq:QuenchHamiltonianEigenstates}). We have:
\begin{align}
\ket{l^{\pm}_{j}}=\frac{e^{\pm i\frac{\pi}{2}(j-1)}}{2i}\sum_{q}\bigg[e^{iq(j-1)}\,G\left(-q\mp\frac{\pi}{2}\right)-\bigg.\nonumber\\
\bigg.e^{-iq(j-1)}\,G\left(q\mp\frac{\pi}{2}\right)\bigg]\ket{q},\label{eq:StateLjinMomentumBasis}\\
\ket{r^{\pm}_{j}}=\frac{e^{\pm i\frac{\pi}{2}(j+1)}}{2i}\sum_{q}\bigg[e^{iq(j+1)}\,G\left(q\pm\frac{\pi}{2}\right)-\bigg.\nonumber\\
\bigg.e^{-iq(j+1)}\,G\left(-q\pm\frac{\pi}{2}\right)\bigg]\ket{q},\label{eq:StateRjinMomentumBasis}
\end{align} 
where:
$G(q)=\sum_{x=1}^{2N}e^{-iqx}\psi(x)$.
Here, we have used the fact that the wave functions of $\ket{l^{\pm}_{j}}$ and $\ket{r^{\pm}_{j}}$ are concentrated around the position $x=j$, so that regions far away from that position do not contribute to the scalar products $\langle q\ket{l^{\pm}_{j}}$ and $\langle q\ket{r^{\pm}_{j}}$. As the wave function of $\ket{l^{\pm}_{j}}$ is concentrated to the right of $x=j$, see Fig.~\ref{fig:FiguraInt02}(b), the expression in Eq.~(\ref{eq:StateLjinMomentumBasis}) holds for any $j$ not close to $2N$. Similarly, the expression in Eq.~(\ref{eq:StateRjinMomentumBasis}) holds for $j$ not close to $1$, as the wave function of $\ket{r^{\pm}_{j}}$ is localized to the left of $x=j$, see Fig.~\ref{fig:FiguraInt02}(d).

Once we have the edge states written in terms of the basis of eigenstates of the quench Hamiltonian, we can easily compute their time evolution as:
\begin{widetext}
\begin{align}
&U(t)\ket{l^{\pm}_{1}}=\frac{1}{2i}\sum_{q}e^{-it\epsilon(q)/J}\bigg[G\left(-q\mp\frac{\pi}{2}\right)-G\left(q\mp\frac{\pi}{2}\right)\bigg]\ket{q},\label{eq:EdgeModeLeftTE1}\\
&U(t)\ket{r^{\pm}_{2N}}=\frac{e^{\pm i\pi(N+1/2)}}{2i}\sum_{q}e^{-it\epsilon(q)/J}\bigg[G\left(q\pm\frac{\pi}{2}\right)-G\left(-q\pm\frac{\pi}{2}\right)\bigg]\ket{q},\label{eq:EdgeModeRightTE1}
\end{align}
\end{widetext}
being the time $t$ expressed in units of $\hbar/J$. Here, we have used that $e^{\pm iq	(2N+1)}=1$, as $q=m\pi/(N+1/2)$ with $m=1,...,2N$.

Now, we use two important properties of the edge modes. On one hand, they are localized in momentum space at $\pm\pi/2$, where the second derivative of the quench Hamiltonian dispersion relation vanishes. On the other hand, they are well localized around that values, as we explained in Chapter 4. In consequence, their time evolution is governed by an almost linear dispersion relation, see Fig.~\ref{fig:FiguraInt03}. The first order Taylor expansion of the dispersion relation of the quench Hamiltonian at momentum $q=\pm\pi/2$ is:
\begin{equation}
\epsilon(q)=-2J\cos q=\pm2J\left(q\mp\frac{\pi}{2}\right)+\mathcal{O}\left(\left(q\mp\frac{\pi}{2}\right)^{3}\right),
\end{equation} 
and therefore we can approximate the dynamical phase in the time evolution as:
\begin{equation}
-it\epsilon(q)/J\approx\mp i2t\left(q\mp\frac{\pi}{2}\right).
\end{equation}
By substituting this approximation in Eq.~(\ref{eq:EdgeModeLeftTE1}) and Eq.~(\ref{eq:EdgeModeRightTE1}), we get the time evolution of $\ket{l^{\pm}_{1}}$ and $\ket{r^{\pm}_{2N}}$:
\begin{widetext}
\begin{align}
&U(t)\ket{l^{\pm}_{1}}=\frac{e^{i\pi t}}{2i}\sum_{q}\bigg[e^{\pm i2tq}\,G\left(-q\mp\frac{\pi}{2}\right)-e^{\mp i2tq}\,G\left(q\mp\frac{\pi}{2}\right)\bigg]\ket{q},\label{eq:EdgeModeLeftTE2}\\
&U(t)\ket{r^{\pm}_{2N}}=\frac{e^{\pm i\pi(N+\frac{1}{2})}e^{i\pi t}}{2i}\sum_{q}\bigg[e^{\pm i2tq}\,G\left(q\pm\frac{\pi}{2}\right)-e^{\mp i2tq}\,G\left(-q\pm\frac{\pi}{2}\right)\bigg]\ket{q},\label{eq:EdgeModeRightTE2}
\end{align}
\end{widetext}
We have used the approximation $\epsilon(q)\approx2J(q-\pi/2)$ in the terms with $G(-q+\pi/2)$ and $G(q-\pi/2)$, as they are concentrated around the momentum $\pi/2$; and the approximation $\epsilon(q)\approx-2J(q+\pi/2)$ in the terms with $G(-q-\pi/2)$ and $G(q+\pi/2)$, as they are localised at the momentum $-\pi/2$.

\begin{figure}[t]
  \centering
    \includegraphics[width=0.485\textwidth]{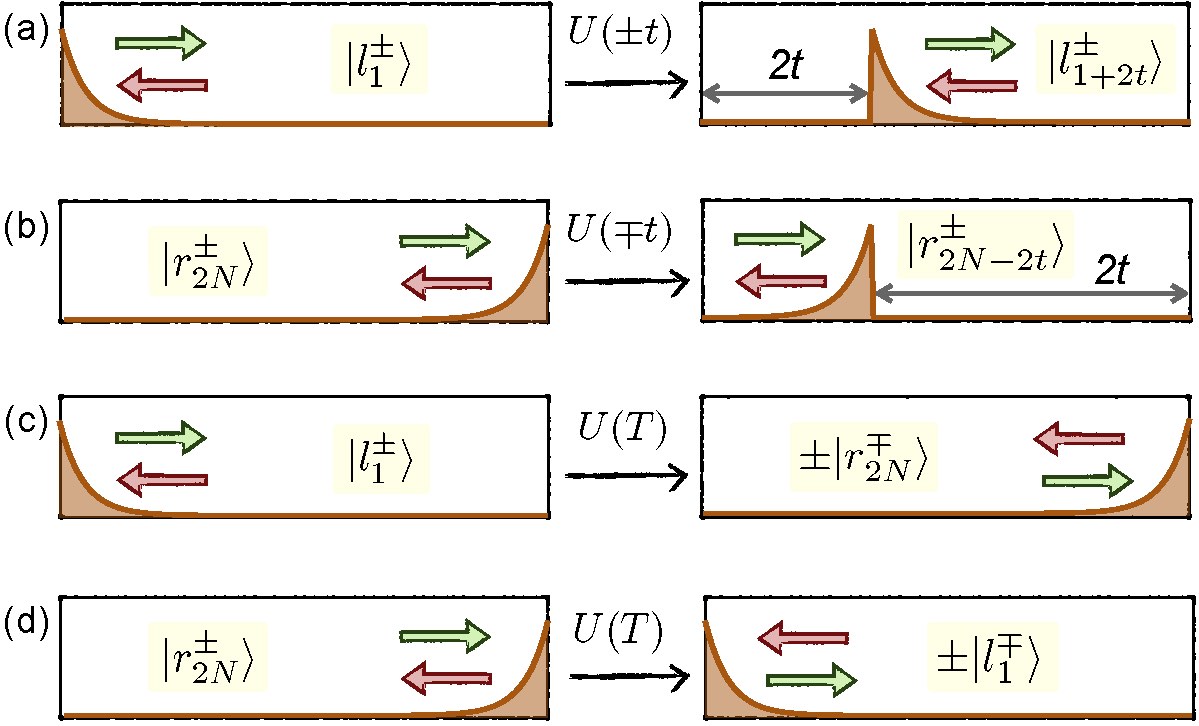}
    \caption[Time evolution of edge modes components.]{\textbf{Time evolution of edge modes components.} (a), (b) The states $\ket{l^{+}_{1}}$ and $\ket{r^{-}_{2N}}$ are localized at the left and right edges of the system and have momentum $\pi/2$ and $-\pi/2$, respectively. Therefore, they can propagate towards the bulk at constant velocities and we have that: $U(t)\ket{l^{+}_{1}}=\ket{l^{+}_{1+2t}}$ and $U(t)\ket{r^{-}_{2N}}=\ket{r^{-}_{2N-2t}}$. In contrast, the states $\ket{l^{-}_{1}}$ and $\ket{r^{+}_{2N}}$ have momentum towards the edges of the system, and thus they evolve from the states $\ket{l^{-}_{1+2t}}$ and $\ket{r^{+}_{2N-2t}}$, respectively. (c), (d) For an evolution time $T=N+1/2$, each state gets its spatial density distribution reflected with respect to the center of the chain and its momentum inverted. In this way, we have: $U(T)\ket{l^{\pm}_{1}}=\pm\ket{r^{\mp}_{2N}}$ and $U(T)\ket{r^{\pm}_{2N}}=\pm\ket{l^{\mp}_{1}}$.}
    \label{fig:FiguraInt04} 
\end{figure}

Comparing Eq.~(\ref{eq:EdgeModeLeftTE2}) with Eq.~(\ref{eq:StateLjinMomentumBasis}) we see that the state $\ket{l^{\pm}_{1}}$ becomes the state $\ket{l^{\pm}_{1+2t}}$ at time $\pm t$; and comparing Eq.~(\ref{eq:EdgeModeRightTE2}) with Eq.~(\ref{eq:StateRjinMomentumBasis}) we see that the state $\ket{r^{\pm}_{2N}}$ becomes the state $\ket{r^{\pm}_{2N-2t}}$ at time $\mp t$. That is:
\begin{align}
&U(\pm t)\ket{l^{\pm}_{1}}=\ket{l^{\pm}_{1+2j}},\label{eq:EvolRule1}\\
&U(\mp t)\ket{r^{\pm}_{2N}}=\ket{r^{\pm}_{2N-2t}}\label{eq:EvolRule2}.
\end{align}
In other words, the state $\ket{l^{+}_{1}}$, with momentum $\pi/2$ and localized at the left edge of the chain, moves towards the bulk after the quench and evolves \textit{into} $\ket{l^{+}_{1+2j}}$ [see Fig.~\ref{fig:FiguraInt04}(a)]. On the contrary, the state $\ket{l^{-}_{1}}$ cannot move in the direction of its momentum $-\pi/2$, as it is already at the left edge of the system. Therefore it evolves \textit{from} the state $\ket{l^{-}_{1+2t}}$ [see Fig.~\ref{fig:FiguraInt04}(a)]. Analogously, the two modes at the right edge of the system, $\ket{r^{+}_{2N}}$ and $\ket{r^{-}_{2N}}$ evolve \textit{from} $\ket{r^{+}_{2N-2t}}$ and \textit{into} $\ket{r^{-}_{2N-2t}}$, respectively [see Fig.~\ref{fig:FiguraInt04}(b)].
\begin{figure*}
  \centering
    \includegraphics[width=0.6\textwidth]{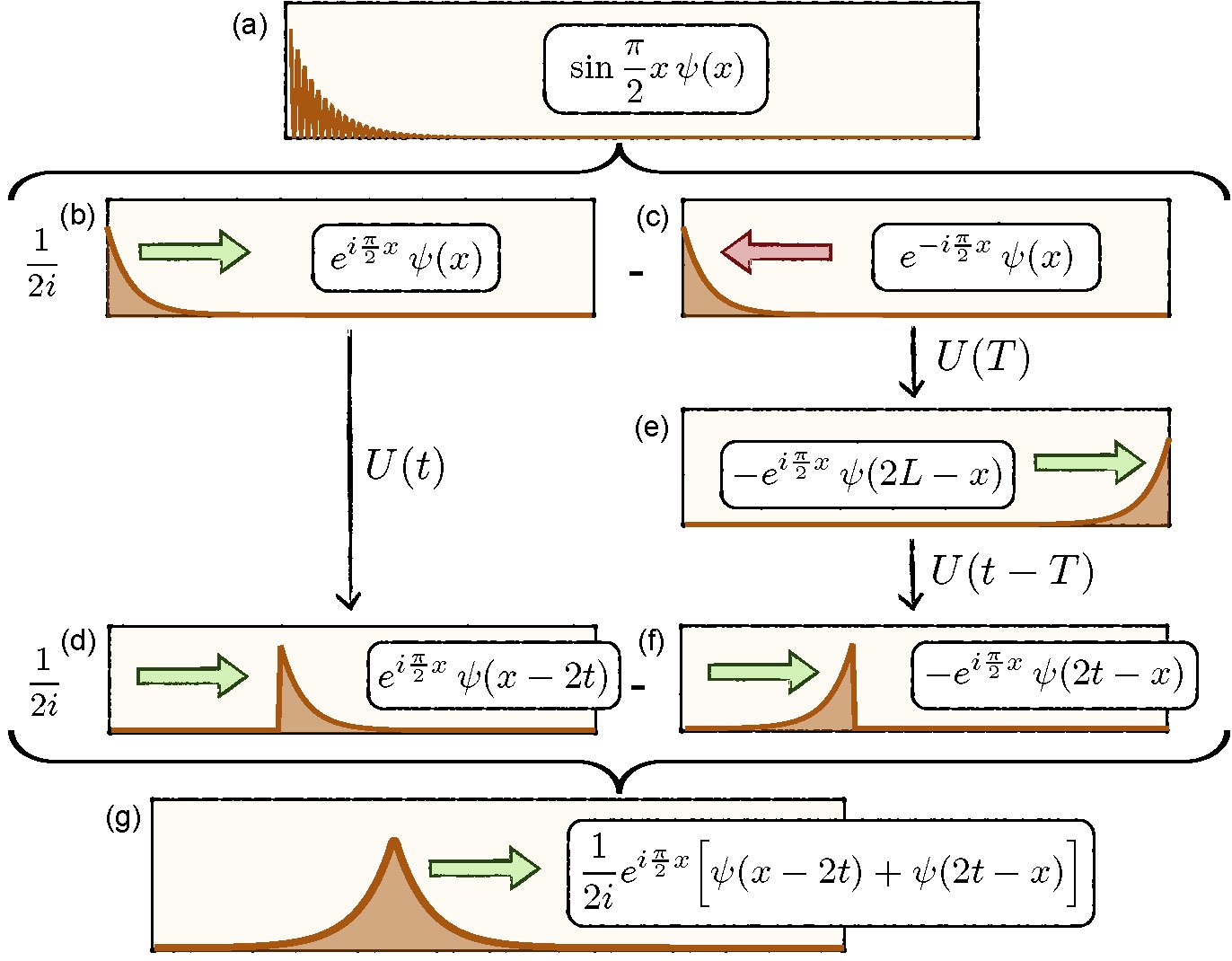}
    \caption[Edge modes free propagation.]{\textbf{Edge modes free propagation.} The initial edge mode located at the left side of the system, (a), is decomposed into two components, (b) and (c), of opposite momenta $\pm\pi/2$. The component with positive momentum, (b), moves to the right and gets its spatial density distribution shifted by a distance $2t$ at time $t$, (d). The component with negative momentum, (c), cannot move to the left as it is already at the edge of the system. We can picture its time evolution in two steps. First, it evolves during time $T$ and gets its spatial density distribution reflected and its momentum inverted, (e). Then, it evolves backwards in time travelling a distance $2(T-t)$ to the left, (f). Finally, we sum the two components and get a wave packet that moves freely through the system at constant velocity, (g). The time evolution of the state localized at the right edge of the system, $\ket{r}$, can be visualise in a similar way.}
    \label{fig:FiguraInt05} 
\end{figure*}
This results hold for $t$ smaller that $N$, so that the wave packets do not have enough time to reach the opposite edge of the chain. If we evaluate Eq.~(\ref{eq:EdgeModeLeftTE2}) and Eq.~(\ref{eq:EdgeModeRightTE2}) for an evolution time $T=N+1/2$, we get:
\begin{align}
&U(T)\ket{l^{\pm}_{1}}=\frac{(-1)^{N}}{2}\sum_{q}\bigg[G\left(-q\mp\frac{\pi}{2}\right)-G\left(q\mp\frac{\pi}{2}\right)\bigg]\ket{q},\label{eq:EdgeModeLeftTE3}\\
&U(T)\ket{r^{\pm}_{2N}}=\frac{\pm1}{2i}\sum_{q}\bigg[G\left(-q\pm\frac{\pi}{2}\right)-G\left(q\pm\frac{\pi}{2}\right)\bigg]\ket{q},\label{eq:EdgeModeRightTE3}
\end{align}
where we have used that $e^{\pm i2Tq}=1$ as $q=m\pi/(N+1/2)$ with $m=1,..,2N$. Comparing Eq.~(\ref{eq:EdgeModeLeftTE3}) and Eq.~(\ref{eq:EdgeModeRightTE3}) with Eq.~(\ref{eq:StateLjinMomentumBasis}) and Eq.~(\ref{eq:StateRjinMomentumBasis}), respectively, we conclude that:
\begin{align}
&U(T)\ket{l^{\pm}_{1}}=\pm\ket{r^{\mp}_{2N}},\label{eq:EvolRule3}\\
&U(T)\ket{r^{\pm}_{2N}}=\pm\ket{l^{\mp}_{1}},\label{eq:EvolRule4}
\end{align}
that is, after time $T$ each state becomes the state of opposite momentum located at the opposite edge of the system [see Fig.~\ref{fig:FiguraInt04}(c) and (d)].

Combining the results in Eq.~(\ref{eq:EvolRule1}), Eq.~(\ref{eq:EvolRule2}) and Eq.~(\ref{eq:EvolRule3}) we can compute the time evolution of the left edge mode:
\begin{align}
&U(t)\ket{l}\overset{a}{=}\frac{1}{2i}\left[U(t)\ket{l^{+}_{1}}-U(t-T)U(T)\ket{l^{-}_{1}}\right]\overset{b}{=}\nonumber\\
&\frac{1}{2i}\left[\ket{l^{+}_{1+2t}}+U(t-T)\ket{r^{+}_{2N}}\right]\overset{c}{=}\frac{1}{2i}\left[\ket{l^{+}_{1+2t}}+\ket{r^{+}_{2t-1}}\right]=\nonumber\\
&\frac{1}{2i}\sum_{x=1}^{2N}e^{i\frac{\pi}{2}x}\left[\psi(x-2t)+\psi(2t-x)\right]\hat{d}_{x}^{\dagger}\ket{0}.
\end{align}
Here, we have first (a) separated $\ket{l}$ into the two different momentum components $\ket{l^{+}_{1}}$ and $\ket{l^{-}_{1}}$, as well as decomposed the time evolution operator that acts on the negative momentum component as the product of two time evolution operators: $U(t)=U(t-T)U(T)$. Then (b) the positive momentum component $\ket{l^{+}_{1}}$ propagates during time $t$ and becomes $\ket{l^{+}_{1+2t}}$, whereas the negative momentum component $\ket{l^{-}_{1}}$ evolves during time $T$ and becomes $\ket{r^{+}_{2N}}$, getting its spatial density distribution reflected and its momentum inverted. Finally (c) the reflected component $\ket{r^{+}_{2N}}$ evolves backwards in time and becomes $\ket{r^{+}_{2t-1}}$. In this way, the two initial components form a wave packet centred at position $2t$ with momentum $\pi/2$ that moves towards the bulk at a constant velocity [see Fig.~\ref{fig:FiguraInt05}].

Analogously, the results in Eq.~(\ref{eq:EvolRule1}), Eq.~(\ref{eq:EvolRule2}) and Eq.~(\ref{eq:EvolRule4}) allow us to compute the time evolution of the right edge mode:
\begin{widetext}
\begin{align}
&U(t)\ket{r}=\frac{1}{2}\left[U(t-T)U(T)\ket{r^{+}_{2N}}+U(t)\ket{r^{-}_{2N}}\right]=\frac{1}{2}\left[U(t-T)\ket{l^{-}_{1}}+\ket{r^{-}_{2N-2t}}\right]=\nonumber\\
&\frac{1}{2}\left[\ket{l^{-}_{1+2(T-t)}}+\ket{r^{-}_{2T-2t}}\right]=\frac{1}{2}\sum_{x=1}^{2N}e^{-i\frac{\pi}{2}x}\left[\psi(x-2N-1+2t)+\psi(2N+1-2t-x)\right]\hat{d}_{x}^{\dagger}\ket{0}.
\end{align}
\end{widetext}
Therefore, the state $\ket{r}$ evolves in time forming a wave packet centred at position $x=2N+1-2t$ with momentum $-\pi/2$ that moves towards the bulk at constant velocity.

From Eq.~(\ref{eq:EvolRule3}) and Eq.~(\ref{eq:EvolRule4}) we can obtain the time evolution of both edge states, $\ket{l}$ and $\ket{r}$, at time $T$:
\begin{align}
U(T)\ket{l}=&\frac{1}{2i}U(T)\big(\ket{l^{+}_{1}}-\ket{l^{-}_{1}}\big)=\nonumber\\
&\frac{1}{2i}\big(\ket{r^{-}_{2N}}+\ket{r^{+}_{2N}}\big)=-i\ket{r},\\
U(T)\ket{r}=&\frac{1}{2}U(T)\big(\ket{r^{+}_{2N}}+\ket{r^{-}_{2N}}\big)=\nonumber\\
&\frac{1}{2}\big(\ket{l^{-}_{1}}-\ket{l^{+}_{1}}\big)=-i\ket{l},
\end{align}
As we see, each edge state becomes the other one after time $T$, so that the same process is repeated over and over.
\begin{figure*}
  \centering
    \includegraphics[width=0.999\textwidth]{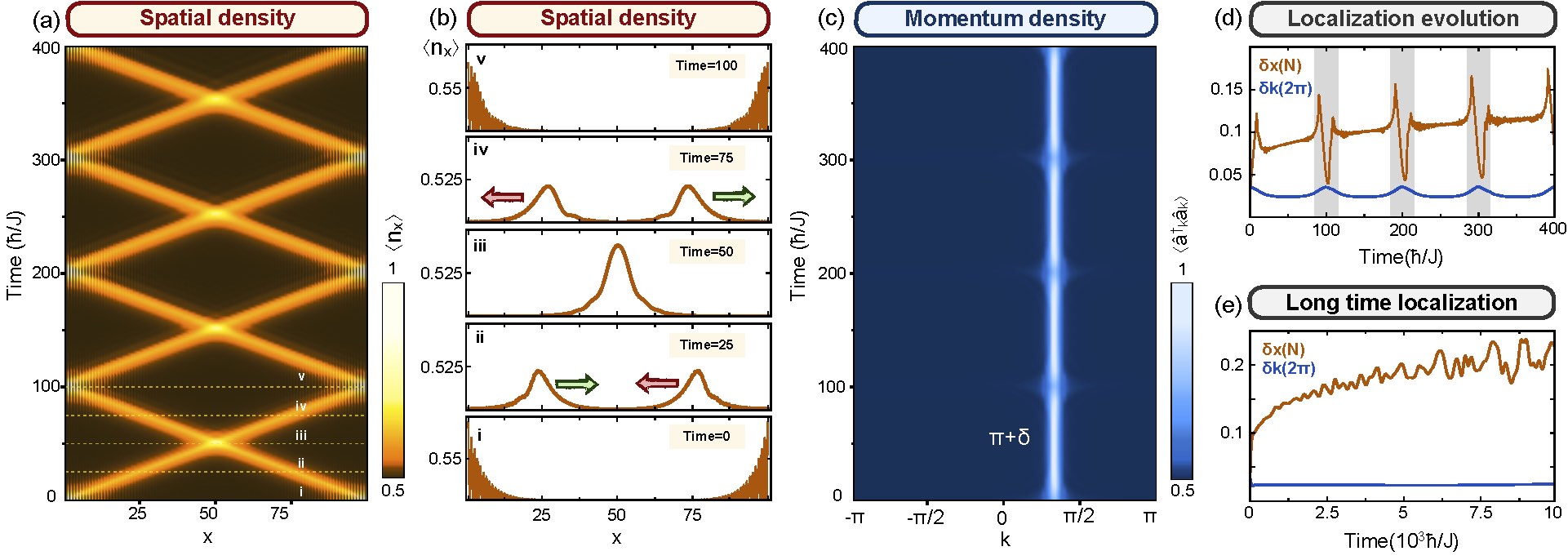}
    \caption[Interference of fractionalized quasiparticles.]{\textbf{Interference of fractionalized quasiparticles.} Time evolution of the many-body state with two fractional quasiparticles after a quench to the critical point. Parameters are $N=100$ and $\delta=-2\pi/3$. The Hamiltonian is quenched from $J'/J=0.9$ to $J'/J=1$. In position space, (a) and (b), the two quasiparticles keep their definition and move towards the bulk in opposite directions on top of a uniform and constant background of density $\nu_{x}=1/2$. They interfere at the center of the system, travelling back and forth with almost no dissipation. In momentum space, (c), quasiparticles keep their good localization at $\pi+\delta=\pi/3$. The spatial and momentum localization lengths remain almost constant in time, (d) and (e), with small oscillations occurring when quasiparticles bounce at the edges of the system (d).}
    \label{fig:FiguraInt06} 
\end{figure*}

\subsection{Interference of quasiparticles}

We have obtained two results: after performing a quench to the critical point \textit{i)} the uniform flat background density corresponding to the many body state $\ket{\Phi_{+}}$ remains constant, and \textit{ii)} the edge states move freely through the system with almost no dissipation, reaching the opposite edges of the system, bouncing and repeating the same process again and again. Combining these two results, we get the time evolution of the many body state $\ket{\Phi_{\text{qp}}}$ in which two quasiparticles exist at the edges of the system, and conclude that the quasiparticles preserve their identity during the time evolution. We have:
\begin{equation} 
\left[\tilde{d}_{x=2t}^{\dagger}\tilde{d}_{x=2t}-\nu_{x}\right]\ket{\Phi_{\text{qp}}(t)}=\frac{1}{2}\ket{\Phi_{\text{qp}}(t)},
\end{equation}
with:
\begin{equation}
\tilde{d}_{x=2t}^{\dagger}=\frac{1}{2i}\sum_{x=1}^{2N}e^{i\frac{\pi}{2}x}\left[\psi(x-2t)+\psi(2t-x)\right].
\end{equation}
Which states that at time $t$ there is a fractionalized quasiparticle with charge $1/2$ over the flat background density occupying the mode $\tilde{d}_{x=2t}^{\dagger}$, localized at position $x=2t$. Analogously, there is another fractionalized quasiparticle at position $x=2N+1-2t$.

In this way, the edge states $\ket{l}$ and $\ket{r}$, located initially at the left and right edges of the system [see Fig.~\ref{fig:FiguraInt06}(b)i], evolve in time forming two wave packets that move towards the bulk keeping their good spatial and momentum localizations [see Fig.~\ref{fig:FiguraInt06}(b)ii]. They meet at the center of the system, interfere constructively with each other [see Fig.~\ref{fig:FiguraInt06}(b)iii] and continue moving until they reach the opposite edge of the lattice [see Fig.~\ref{fig:FiguraInt06}(b)iv]. At this point the two initial states are restored [see Fig.~\ref{fig:FiguraInt06}(b)v], so that the same process is repeated over and over [see Fig.~\ref{fig:FiguraInt06}(a)].

This dynamics is a consequence of the simultaneous good spatial and momentum localizations of the edge modes, together with the fact that, in the critical point, the dispersion relation of the Hamiltonian is almost linear at momentum $k=\pi+\delta$, where the edge modes are located in momentum space.
The linear dispersion relation approximation holds for long evolution times, as the edge states remain localized at momentum $k=\pi+\delta$ [see Fig.~\ref{fig:FiguraInt06}(c)].

In Fig.~\ref{fig:FiguraInt06}(d) we show how the spatial and momentum localization lengths of the edge modes evolve in time. At the beginning of the time evolution the spatial localization length increases, as the wave packet that moves along the system consists of the sum of two edge mode wave functions [see Fig.~\ref{fig:FiguraInt04}(g)], and thus has a larger spatial extension. In consequence, the momentum localization length decreases, keeping the position-momentum uncertainty constant. This localization lengths remain almost constant, with oscillations every time the quasiparticles bounce at the edges of the system [grey regions in Fig.~\ref{fig:FiguraInt06}(d)]. For long evolution times, the edge modes remain well localized in momentum space, whereas their spatial localization gets worse. Nevertheless, the spatial localization length increases slowly in time, being $\delta x\approx0.2$ after the edge modes have travelled the whole system a hundred times [see Fig.~\ref{fig:FiguraInt06}(e)].\\

This research was funded by the Deutsche Forschungsgemeinschaft (DFG, German Research Foundation) via Research Unit FOR 2414 under project number 277974659.


\begin{thebibliography}{56}
\bibitem{Ryu2010} S.~Ryu, A.~P.~Schnyder, A.~Furusaki and A.~W.~W.~Ludwig, New J. Phys. \textbf{12}, 065010 (2010).
\bibitem{Altland1997} A.~Altland and M.~R.~Zirnbauer, Phys. Rev. B \textbf{55}, 1142 (1997).
\bibitem{Jackiw1976} R.~Jackiw and C.~Rebbi, Phys. Rev. D \textbf{13}, 3398 (1976).
\bibitem{Goldstone1981} J.~Goldstone and F.~Wilczek, Phys. Rev. Lett. \textbf{47}, 986 (1981).
\bibitem{Su1979} W.~P.~Su, J.~R.~Schrieffer and A.~J.~Heeger, Phys. Rev. Lett. \textbf{42}, 1698 (1979).
\bibitem{Heeger1988} A.~J.~Heeger, S.~Kivelson,  J.~R.~Schrieffer and W.~P.~Su, Rev. Mod. Phys. \textbf{60}, 781 (1988).
\bibitem{Sierra2014} G.~Sierra, J. Phys. A: Math. Theor. \textbf{47}, 325204 (2014).
\bibitem{Teo2010} J.~C.~Y.~Teo and C.~L.~Kane, Phys. Rev. B \textbf{82}, 115120 (2010).
\bibitem{Kivelson2002} S.~A.~Kivelson, Synthetic Metals \textbf{125}, 99 (2002).
\bibitem{Bloch2012} I.~Bloch J.~Dalibard and S.~Nascimb{\`e}ne, Nat. Phys. \textbf{8}, 267 (2012).
\bibitem{Bakr2009} W.~S.~Bakr, J.~I.~Gillen, A.~Peng, S.~F{\"o}lling and M. Greiner, Nature \textbf{462}, 74 (2009).
\bibitem{Sherson2010} J.~F.~Sherson, C.~Weitenberg, M.~Endres, M.~Cheneau, I.~Bloch and S.~Kuhr, Nature \textbf{467}, 68 (2010).
\bibitem{Jotzu2014} G.~Jotzu, M.~Messer, R.~Desbuquois, M.~Lebrat, T.~Uehlinger, D.~Greif and T.~Esslinger, Nature \textbf{515}, 237 (2014).
\bibitem{Aidelsburger2015} M.~Aidelsburger, M.~Lohse, C.~Schweizer, M.~Atala, J.~T.~Barreiro, S.~Nascimb{\`e}ne, N.~R.~Cooper, I.~Bloch and N.~Goldman, Nat. Phys. \textbf{11}, 162 (2015).
\bibitem{Duca2015} L.~Duca, T.~Li, M.~Reitter, I.~Bloch, M.~Schleier-Smith and U.~Schneider, Science \textbf{347}, 288 (2015).\bibitem{Flaeschner2016} N.~Fl{\"a}schner, B.~S.~Rem, M.~Tarnowski, D.~Vogel, D.-S.~L{\"u}hmann, K.~Sengstock and C.~Weitenberg, Science \textbf{352}, 1091 (2016).
\bibitem{Mancini2015} M.~Mancini, G.~Pagano, G.~Cappellini, L.~Livi, M. Rider, J.~Catani, C.~Sias, P.~Zoller, M.~Inguscio, M.~Dalmonte and L.~Fallani, Science \textbf{349}, 1510 (2015).
\bibitem{Stuhl2015} B.~K.~Stuhl, H.-I.~Lu, L.~M.~Aycock, D.~Genkina and I.~B.~Spielman, Science \textbf{349}, 1514 (2015).
\bibitem{Atala2013} M.~Atala, M.~Aidelsburger, J.~T.~Barreire, D.~Abanin, T.~Kitagawa, E.~Demler and I.~Bloch, Nat. Phys. \textbf{9}, 795 (2013).
\bibitem{Dalibard2011} J.~Dalibard, F.~Gerbier, G.~Juzeliunas and P.~{\"O}hberg, Rev. Mod. Phys. \textbf{83}, 1523 (2011).
\bibitem{Goldman2012} N.~Goldman, J.~Beugnon and F.~Gerbier, Phys. Rev. Lett. \textbf{108}, 255303 (2012).
\bibitem{Goldman2013} N.~Goldman, J.~Beugnon and F.~Gerbier, Eur. Phys. J. Special Topics \textbf{217}, 135 (2013).
\bibitem{Goldman2013b} N.~Goldman, J.~Dalibard, A.~Dauphin, F.~Gerbier, M.~Lewenstein, P.~Zoller and I.~B.~Spielman, PNAS \textbf{110}, 6736 (2013).
\bibitem{Delplace2011} P.~Delplace, D.~Ullmo and G.~Montambaux, Phys. Rev. B \textbf{84}, 195452 (2011).
\bibitem{Aidelsburger2011} M.~Aidelsburger, M.~Atala, S.~Nascimb{\`e}ne, S.~Trotzky, Y.-A.~Chen and I.~Bloch, Phys. Rev. Lett. \textbf{107}, 255301 (2011).
\bibitem{Aidelsburger2013} M.~Aidelsburger, M.~Atala, M.~Lohse, J.~T.~Barreiro, B.~Paredes and I.~Bloch, Phys. Rev. Lett. \textbf{111}, 185301 (2013).
\bibitem{Buchhold2012} M.~Buchhold, D.~Cocks and W.~Hofstetter, Phys. Rev. A \textbf{85}, 063614 (2012).

\end{thebibliography}
\end{document}